\documentclass[final,5p,times,twocolumn]{elsarticle}

\usepackage{amssymb}
 \usepackage{amsthm}
 \usepackage{amsmath}

 \usepackage{lineno}

\usepackage[normalem]{ulem}

\usepackage{transparent}

\usepackage{booktabs}
\renewcommand{\arraystretch}{1.2}
\newcommand{\ra}[1]{\renewcommand{\arraystretch}{#1}}

\usepackage{longtable}

\usepackage{multirow}

\usepackage{enumitem}

\usepackage{float}

\usepackage{hyperref}

\usepackage{cuted}

\usepackage{xcolor}
\newcommand{\CBa}{\color{black}}
\newcommand{\CBb}{\color{black}}

\newcommand{\CBd}{\color{black}}
\newcommand{\CBe}{\color{black}}
\newcommand{\CBf}{\color{black}}
\newcommand{\CBg}{\color{black}}
\newcommand{\CBh}{\color{black}}

\usepackage{pdflscape}
\usepackage{rotating}
\usepackage{rotating}

\usepackage{subcaption}

\newtheorem{remark}{Remark}
\newtheorem{example}{Example}

\journal{Annual Reviews in Control}

\usepackage{caption}

\begin{document}

\begin{frontmatter}



\title{Partitioning techniques for non-centralized predictive control: \\ A systematic review {\CBe and novel theoretical insights}\tnoteref{label1}}
\tnotetext[label1]{This project has received funding from the European Research Council (ERC) under the European Union's Horizon 2020 research and innovation program (Grant agreement No. 101018826) -- Project CLariNet.}


\author[a]{Alessandro Riccardi\corref{label2}}
\author[a]{Luca Laurenti}
\author[a]{Bart De Schutter}
\affiliation[a]{organization={ Delft Center for Systems and Control (DCSC), Delft University of Technology },
	city={Delft},
	country={The Netherlands}}
\cortext[label2]{Corresponding author: \tt{a.riccardi@tudelft.nl}}
\begin{abstract} 
    The partitioning problem is of central relevance for designing and implementing non-centralized Model Predictive Control (MPC) strategies for large-scale systems. These control approaches include decentralized MPC, distributed MPC, hierarchical MPC, and coalitional MPC. Partitioning a system for the application of non-centralized MPC consists of finding the best definition of the subsystems, and their allocation into groups for the definition of local controllers, to maximize the relevant performance indicators. The present survey proposes a novel systematization of the partitioning approaches in the literature in five main classes: optimization-based, algorithmic, community-detection-based, game-theoretic-oriented, and heuristic approaches. A unified graph-theoretical formalism, a mathematical re-formulation of the problem in terms of mixed-integer programming, the novel concepts of predictive partitioning and multi-topological representations, and a methodological formulation of quality metrics are developed to support the classification and further developments of the field. We analyze the different classes of partitioning techniques, and we present an overview of their strengths and limitations, which include a technical discussion about the different approaches. Representative case studies are discussed to illustrate the application of partitioning techniques for non-centralized MPC in various sectors, including power systems, water networks, wind farms, chemical processes, transportation systems, communication networks, industrial automation, smart buildings, and cyber-physical systems. An outlook of future challenges completes the survey.    
\end{abstract}




\end{frontmatter}

\tableofcontents

%



\section{Introduction}


{
\subsection{Motivation}
Modern systems are increasingly characterized by architectural scales and implementation complexities that challenge the implementation of centralized control strategies \cite{siljak_LargeScaleDynamicSystems_2008,kordestani_RecentSurveyLargescale_2021}. This trend is supported by the advancements and availability of information transmission networks, as well as by the wide accessibility of computing resources \cite{kamel_UltraDenseNetworksSurvey_2016}. {\CBf When the scale of a system grows, it is common and advisable to structure it as a collection of autonomous interconnected components (subsystems). These subsystems should coordinate or be coordinated to achieve a common goal.} To this aim, these entities necessitate local computing power, and communication and negotiation abilities: this is why, when these features are available, {\CBf these advanced subsystems are usually defined as control agents. A schematic representation of a network of control agents is proposed in Fig.\ \ref{fig:Agents}}. Consequently, modern systems constituted by multiple agents having scales that exceed specific (hardware) operational thresholds are commonly referred to as large-scale multi-agent systems (LS-MASs) \cite{dorri_MultiagentSystemsSurvey_2018}. Examples of LS-MASs can be found in infrastructural systems such as power generation and distribution networks \cite{kundur_PowerSystemStability_2022,javid_FutureDistributionNetworks_2024,rakhshani_IntegrationLargeScale_2019,poullikkas_ComparativeOverviewLargescale_2013}; urban and freeway networks \cite{siri_FreewayTrafficControl_2021}; railway and subway networks \cite{louf_ScalingTransportationNetworks_2014}; water distribution networks \cite{bello_SolvingManagementProblems_2019}; oil and gas distribution networks; large groups of mobile robots such as swarms of UAVs \cite{zhou_UAVSwarmIntelligence_2020}, or of terrestrial and maritime autonomous vehicles; large plants for chemical processing \cite{metzger_SurveyApplicationsAgent_2011}, which might also integrate autonomous energy generation; large industrial networks \cite{galloway_IntroductionIndustrialControl_2013}; and satellite constellations \cite{curzi_LargeConstellationsSmall_2020}; where this list of applications keeps growing and evolving with the introduction of new technologies.

{\CBf Conventional control methodologies such as proportional-integral control and pole placement \cite{ogata_ModernControlEngineering_2022}, loop-shaping and h-infinity synthesis, \cite{skogestad_MultivariableFeedbackControl_2001}, or feedback linearization \cite{khalil_NonlinearSystems_2002}} are not directly applicable to LS-MASs because of the presence of a large number of input-output channels and the large spatial distribution of such networks, which complicate centralized controller design and parameter tuning. 
Therefore, deployment of non-centralized control strategies \cite{siljak_DecentralizedControlComplex_1991,bakule_DecentralizedControlOverview_2008a} is necessary for LS-MASs, and the level of sophistication of such approaches is tightly related to the availability of reliable communication channels and local computing power. 

\begin{figure}[t]
	\centering
	\includegraphics[width=.7\linewidth]{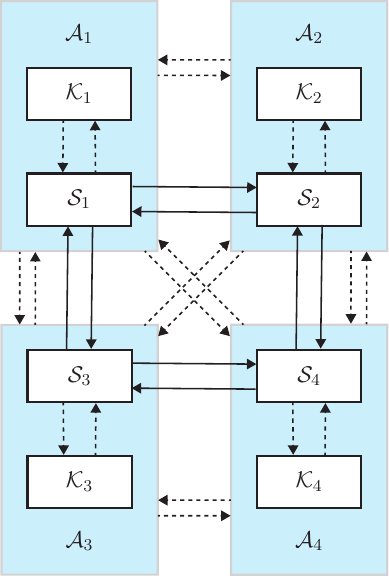}
	\caption{\CBf A network of control agents. Subsystems are indicated by $\mathcal{S}$, local controllers by $\mathcal{K}$, and control agents by $\mathcal{A}$. The solid lines represent the interactions at the physical level, i.e.\ the dynamical couplings; instead, the dashed lines represent interactions at the information level.}
	\label{fig:Agents}
\end{figure}

\begin{figure*}[t]
	\centering
	\includegraphics[width=.95\linewidth]{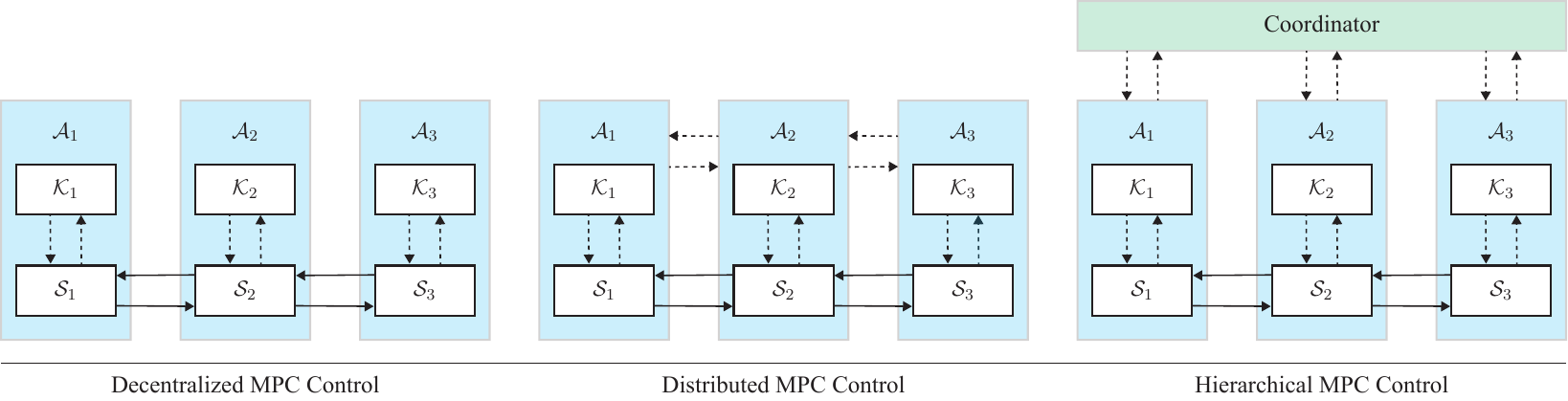}
	\caption{\CBf Main categories of non-centralized control architectures. In decentralized control, there is no information-level interaction among control agents. In distributed control, the information-level interaction is horizontal, i.e.\ each control agent can communicate with the others. In hierarchical control, the information-level interaction among control agents is vertical, i.e.\ they should, in principle, communicate with the coordinator. Mixed approaches are also possible.
    }
	\label{fig:Non-centralized_Architectures}
\end{figure*}

\subsection{Non-centralized MPC: Control architectures}

One of the most advanced modern control strategies is model predictive control (MPC) \cite{rawlings_ModelPredictiveControl_2017b}, which integrates the use of a mathematical model of the system dynamics with optimal control methodologies to compute predictive control actions that optimize performance while guaranteeing the stability of the controlled system, as well as the respect of operational constraints \cite{machowski_PredictiveControlConstraints_2002,mesbah_StochasticModelPredictive_2016}, according to the receding horizon paradigm. The MPC framework has also significantly evolved thanks to its design flexibility, which allows a relatively easy development of non-centralized predictive control strategies (NCen-MPC) \cite{christofides_DistributedModelPredictive_2013b,maestre_DistributedModelPredictive_2014}, i.e.\ of MPC strategies in which the computation of the control action for the overall system is not performed by a single central unit, but divided across control agents. The traditional classification of these strategies \cite{scattolini_ArchitecturesDistributedHierarchical_2009} comprehends decentralized MPC (Dec-MPC), distributed MPC (DMPC), and hierarchical MPC (HMPC). A conceptual representation of these architectures is proposed in Fig.\ \ref{fig:Non-centralized_Architectures}. More recently, a novel NCen-MPC methodology incorporating concepts from game theory has emerged, called coalitional predictive control (Coal-MPC) \cite{fele_CoalitionalControlCooperative_2017b}. In this survey, we will abbreviate centralized MPC as CMPC, to distinguish it from NCen-MPC. A list of these abbreviations is reported in Tab.\ \ref{tab:abbreviations}. The single common characteristic of all NCen-MPC approaches is that they assume to operate in a network of agents, where, for each individual subsystem, a local optimization problem is solved. Then, the various techniques are distinguished according to how they handle communication and coordination of the local control actions. 


When referring to NCen-MPC techniques, the simplest coordination technique is Dec-MPC, in which there is no communication among agents, but the effect of neighboring subsystems on local dynamics is generally assumed to be contained in invariant sets, thus allowing stable operation of such networks while preserving privacy, security, and resilience since there is no information sharing. A communication and coordination protocol is instead at the basis of DMPC approaches, where the agents in the network usually share their measurements or predicted evolution of local variables with neighbors, thus allowing for iterative or non-iterative adjustments of local control actions. In the context of linear systems, this distributed control approach can achieve global performance close to CMPC while drastically reducing computation times, and allowing real-time operation of the networks where centralized predictive control would not be possible. In HMPC, the control architecture is structured across multiple vertical layers, with at least the presence of a global coordinator and a set of local controllers. These strategies pose as an alternative to DMPC, and can enhance global coordination, as well as network resilience, introduce privacy features, or allow for multi-time-scale operation of different network models at different aggregation layers. Finally, the Coal-MPC strategy arises as the result of the combination of predictive control with game theory. In fact, in Coal-MPC, the network is seen as a collection of agents that participate in a cooperative game with the objective of maximizing the global collective outcome, which is the global operation cost of the network. 

In conclusion, NCen-MPC strategies allow for the introduction of complex control features, such as advanced algorithmic coordination procedures, plug-and-play capabilities, and privacy and security preservation strategies, into LS-MASs. At the same time, NCen-MPC strategies can ensure stable real-time control of LS-MASs while preserving the optimality of their operation as much as possible.

\subsection{The partitioning problem}

The underlying assumption of the above discussion about NCen-MPC of LS-MASs is that the network is provided as a collection of agents with full autonomy. While this assumption may seem simple to satisfy, this is not always true in practice. In fact, the definition of the agents themselves may be challenging, especially for large and interconnected networks. Additionally, even if the network is given as a collection of individual agents, it might be more convenient for network operation to aggregate them into bigger entities. {\CBf These two distinct classes of problems, i.e.\ the definition of the agents of the network and the problem of their aggregation,} fall both into the category of network partitioning \cite{siljak_DecentralizedControlComplex_1991,chanfreut_SurveyClusteringMethods_2021}.

Formally speaking, the partitioning problem consists of finding the optimal allocation of a group of elements into given sets according to a given metric. If the network $\mathcal{N}$ is provided as a collection of agents $N_{\mathcal{A}}$, i.e.\ $\mathcal{N} = \{\mathcal{A}_1,\ldots,\mathcal{A}_{N_{\mathcal{A}}}\}$, and we have a number $N_{\mathcal{C}}$ of possible sets for the allocation, whose quality is defined by a cost function $h(\cdot)$, then the optimal partitioning problem consists in finding the set $\mathcal{P}$ (i.e.\ the partition) defined as $\mathcal{P} = \{\mathcal{C}_1,\ldots,\mathcal{C}_{N_{\mathcal{C}}}\}$, where the elements $\mathcal{C}_i$ are groups of agents $\mathcal{A}_j$, such that the quality measure $h(\mathcal{P})$ is optimized. {\CBf On the other hand, if the network $\mathcal{N}$ is provided as a monolithic system that does not show any natural decomposition, the partitioning problem consists of selecting several subsystems $N_{\mathcal{A}}$ for which control agents can be defined, which allows to interpret the network as a collection of agents $\mathcal{N} = \{\mathcal{A}_1,\ldots,\mathcal{A}_{N_{\mathcal{A}}}\}$. Also in this case, the subsystem selection is generally guided by a cost function $h(\cdot)$}. Both these problems are known to be NP-hard \cite{karp_ReducibilityCombinatorialProblems_1972,sandholm_CoalitionStructureGeneration_1999,brandes_MaximizingModularityHard_2006}.

When the partitioning problem is applied to NCen-MPC, several further features can be developed and extended for both the partitioning and the MPC. Many questions may arise, such as: What is the best definition for the individual agents? How can agents be allocated optimally into sets to maximize the performance of the NCen-MPC architecture? How can the partitioning strategy handle topological changes in the network or different operating conditions? These are a few examples of profound technical challenges that researchers in this field have encountered in the last decades, finding answers and new open problems. 

Many of the partitioning strategies that will be presented in this survey are borrowed from other scientific sectors, such as network and graph theory, machine learning, or computer science in general. A general overview of clustering methodologies applied to distributed network control can be found in \cite{chanfreut_SurveyClusteringMethods_2021}, which can serve as a general reference for these methods, while the current survey is tailored specifically for NCen-MPC. We also refer to the work \cite{xu_SurveyClusteringAlgorithms_2005} to explore further general clustering methodologies such as $k$-means, fuzzy $c$-means, and hierarchical clustering. Other general approaches that have been applied to partitioning for NCen-MPC are community detection methodologies \cite{fortunato_CommunityDetectionGraphs_2010a,fortunato_CommunityDetectionNetworks_2016}, such as modularity maximization and spectral algorithms; and coalition formation approaches \cite{apt_GenericApproachCoalition_2009}, which have led to the development of game-theory-based MPC architectures.

\subsection{Survey objectives and contributions}

Under these considerations, the present survey has two main overarching goals:
\begin{enumerate}
    \item Unifying in a common framework all the results currently present in the literature addressing the partitioning problem for NCen-MPC.
    \item Laying foundations for further systematic developments of this field.
\end{enumerate}
These two objectives are achieved through the following series of steps: a systematization of fundamental notions for graph representation of dynamical systems and networks; the introduction of precise key performance indicators that are comparable across strategies and application domains, as well as a precise assessment methodology of the quality of a partition; a categorization of the known partitioning strategies for NCen-MPC in terms of methodology, partitioning objective, and relative control strategy; a discussion of the main partitioning methodologies to highlight their strengths and limitations; a brief technical discussion of each partitioning technique found in the literature; and a classification of the current application domains of the partitioning techniques.

{\CBf Further, we extend the analysis and classification of the partitioning techniques with novel theoretical insights, which are: the introduction of multi-topological graph representations to model variable topologies, and their link to hybrid systems; a formal definition of the partitioning problem for performance optimization in terms of a bi-level mixed-integer program (MIP); and a re-definition of the problem of time-varying partitioning, introducing the concept of predictive partitioning for control.}


Given the extension of this survey, and the amount of different topics explored in detail, we provide an overview of its organization in Sec.\ \ref{sec:organization} below, briefly describing the contents and the objectives of each section. 


\begin{table}[t]\CBf 
	\centering
	\ra{1.3}
		\begin{tabular}{l|l}
        \toprule[1.5pt]
        MPC & Model Predictive Control \\
        CMPC & Centralized Model Predictive Control \\
        NCen-MPC & Non-Centralized Model Predictive Control \\
        Dec-MPC & Decentralized Model Predictive Control \\
        DMPC & Distributed Model Predictive Control \\
        HMPC & Hierarchical Model Predictive Control \\
        Coal-MPC & Coalitional Model Predictive Control \\
        NLin-MPC & Nonlinear Model Predictive Control \\
        LS-MAS & Large-Scale Multi-Agent System \\
        MIMO & Multiple-Input Multiple-Output \\
        \bottomrule[1.5pt]
		\end{tabular}%
	\caption{List of abbreviations}
	\label{tab:abbreviations}
\end{table}



}



%
{
\section{Organization of the Survey}\label{sec:organization}

In this section, we present the structure of the survey, briefly describing the content of each section. This will provide the reader with an organic view of the material presented, and will help to navigate the content, having a general knowledge of all the topics that will be discussed throughout the survey. 
	
\paragraph{Graph representations}
	
Most partitioning techniques, both involving NCen-MPC or other control strategies, are based on abstract representations of the underlying system dynamics \cite{siljak_DecentralizedControlComplex_1991}. This representation is generally provided in the form of a graph \cite{diestel_GraphTheory_2017}; therefore, it is natural to start the discussion about partitioning techniques by introducing graph representations in Sec.\ \ref{sec:graph-representations}. {\CBf In this section, we classify the graph representations used in partitioning and presented in Fig.\ \ref{fig:graph-representations}. This classification is supported by a technical discussion of each type of representation in dedicated subsections.}

\paragraph{Partitioning for predictive control}

Once the abstract representation of the network is available, the partitioning problem for NCen-MPC can be formally introduced and discussed in Sec.\ \ref{sec:partitioning-predictive-control}. {\CBf In this section, we discuss the general problem definition and its common characteristics usually present in the partitioning techniques. In addition, we provide metrics and an evaluation methodology to assess the quality of a partition, and we complete the discussion by introducing the novel concept of predictive partitioning as a component of the MPC formulation.}


\begin{figure}[t]
	\centering
	\includegraphics[width=.7\linewidth]{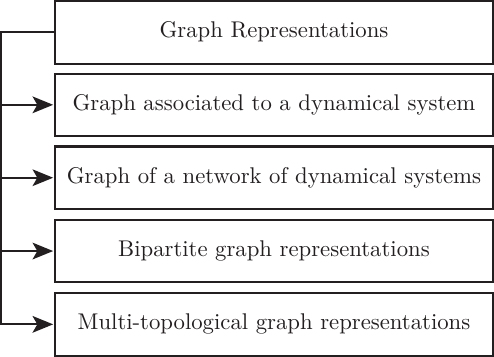}
	\caption{Graph representations used in partitioning for NCen-MPC.}
	\label{fig:graph-representations}
\end{figure}

\paragraph{Classification of the partitioning techniques} In Sec.\ \ref{sec:analysis}, we will provide a classification of the partitioning methodologies for the application of NCen-MPC according to three criteria: 1) the general partitioning class; 2) the subclass defined by the main structure of the method or by its objective; and 3) the control architecture to which it has been applied. The classification performed according to the first two criteria is proposed in Fig.\ \ref{fig:classification-partitioning}, where the first level of the classification tree defines the main class, and the second level defines the subclass. {\CBf The main theoretical characteristics as well as the strengths and limitations of the five main partitioning classes are discussed in Sec.\ \ref{subsec:classification-main-branch}, for the subclasses in Sec.\ \ref{subsec:classification-sub-class}, and for the methodologies in Sec.\ \ref{sucsec:classification-methodology}. Finally, in Sec.\ \ref{sucsec:classification-control} we classify the techniques according to the control methodology for which they have been designed.}

\begin{figure*}[t]
	\centering
	\includegraphics[width=\linewidth]{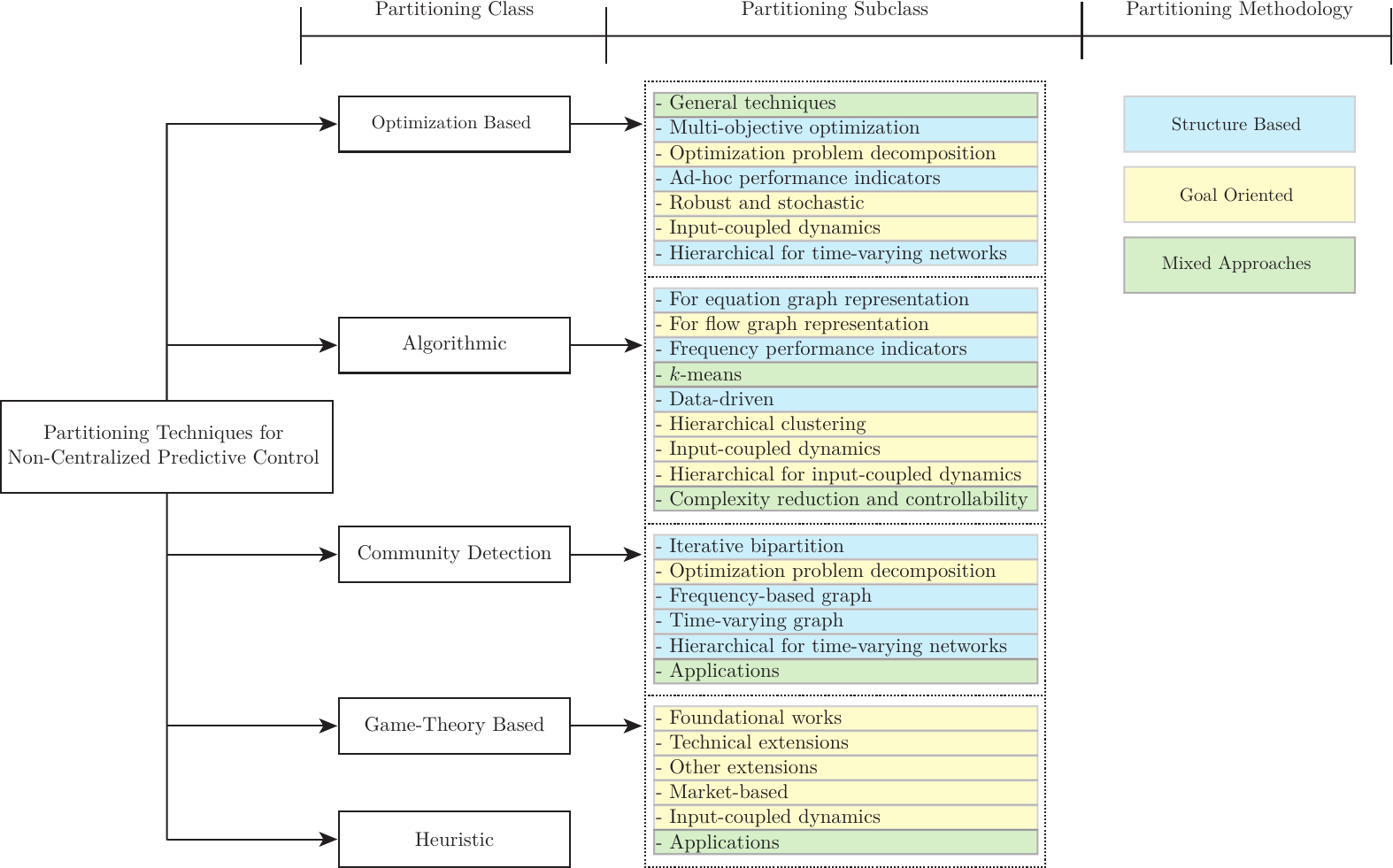}
	\caption{Categorization of the partitioning techniques in classes and subclasses. {\CBf The methodologies in each subclass can be further distinguished between the approaches based on the structure of the network, and the ones oriented at achieving a given objective, whether it is a control or another functional specification. }}
	\label{fig:classification-partitioning}
\end{figure*}

\paragraph{Analysis of the individual partitioning techniques}

Once the classification of the partitioning strategies has been presented, and the main characteristics of each class and subclass have been highlighted, we deepen the technical discussion by providing further details about the methods in each class in Fig.\ \ref{fig:classification-partitioning}. Therefore, an extensive analysis of the individual methodologies in the literature can be found in the dedicated sections, which are: Sec.\ \ref{sec:optimization-based} for optimization-based partitioning; Sec.\ \ref{sec:algorithmic} for algorithmic partitioning; Sec.\ \ref{sec:modularity-based} for community-detection-based partitioning; Sec.\ \ref{sec:coalitional} for game-theory-based partitioning; and Sec.\ \ref{sec:heuristic} for heuristic partitioning.  

\paragraph{Applications}

In Sec.\ \ref{sec:applications}, we discuss the main case studies that have been used in the literature about partitioning for NCen-MPC. These are divided by application sector, and, when possible, we also provide reference systems with further details about the systems considered. 

\paragraph{Conclusions and future work} The overall discussion of the main topic of the survey is completed in Sec.\ \ref{sec:conclusions} with final considerations about the state of this research field, and with recommendations for future work,
identifying the current research gaps and potential new directions to explore.

}
\section{Graph Representations} \label{sec:graph-representations}
{\CBd 
At the basis of almost all partitioning approaches, there is a graph representation of the system to be decomposed. Accordingly, specific graph representations can be deployed when defining a partitioning strategy for applying an NCen-MPC method. These representations belong to three main categories: 1) graphs equivalent to dynamical systems; 2) graphs representing networks of dynamical systems; and 3) graph representations of an optimization problem. In this section, we first introduce graph theory terminology that will be used throughout the article. Then, we present the classes of graphs introduced above. We close the section by conceptually reformulating the graph representation of a network of dynamical systems linking multi-topological graphs and hybrid systems.
}

\subsection{Fundamentals of graph theory} \label{subsec:graph-terminology} 
{\CBd
A graph \cite{diestel_GraphTheory_2017} is an ordered pair of sets $\mathcal{G}=(\mathcal{V},\mathcal{E})$ where $\mathcal{V} = \{1,\ldots, n\}$ is the set of $n$ vertices (or nodes), and  $\mathcal{E}\subseteq\mathcal{V}\times\mathcal{V}$ is the set of the edges (or arcs, links). The edges are associated to the vertices through an {\CBe $n\times n$} binary adjacency matrix $A^{\text{adj}}$, where $A^{\text{adj}}_{(i,j)} = 1$ if and only if an edge $\epsilon_{ij} = (i,j) \in\mathcal{E}$ exists. Therefore, the topology of the graph is specified by the adjacency matrix $A^{\text{adj}}$, and the set of the edges can also be written as $\mathcal{E} = \{(i,j)\,|\,[i,j\in\mathcal{V}]\wedge [A^{\text{adj}}_{(i,j)} = 1]\}$. A subgraph of $\mathcal{G}$ is a graph $\mathcal{S}_\ell=(\mathcal{V}_\ell,\mathcal{E}_\ell)$ representing a part of $\mathcal{G}$. The set of vertices $\mathcal{V}_\ell$ is a subset of $\mathcal{V}$, i.e.\ $\mathcal{V}_\ell\subseteq\mathcal{V}$, and the set of the edges is $\mathcal{E}_\ell = \{(i,j)\,|\,[i,j\in\mathcal{V}_\ell]\wedge [A^{\text{adj}}_{(i,j)} = 1]\}$, where the topology is still specified by the relevant entries of $A^{\text{adj}}$. For a directed graph $\mathcal{G}$, an edge $\epsilon_{ij} = (i,j)$ denotes an arrow starting from node $i$ and ending in node $j$. A graph $\mathcal{G}$ is weighted if a weighting matrix $W^{\text{adj}}$ assigning to each edge a number is specified in addition to $A^{\text{adj}}$. For each vertex $i\in\mathcal{V}$ we denote by $d_i$ its degree, i.e.\ the number of edges entering or exiting that vertex. In directed graphs, we can specify an in-degree ($d_{i}^{\text{in}}$) and an out-degree ($d_{i}^{\text{out}}$), if the edge is respectively ending or starting in the vertex $i$. For a vertex $i$, the neighborhood of $i$ is the set of all vertices connected to it, and we denoted it by $\mathcal{N}_i = \{j\in\mathcal{V} \,\,|\,\,[(i,j)\vee(j,i)]\in\mathcal{E}\}$. For a subgraph $\mathcal{S}_\ell=(\mathcal{V}_\ell,\mathcal{E}_\ell)$, the frontier is its set of nodes that are connected to nodes outside the subgraph, i.e.\ $\mathcal{F}_\ell = \{i\in\mathcal{V}_\ell\,\,|\,\,[(i,j)\vee (j,i)]\in\mathcal{E}, j\in\mathcal{V}\setminus\mathcal{V}_\ell\}$. These fundamental concepts will be extended throughout the survey for specific topics when necessary.}

\subsection{Graph associated to a dynamical system} \label{subsec:equivalent-graph} 
{\CBd 
The most direct and intuitive graph representation of a dynamical system is the so-called associated graph. According to \cite{siljak_DecentralizedControlComplex_1991}, the earliest formulation of this type of graph representation for linear systems can be traced back to \cite{lin_StructuralControllability_1974}.
We start by presenting associated graph representations for linear discrete-time systems, where the same formulation proposed in \cite{siljak_DecentralizedControlComplex_1991} for the continuous-time version holds. Consider the dynamics:
\begin{equation} \label{eq:linear-system}
	\mathcal{S}: \left\{
	\begin{matrix}
		x(k+1) = A x(k) + B u(k) \\
		y(k) = C x(k) \hfill 
	\end{matrix}
	\right.
\end{equation}
where $x\in\mathbb{R}^{n_x}$, $u\in\mathbb{R}^{n_u}$, $y\in\mathbb{R}^{n_y}$ are respectively the state, input, and output of the system; and $A$, $B$, $C$ are matrices of appropriate dimensions. The graph $\mathcal{G}=(\mathcal{V},\mathcal{E})$ associated to \eqref{eq:linear-system} is constructed by first defining one node for each variable, which provides the set of vertices $\mathcal{V} = \{x_1,\ldots,x_{n_x},u_1,\ldots,u_{n_u},y_1,\ldots,y_{n_y}\}$, where this set can be considered as the union of the sets for the individual state, input, and output variables, i.e.\ $\mathcal{V} = \mathcal{V}_x \cup \mathcal{V}_u \cup\mathcal{V}_y$, $|\mathcal{V}| = n_x+n_u+n_y$. Then, the set of edges $\mathcal{E}$ is built looking at the nonzero entries of matrices $A$, $B$, $C$, and as before, it can be thought of as the union of three different sets $\mathcal{E} = \mathcal{E}_{ux} \cup \mathcal{E}_{xx} \cup \mathcal{E}_{xy}$. These sets of edges define the interactions among variables, and are derived respectively as $\mathcal{E}_{ux} = \{(i,j)\,\,|\,\,i\in\mathcal{V}_u,j\in\mathcal{V}_x,B_{(i,j)}\neq0\}$, $\mathcal{E}_{xx} = \{(i,j)\,\,|\,\,i,j\in\mathcal{V}_x,A_{(i,j)}\neq0\}$, $\mathcal{E}_{xy} = \{(i,j)\,\,|\,\,i\in\mathcal{V}_x,j\in\mathcal{V}_y,C_{(i,j)}\neq0\}$. This graph $\mathcal{G}$ associated with the dynamics \eqref{eq:linear-system} is static because the dynamical system is time-invariant. Moreover, the graph represents the interactions among the variables in the system. A measure of this interaction is provided by the weighting matrix that can be constructed considering the entries of matrices $A$, $B$, $C$: 
\begin{equation}\label{eq:weighting-matrix}
	W^{\textrm{adj}} = \begin{bmatrix}
		A & B & 0 \\
		0 & 0 & 0 \\
		C & 0 & 0
	\end{bmatrix}
\end{equation}

A more recent evolution in the associated graph representation is found in  \cite{riccardi_GeneralPartitioningStrategy}, where the following nonlinear dynamics is considered:
\begin{equation} \label{eq:nonlinear-system}
	\mathcal{S}: \left\{
	\begin{matrix}
		x(k+1) = f(x(k),u(k)) \\
		y(k) = h(x(k)) \hfill 
	\end{matrix}
	\right. 
\end{equation}
The scope in \cite{riccardi_GeneralPartitioningStrategy} is to obtain a weighted and time-varying representation $\mathcal{G}(k) = (\mathcal{V},\mathcal{E}(k))$ of the system \eqref{eq:nonlinear-system}. To this aim, using the same vertices definition introduced for \eqref{eq:linear-system}, the following weighting function is defined:
\begin{equation}\label{eq:weights-edges}
	w_{(i,j)}(k) = \left\{
	\begin{matrix}
		\frac{\partial f_j(x(k),u(k))}{\partial i} & \textrm{for} & i\in\mathcal{V}_u\cup\mathcal{V}_x, j\in\mathcal{V}_x \\
		0  & \textrm{for} & i\in\mathcal{V}, j\in\mathcal{V}_u \hfill\\
		\frac{\partial h_j(x(k))}{\partial i} & \textrm{for} & i\in\mathcal{V}_x, j\in\mathcal{V}_y \hfill
	\end{matrix}
	\right.
\end{equation}
Accordingly, a time-varying set of edges $\mathcal{E}(k)$ is defined as:
\begin{align} \label{eq:edges-nonlinear}
	\mathcal{E}(k) = \{(i,j)\,\,|\,\,i,j\in\mathcal{V},w_{(i,j)}(k)\neq0\}  
\end{align} 
This time-varying graph can capture the instantaneous interactions among the system variables at each time step. In the most general case, a different topological representation exists at each time step. Accordingly, a different choice of graph partition might be the best option for non-centralized predictive control. However, such an approach is computationally demanding. 
}
{\CBf
\begin{example} \label{ex:subsystems}
We consider the following linear discrete-time system to show how to construct the graph associated with a dynamical system. Consider the system:
\begin{equation}
    x(k+1) = A x(k) + B u(k)
\end{equation}
with $x\in\mathcal{X}\subseteq\mathbb{R}^{10}$, $u\in\mathcal{U}\subseteq\mathbb{R}^{3}$, where the matrices $A$ and $B$ are defined by the entries
\begin{equation} \label{eq:example_linear}
\begin{matrix}
    a_{2,1} = 0.5 & a_{6,1} = 0.1 & a_{8,2} = 0.84 & a_{9,2} = 0.57 \\
    a_{8,4} = 0.54 & a_{9,5} = 0.91 & a_{2,6} = 0.98 & a_{3,6} = 0.96\\
    a_{5,6} = 0.8 & a_{6,7} = 0.6& a_{2,8} = 0.31&  b_{4,1} = 0.04 \\
    b_{9,1} = 0.6& b_{10,1} = 0.63& b_{2,2} = 0.02& b_{4,2} = 0.6\\
    b_{10,2} = 0.11& b_{1,3} = 0.19& b_{2,3} = 0.03 \\
\end{matrix}
\end{equation}
and zero elsewhere. According to the definition of a graph $\mathcal{G}$ associated with a dynamical system, we define the set of vertices $\mathcal{V}=\{u_1,\ldots,u_3,x_1,\ldots,x_{10}\}$, while the nonzero entries of matrices $A$, $B$ define the edges in $\mathcal{E}$ of the graph and their weights in the matrix $W^{\textrm{adj}}$. The representation of this graph is given in Fig.\ \ref{fig:network_graph}. This example will be continued in Sec.\ \ref{subsec:terminology} to show how to select subsystems for constructing control agents.
\end{example}
\begin{figure}[t]
	\centering
	\includegraphics[width=.9\linewidth]{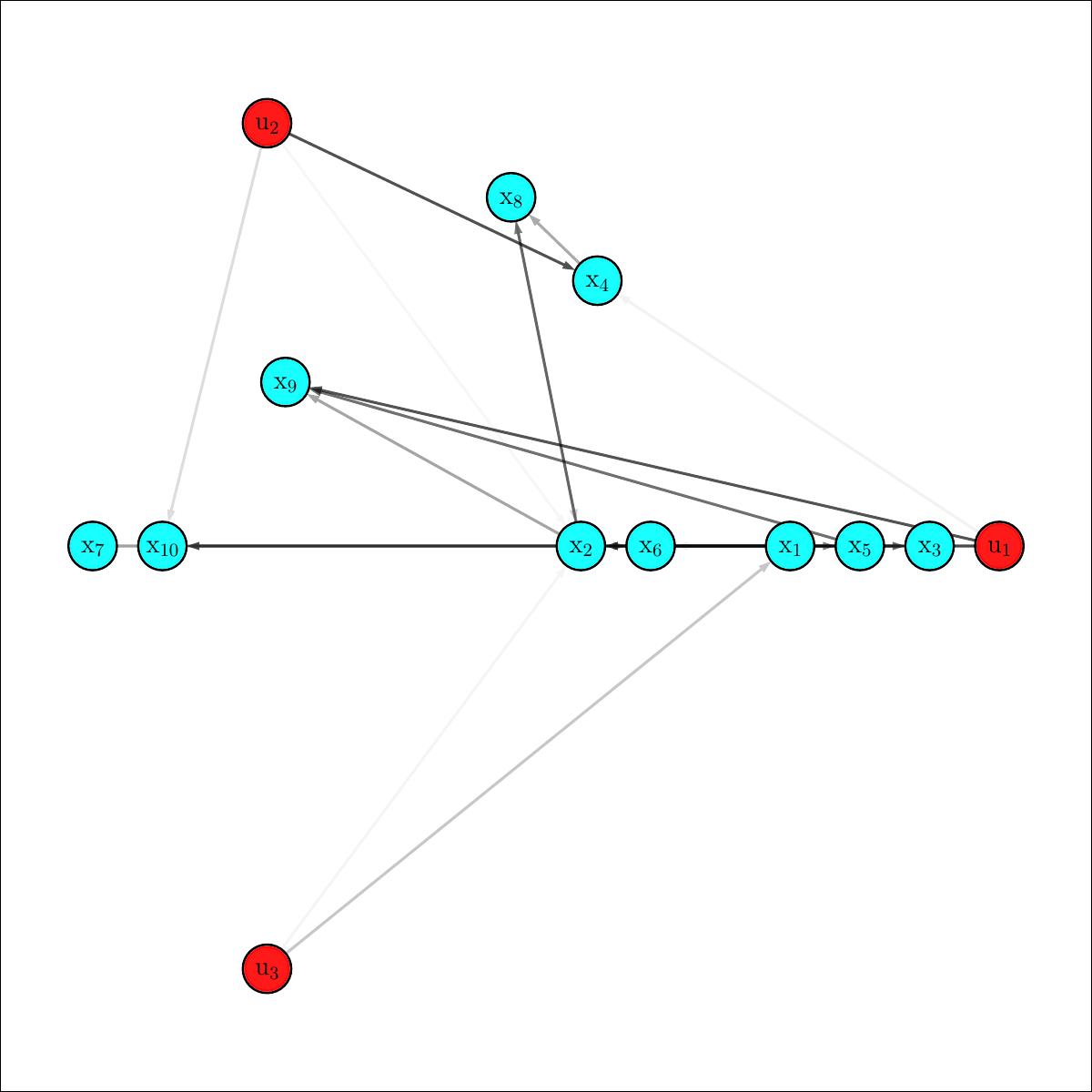}
	\caption{\CBf Graph associated with the dynamical system \eqref{eq:example_linear}. The vertices are the system variables and are colored in red if they are inputs and cyan if they are states. The arrows represent the edges, and their opacity the strength of interaction, i.e.\ the weight, defined by the entries of matrices $A$ and $B$.}
	\label{fig:network_graph}
\end{figure}

}

\subsection{Graph representation of a network of systems} \label{subsec:agent-graph} 
\begin{figure}[t]
	\centering
	\includegraphics[width=.9\linewidth]{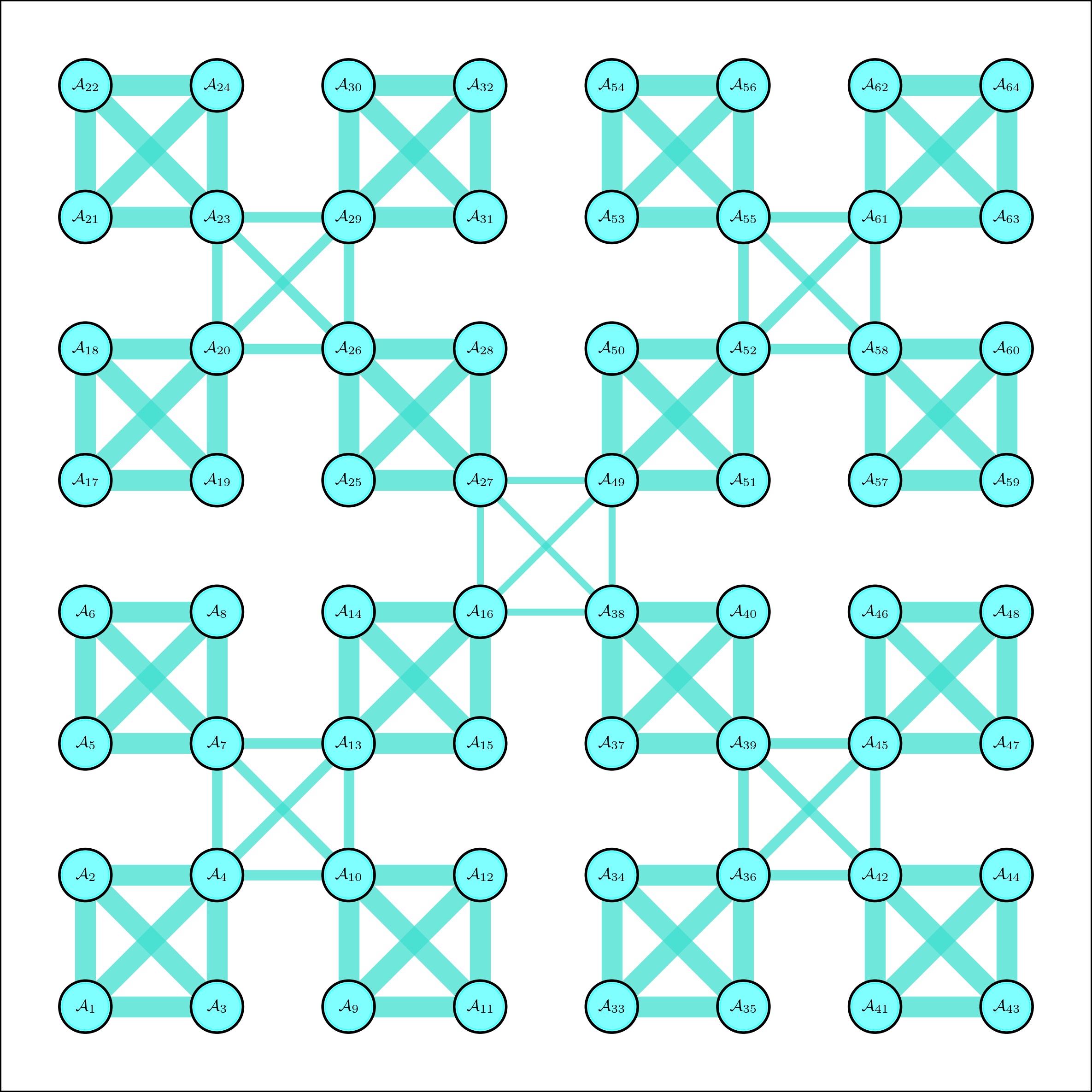}
	\caption{\CBf Graph representation of a modular network with 64 agents. The width of the edges represents the strength of the interaction among the agents. This network exhibits a repeating modular pattern.}
	\label{fig:modular_network}
\end{figure}

\begin{figure}[t]
	\centering
	\includegraphics[width=.9\linewidth]{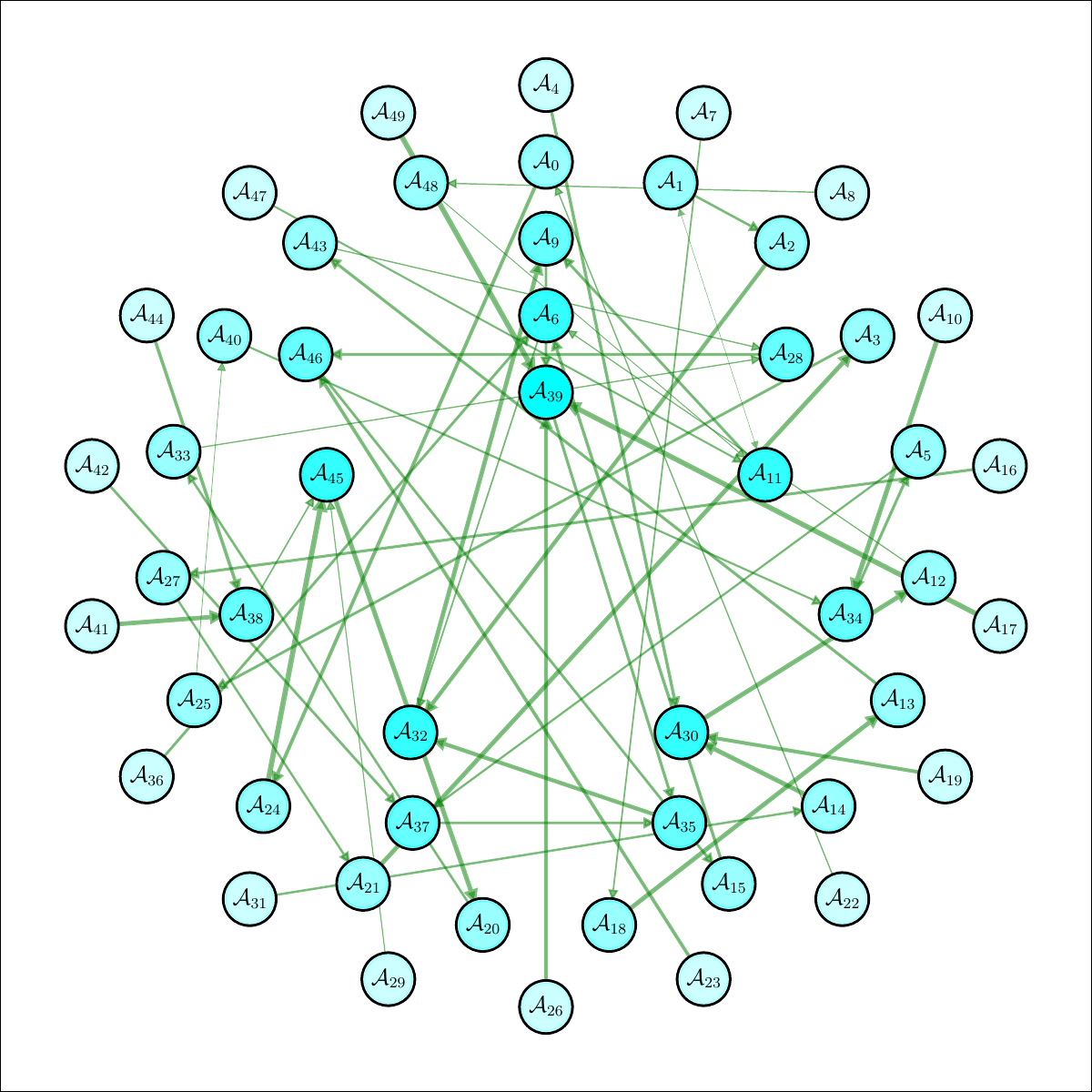}
	\caption{\CBf Graph representation of a random network with 50 agents. The nodes are sorted according to their degree, which is also reflected in the strength of their color. The randomly generated topology is detailed in Tab.\ \ref{tab:random_topology}.}
	\label{fig:random_network}
\end{figure}

{\CBd
A different type of graph representation is considered when the dynamical system is a network admitting a natural decomposition into fundamental subsystems interacting through their dynamics. {\CBe In this case, the network admits a graph representation  $\mathcal{G}$ where the individual subsystems constitute the elements of the set of vertices $\mathcal{V} = \{\mathcal{S}_1,\ldots,\mathcal{S}_{N_{\mathcal{S}}}\}$}. The set of edges $\mathcal{E}$ is defined by state-to-state interactions. Accordingly, to each subsystem $\mathcal{S}_i$ are associated a local state $x_{\mathcal{S}_i}\in\mathbb{R}^{n_{x_{\mathcal{S}_i}}}$ and input $u_{\mathcal{S}_i}\in\mathbb{R}^{n_{u_{\mathcal{S}_i}}}$. The neighbors of a node of the network, i.e.\ of a subsystem $\mathcal{S}_i$, is the set $\mathcal{N}_{\mathcal{S}_i} = \{\mathcal{S}_j\,\,|\,\,(i,j)\in\mathcal{E}\}$. The definition of an output vector $y_{\mathcal{S}_i}\in\mathbb{R}^{n_{y_{\mathcal{S}_i}}}$ 
can also be included, but it will be omitted in the following for simplicity. In other words, for a general nonlinear system of the form \eqref{eq:nonlinear-system}, there exists a natural subdivision of the state and input vectors such that every individual subsystem is described by:
\begin{equation} \label{eq:subsystem-nonlinear}
	\mathcal{S}_i: 
		x_{\mathcal{S}_i}(k+1) = f_{\mathcal{S}_i}(x_{\mathcal{S}_i}(k),(x_{\mathcal{S}_j}(k))_{\mathcal{S}_j \in \mathcal{N}_{\mathcal{S}_i}},u_{\mathcal{S}_i}(k)) 
\end{equation}
This type of representation has been extensively used in partitioning for non-centralized predictive control, especially in the form of linear interacting systems, 
where each subsystem takes the form:
\begin{equation} \label{eq:subsystem-linear}
	\mathcal{S}_i: \left\{
	\begin{matrix}
		x_{\mathcal{S}_i}(k+1) = A_{\mathcal{S}_i}x_{\mathcal{S}_i}(k)+B_{\mathcal{S}_i}u_{\mathcal{S}_i}(k) + w_{\mathcal{S}_i}(k)\\
		\displaystyle w_{\mathcal{S}_i}(k) = \sum_{\mathcal{S}_j \in \mathcal{N}_{\mathcal{S}_i}} A_{\mathcal{S}_{ij}}x_{\mathcal{S}_j}(k)\hfill 
	\end{matrix}
	\right. 
\end{equation}
Each subsystem $\mathcal{S}_i$ is affected only by its local input, and is coupled to its neighbors through dynamic interactions defined by matrices $A_{\mathcal{S}_{ij}}$. This coupling is seen by subsystem $\mathcal{S}_i$ as an exogenous signal $w_{\mathcal{S}_i}$ whose nature is determined by the coordination protocol used in the control strategy, i.e.\ it is considered a disturbance in decentralized control, or it is known or measurable for cooperative strategies. 
Further details about this topic are given in Sec.\ \ref{subsec:multi-topological} where multi-topological representations are introduced. 
\begin{remark}
	From the discussion above, it is clear that each subsystem defined by \eqref{eq:subsystem-nonlinear} can itself be seen as a graph as described in Sec.\ \ref{subsec:equivalent-graph}. A possible algorithmic approach to link the graph associated with a dynamical system and the graph associated with a network of dynamical systems is proposed in \cite{riccardi_GeneralPartitioningStrategy}. 
\end{remark}
\begin{remark}
	In the definition of subsystem \eqref{eq:subsystem-nonlinear}, we assumed that each $\mathcal{S}_i$ is driven only by its local input $u_{\mathcal{S}_i}$. There is, however, the mathematical possibility that dynamics \eqref{eq:subsystem-nonlinear} may be driven also by $u_{\mathcal{S}_j}$ with $\mathcal{S}_j \in \mathcal{N}_{\mathcal{S}_i}$. The resulting networks are constituted by input-coupled subsystems. We decided to treat these networks in separate subsections.
\end{remark}
}
\begin{table*}[t]
\CBf
	\centering
	\ra{1.3}
    \resizebox{\textwidth}{!}{%
		\begin{tabular}{cccccccccc}
        \toprule[1.5pt]
        $w_{1,25}=0.53$ & 
        $w_{2,3}=0.36$ & 
        $w_{2,12}=0.01$ & 
        $w_{3,33}=0.60$ & 
        $w_{4,26}=0.41$ & 
        $w_{5,31}=0.47$ & 
        $w_{6,38}=0.32$ & 
        $w_{7,33}=0.24$ & 
        $w_{8,19}=0.24$ & 
        $w_{9,49}=0.20$ \\ 
        $w_{10,40}=0.36$ & 
        $w_{11,35}=0.72$ & 
        $w_{12,2}=0.01$ & 
        $w_{12,10}=0.42$ & 
        $w_{13,7}=0.17$ & 
        $w_{14,44}=0.44$ & 
        $w_{15,31}=0.67$ & 
        $w_{16,7}=0.46$ & 
        $w_{17,28}=0.42$ & 
        $w_{18,40}=0.76$ \\ 
        $w_{19,14}=0.67$ & 
        $w_{20,31}=0.55$ & 
        $w_{21,34}=0.37$ & 
        $w_{22,4}=0.66$ & 
        $w_{23,1}=0.20$ & 
        $w_{24,47}=0.51$ & 
        $w_{25,46}=0.78$ & 
        $w_{26,41}=0.10$ & 
        $w_{27,40}=0.60$ & 
        $w_{28,22}=0.35$ \\
        $w_{29,47}=0.43$ & 
        $w_{30,46}=0.16$ & 
        $w_{31,13}=0.68$ & 
        $w_{32,15}=0.34$ & 
        $w_{33,10}=0.66$ & 
        $w_{34,29}=0.19$ & 
        $w_{35,6}=0.43$ & 
        $w_{36,33}=0.60$ & 
        $w_{37,7}=0.41$ & 
        $w_{38,36}=0.40$ \\ 
        $w_{39,46}=0.23$ & 
        $w_{40,36}=0.44$ & 
        $w_{41,35}=0.31$ & 
        $w_{42,39}=0.66$ & 
        $w_{43,38}=0.39$ & 
        $w_{44,29}=0.19$ & 
        $w_{45,39}=0.49$ & 
        $w_{46,21}=0.69$ & 
        $w_{47,16}=0.40$ & 
        $w_{48,12}=0.29$ \\ 
        $w_{49,12}=0.13$ & 
        $w_{50,40}=0.77$   \\
        \bottomrule[1.5pt]
        \end{tabular}
        }
        \caption{\CBf Randomly generated topology of the network in Fig.\ \ref{fig:random_network}. The entries $w_{i,j}$ are the $i$-th row and $j$-th column of the weighted adjacency matrix $W^{\textrm{adj}}$.}
        \label{tab:random_topology}
\end{table*}
{\CBe 
\begin{example}\label{ex:agents}
In this example, we propose two different network representations of control agents, one having a modular topology, the other having a random one. According to the discussion above, a control agent will incorporate subsystem dynamics and all the control, communication, coordination, and algorithmic requirements for deploying an NCen-MPC strategy.   

A network can be considered modular if it exhibits a high level of modularity, which can be quantified using the modularity metric, but also visually because it will present recurring patterns. An example of such a network with 64 control agents is reported in Fig.\ \ref{fig:modular_network}, where the recurring structure of 4 and 16 agents is evident. The topology of this network is defined as follows: from the thickest to the thinnest lines, the bidirectional interactions have a strength of $w_{i,j} = 0.1, 0.01, 0.001$.  

The second network has 50 control agents and a randomly generated topology, which is reported in Tab.\ \ref{tab:random_topology}, and which shows the presence of directed arcs. The network representation is proposed in Fig.\ \ref{fig:random_network}.  

We will use these modular and random networks in Sec.\ \ref{subsec:solution-methodologies} to show an application of optimization-based and algorithmic partitioning approaches and the evaluation methodology for the quality of a partition.  
\end{example}
}


\subsection{Bipartite graph representations}\label{subsec:io-graph} 
{\CBd 
In a bipartite graph \cite{diestel_GraphTheory_2017}, the set of the nodes is divided into two groups $\mathcal{V}=\mathcal{V}_a \cup \mathcal{V}_b$, $\mathcal{V}_a \cap \mathcal{V}_b = \emptyset$, and all the edges start in one group and end in another. This type of graph representation has two main use cases in partitioning for non-centralized control. In the first case, a bipartite graph is used to represent the relations between the variables and the constraints of an optimization problem, e.g.\ as done in \cite{tang_OptimalDecompositionDistributed_2018}. This approach is used to decompose the optimization problem by minimizing the number of complicating\footnote{\CBe Complicating constraints are those that introduce an interdependence into subproblems, thus affecting (complicating) the separability of the original problem. In this discussion, complicating constraints are those that involve variables of different subsystems.} constraints that are removed in the distributed solution of the problem. In the second case, a bipartite graph is used to represent the input-output paths of the network, as done in \cite{tang_NetworkDecompositionDistributed_2018, wang_DistributedModelPredictive_2023}. In these approaches, the relationships between output and input variables are made explicit.
Then, among all possible paths between each pair, the shortest is chosen. Accordingly, partitioning is used to minimize the interactions between input and output dynamics, an approach that is conceptually similar to the quantification of input-output interactions in MIMO systems using an RGA matrix \cite{skogestad_MultivariableFeedbackControl_2001}.


}

\begin{figure}[t]
	\centering
	\includegraphics[width=.7\linewidth]{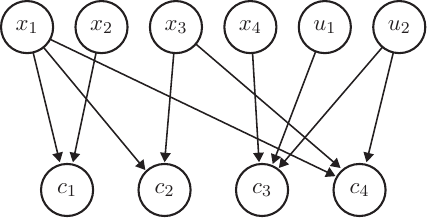}
	\caption{\CBf The bipartite graph representation of a set of complicating constraints. The graph is constituted by two sets of node, one for the optimization variables, one for the problem constraints. The arcs represents the participation of a variable in a constraint. The graph is bipartite because it is formed by two sets of nodes for which arcs only go from one set to another. }
	\label{fig:bipartite_graph}
\end{figure}

{\CBe 
\begin{example}
In this example, we propose the use of a bipartite graph representation for a network subject to complicating constraints\footnote{\CBf  The reader can refer to Tab.\ \ref{tab:classification} for examples of bipartite representations used to capture input-output interactions in MIMO systems.}. Consider the following optimization problem representing MPC optimization at a generic time step $k$ for a network of linear systems, each with two states and one input, with no dynamical coupling, but subject to complicating constraints:
\begin{equation}
    \begin{matrix}
        \displaystyle \min_{\tilde{x},\tilde{u}} & J(\tilde{x},\tilde{u}) \hfill \\
        \textrm{s.t.} & x(k+1) = Ax(k) +Bu(k) \hfill \\
                    c_1(k): & x_1(k)+x_2(k) \leq 0 \hfill \\
                    c_2(k): & x_1(k)+x_3(k) \leq 0 \hfill \\
                    c_3(k): & u_1(k)+u_2(k)+x_4(k)+5 \leq 0 \hfill \\
                    c_4(k): & u_2(k)+x_1(k)+x_3(k) \leq 0 \hfill \\
    \end{matrix}
\end{equation}
where $\tilde{x},\tilde{u}$ represent state and input sequences over an optimization horizon $N$, and $k=0,\ldots,N$. The constraints $c_i$ introduce interactions among the subsystems of the network, which can be, e.g.\, interpreted as a set of specifications on shared resources. This coupling can be captured by the bipartite graph in Fig.\ \ref{fig:bipartite_graph}, where the nodes in one set are the variables, the nodes in the other set are the constraints, and the arcs represent the participation of variables in constraints. To partition a network subject to complicating constraints, it is possible to develop algorithmic procedures to maximize the effect of constraint couplings among cooperating agents in the same coalition, and to minimize the coupling between agents in different coalitions. These inter-coalition constraints can be ignored in solving local problems at first, and they can then be accounted for in a later step of the network optimization.
\end{example}
}

\subsection{Multi-Topological network representations}\label{subsec:multi-topological}

\begin{figure}[t]
	\centering
	\centering
	\includegraphics[width=\linewidth]{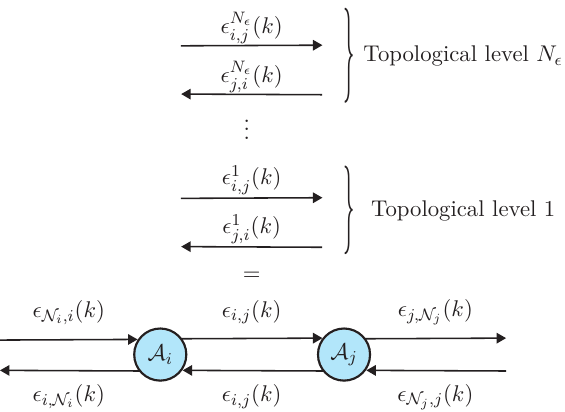}
	\caption{General representation of the connections between agents $i$ and $j$ at a time-step $k$. The topology describing the dynamical coupling among agents at a time step $k$ results from multiple topological levels all acting simultaneously on the network.}
	\label{fig:connection}
\end{figure}

Consider a network of dynamical systems where the connections are determined by a variable topology whose nature will be specified later in this section. The presence of a link introduces a directed relationship between two subsystems that represents a dynamic coupling as described in Fig.\ \ref{fig:connection}. {\CBd Concerning the representation of a network of systems in Sec.\ \ref{subsec:agent-graph}, here we consider only macro links connecting subsystems, thus omitting the subscript notation related to the topological representation of the interactions among variables of different subsystems. Moreover, we index each subsystem $\mathcal{S}_i$ with the letter $i$, so that the network of subsystems is made by the set of nodes $\mathcal{V} = \{1,\ldots,N_{\mathcal{S}}\}$. These choices simplify the presentation of the following concepts.
}
The existence of a link between the subsystem $i$ and $j$ at a time step $k$ can be represented by the binary variable $\epsilon_{ij}(k)$ such that:
\begin{equation}
	\epsilon_{ij}(k) = \left\{
	\begin{matrix}
		1 & \text{if $\mathcal{S}_i$ is connected to $\mathcal{S}_j$ at time step $k$} \hfill \\
		0 &  \text{otherwise} \hfill\\
	\end{matrix}
	\right.
\end{equation}
The collection of these links determines the topology of the network.
In the context of control systems, a link representing a dynamical coupling in this network can have three different natures:
\begin{itemize}
	\item The existence of the link depends on the input-state configuration of the network, i.e.\ the network has an input-state-dependent topology. This happens when the dynamical coupling is determined by the regions of the input-state configuration of the system, such as in PWA dynamics \cite{heemels_EquivalenceHybridDynamical_2001,bemporad_ControlSystemsIntegrating_1999}.
	\item The link can be activated or deactivated as a part of the control strategy of the network, i.e.\ it is a decision variable.
	\item The link activation is driven by an external function, either known or unknown. 
\end{itemize}
A possible topology can co-exist for each of the above-mentioned link classes. Consequently, the overall topology of the network will result from the composition of these superposed topologies, i.e.\ a multi-topological network representation, as in Fig.\ \ref{fig:connection}.  

In the general case, we assume a number of $N_{\epsilon}$ distinct topological levels characterizing the network. We associate a binary variable $\epsilon_{ij}^q(k)$ representing the connection between areas $i$ and $j$ in the topological level $q$ at time step $k$. According to the nature of the topology with which this variable is associated, it can be an input-state-dependent variable, a decision variable, or a signal. Since all binary variables must be equal to one for a connection to exist, the state of variable $\epsilon_{ij}(k)$ is directly determined by the product:
\begin{equation}
	\epsilon_{ij}(k) = \prod_{q = 1}^{N_{\epsilon}} \epsilon_{ij}^q(k)
\end{equation}
Incorporating binary variables $\epsilon_{ij}(k)$ in the network description is straightforward. For this, consider the network of nonlinear systems: 
\begin{equation}
	x(k+1) = f(x(k),u(k))
\end{equation}
and assume it admits a decomposition in $N_{\mathcal{S}}$ subsystems according to the discussion in Sec.\ \ref{subsec:agent-graph}. Then, their time-varying topological dynamics is:
\begin{align}
	& x_i(k+1)= f_i(x_i(k),u_i(k),\omega_i(k))\\
	& \omega_{ij}(k) = \epsilon_{ji}(k) x_{j}(k) \qquad \forall j \in\mathcal{N}_{i}
\end{align}
where $x_i\in\mathbb{R}^{n_{x_i}}$, $u_i\in\mathbb{R}^{n_{u_i}}$ are the state and input vectors of subsystem $i$; and the vector $\omega_{i}$ constituted by the elements $\omega_{ij}$ incorporates all topologically defined dynamical couplings of subsystem $i$ with the its neighborhood $\mathcal{N}_{i}$.

\subsection{Multi-Topological representations and hybrid systems}
When applying the concept of multi-topological time-varying representations to networks of linear systems, the result is a hybrid network system \cite{tabuada_VerificationControlHybrid_2009}. For the sake of simplicity, and without any loss of generality, in what follows, we consider the case of three topological levels of different nature, but the more general case of $N_\epsilon>3$ topological levels follows similarly. In particular, the network is described as:
\begin{align} \label{eq:PWA-network}
	& x_{i}(k+1) = A_{ii} x_{i}(k) + B_{ii} u_{i}(k) + \omega_{i}(k) \\
	& \omega_{i}(k) = \sum_{j \in\mathcal{N}_{i}} \epsilon_{ij}(k) A_{ij} x_{j}(k) \\
	& \epsilon_{ij}(k) = \epsilon_{ij}^{1}(k) \epsilon_{ij}^{2}(k) \epsilon_{ij}^{3}(k) \\
	& \text{s.t.} \quad \epsilon_{ij}^{1}(k) = 1 \Leftrightarrow  \quad  \begin{bmatrix}
		x_{j}(k) \\ u_{j}(k)
	\end{bmatrix} \in\Omega_{j}^{\textrm{a}}  \label{eq:condition-link}
\end{align}
where $A_{ii}\in\mathbb{R}^{n_{x_i}\times n_{x_i}}$, $B_{ii}\in\mathbb{R}^{n_{x_i}\times n_{u_i}}$, $A_{ij}\in\mathbb{R}^{n_{x_i}\times n_{x_j}}$; $\epsilon^{1}$ is the logical variable related to the input-state-dependence of a link; $\Omega^{\textrm{a}}$ is the convex polyhedron for which the link $\epsilon^{1}$ is activated; $\epsilon^{2}$ is a control action; and $\epsilon^{3}$ an external signal affecting the topology.

This multi-topological network description admits a reformulation into Mixed-Logical Dynamical (MLD) form \cite{bemporad_ControlSystemsIntegrating_1999}, allowing the direct application of MPC control. To this, assume that the directed dynamical coupling of the $j$-th system is defined over the polytope $\Omega_{j}^{\textrm{a}} = \left\{\begin{bmatrix}
	x_j^{\intercal}; u_j^{\intercal}
\end{bmatrix}^{\intercal} : S_j^{\textrm{a}}x_j + R_j^{\textrm{a}} u_j \leq T_j^{\textrm{a}}\right\}$, and we compute the constant $M_j^* \stackrel{\vartriangle}{=} \max_{\Omega_j} S_j^{\textrm{a}}x_j + R_j^{\textrm{a}} u_j - T_j^{\textrm{a}}$. Then, we introduce auxiliary variables $z^1$, $z^2$, $z^3$ for each edge of the graph, with $i,j\in\mathcal{V}$:
\begin{align}
	& A_{ij} \epsilon_{ij}^1(k)x_j(k) = z_{ij}^1(k) \\
	& \epsilon_{ij}^2(k)z_{ij}^1(k) = z_{ij}^2(k) \\
	& \epsilon_{ij}^3(k)z_{ij}^2(k) = z_{ij}^3(k)
\end{align}
and the set of constraints that ensure the satisfaction of the logical conditions, and the correct definition of auxiliary variables:
\begin{align} \label{eq:MLD-contraints}
	& S_j^{\textrm{a}} x_j(k) - T_j^{\textrm{a}} \leq M^{*}_j(1 - \epsilon_{ij}^1(k)) \\
	& z_{ij}^1(k) \leq M_{j} \epsilon_{ij}^1(k)\\
	& z_{ij}^1(k) \geq m_{j} \epsilon_{ij}^1(k)\\
	& z_{ij}^1(k) \leq  A_{ij}x_j(k) - m_{j}(1 - \epsilon_{ij}^1(k)) \\
	& z_{ij}^1(k) \geq  A_{ij}x_j(k) - M_{j}(1 - \epsilon_{ij}^1(k)) \\
	& z_{ij}^\ell(k) \leq M_{j} \epsilon_{ij}^\ell(k)\\
	& z_{ij}^\ell(k) \geq m_{j} \epsilon_{ij}^\ell(k)\\
	& z_{ij}^\ell(k) \leq  z_{ij}^{\ell-1}(k) - m_{j}(1 - \epsilon_{ij}^\ell(k)) \\
	& z_{ij}^\ell(k) \geq  z_{ij}^{\ell-1}(k)  - M_{j}(1 - \epsilon_{ij}^\ell(k))
\end{align}
for $\ell = 2,3$, allowing the definition of constraints related to the second and third variables; and $M_{j} = - m_j = \max_{\Omega_{j}} A_{ij}x_j$ are constants. 
The resulting system dynamics is then:
\begin{align} \label{eq:MLD}
	& x_i(k+1) = A_{ii} x_i(k) + B_{ii} u_i(k) + \omega_i(k) \\
	& \omega_i(k) = \sum_{j \in\mathcal{N}_i} z_{ij}^3(k)  \label{eq:MLD-link}
\end{align}
The equations \eqref{eq:MLD-contraints}-\eqref{eq:MLD-link} constitute the MLD form of \eqref{eq:PWA-network}-\eqref{eq:condition-link}.
\begin{remark}
	{\CBe The procedure to obtain multi-topological representations presented in this section is also valid for more complex classes of systems, other than the linear ones. However, for a general nonlinear system, obtaining an MLD representation might not be possible, and more complex approaches to incorporate variable topologies into the dynamics could be required.}
\end{remark}
\begin{remark}
	The existence of an input-state-dependent link between two areas can also be based on the configuration of both areas. In this case, the condition \eqref{eq:condition-link} must include variables of both areas.
\end{remark}

\section{Partitioning for Predictive Control} \label{sec:partitioning-predictive-control}

{\CBd This section introduces the main ideas behind the partitioning problem for non-centralized predictive control. To this aim, we first define a specific terminology for the network components, then we introduce the metrics and the evaluation methodology to assess the quality of a partition, and finally, we present a formulation of the partitioning problem for the maximization of the performance of the control architecture. This section aims to provide the reader with a clear perspective of what partitioning optimally means, and the consequent effect on the non-centralized control architecture. 
}

\subsection{The general partitioning problem} \label{subsec:terminology}
{\CBd 
Consider a network described by the nonlinear dynamics \eqref{eq:nonlinear-system}, and denoted by $\mathcal{N}$. The state, input, and output vectors are respectively $x\in\mathbb{R}^{n_x}$, $u\in\mathbb{R}^{n_u}$, and $y\in\mathbb{R}^{n_y}$. The act of partitioning consists in finding a subdivision of the vectors $x, u, y$ into a number $N_{\mathcal{C}}$ of subvectors $x_i\in\mathbb{R}^{n_{x_i}}$, $u_i\in\mathbb{R}^{n_{u_i}}$, and $y_i\in\mathbb{R}^{n_{y_i}}$ for $i = 1,\ldots,N_{\mathcal{C}}$, and of the respective vector fields into $f_i$, $h_i$, which describe the local subsystem dynamics:
\begin{equation} \label{eq:nonlinear-subsystem}
	\mathcal{C}_i: \left\{
	\begin{matrix}
		x_i(k+1) = f_i(x_i(k),u_i(k),w_i(k)) \\
		y_i(k) = h_i(x_i(k),w_i(k)) \hfill 
	\end{matrix}
	\right. 
\end{equation}
where $w_i(k)$ represents the coupling of subsystem $i$ with its neighboring subsystems $j\in\mathcal{N}_i$. The partition of the network is thus constituted by the set of subsystem dynamics:
\begin{equation} \label{eq:partition}
	\mathcal{P} = \{\mathcal{C}_1,\ldots,\mathcal{C}_{N_{\mathcal{C}}}\}
\end{equation} 
Depending on the context, we call these groups $\mathcal{C}_j$ sets or collections of subsystems, clusters, or coalitions. {\CBe This general formulation of the partitioning problem is generally too broad to be considered directly in defining a partitioning strategy}. Instead, this setting has several simplified reformulations, most notably the ones reported next.

\paragraph{Complete non-overlapping partitioning} 
In \eqref{eq:nonlinear-subsystem}, there is no limitation on the structure of local vectors and dynamics. However, the prevalent setting in partitioning for non-centralized predictive control is to assume that the partitioning is complete and non-overlapping,
and it covers the entirety of the original dynamics. Using set notation, a complete non-overlapping partitioning $\mathcal{P}$ is such that:
\begin{equation}
	\bigcup_{i = 1}^{N_{\mathcal{C}}} \mathcal{C}_i = \mathcal{P} \quad \textrm{and} \quad \bigcap_{i = 1}^{N_{\mathcal{C}}} \mathcal{C}_i = \emptyset \quad \textrm{with} \quad \mathcal{C}_i \neq0 \,\,\forall i
\end{equation}
A complete non-overlapping partitioning allows the straightforward definition of local controllers and coordination protocols, making it the preferred choice in non-centralized control. Overlapping partitionings, on the other hand, are generally used to achieve performance or resilience improvements in the network.  

\paragraph{Coupling through state dynamics} 
The coupling term $w_i(k)$ in \eqref{eq:nonlinear-subsystem} can, in general, comprehend both state and input interactions with neighbors, i.e.\ $w_i(k) = [(x_{\mathcal{S}_j}(k))_{\mathcal{S}_j \in \mathcal{N}_{\mathcal{S}_i}};(u_{\mathcal{S}_j}(k))_{\mathcal{S}_j \in \mathcal{N}_{\mathcal{S}_i}}]$. However, in most settings, only state couplings are considered, yielding the vector field $f_i(x_i(k),(x_{\mathcal{S}_j}(k))_{\mathcal{S}_j \in \mathcal{N}_{\mathcal{S}_i}},u_i(k))$.
This approach is the most intuitive and represents most real-world scenarios in which a local controller would be designed to steer local dynamics through the input channel $u_i$ without directly interfering with the neighbor dynamics $x_j$. Moreover, it is often assumed that the output function depends only on the local state, thus taking the form $h_i(x_i(k))$. However, even if this is the most used setting in the partitioning literature, we acknowledge the presence and relevance of studies for input-coupled subsystems. We decided to treat these approaches separately in Sec,\ \ref{subsec:opt-input-coupling}, \ref{subsec:alg-input-coupling}, \ref{subsec:alg-hierarchical-input-coupling}, and \ref{subsec:coal-input-coupling}.
In fact, studies for input-coupled subsystems generally consider a small number of subsystems, or neglect the existence of delays in the input coupling that would introduce dynamics in the interaction among subsystems. 

\paragraph{Fundamental subsystems} 
A common assumption in partitioning for non-centralized predictive control is that the network $\mathcal{N}$ in \eqref{eq:nonlinear-system} admits a natural decomposition into a number $N_{\mathcal{S}}$ of atomic or fundamental subsystems that cannot be further divided {\CBe for the definition of local controllers}. Moreover, fundamental subsystems are coupled exclusively by state dynamics interactions, as formalized in \cite{riccardi_GeneralPartitioningStrategy}. Therefore, the network is given as a collection of subsystems  $\mathcal{N} = \{\mathcal{S}_1,\ldots,\mathcal{S}_{N_{\mathcal{S}}}\}$. In this network setting, partitioning consists in grouping the subsystems $\mathcal{S}_i$ into a number $N_{\mathcal{C}}\leq N_{\mathcal{S}}$ of bigger units $\mathcal{C}_j$, i.e.\ using the notation \eqref{eq:partition} in defining the partition $\mathcal{P} = \{\mathcal{C}_1,\ldots,\mathcal{C}_{N_{\mathcal{C}}}\}$. Two extreme partitions are possible, one where each group is an individual subsystem, i.e.\ $\mathcal{P}\equiv \mathcal{N}$, $N_{\mathcal{C}} = N_{\mathcal{S}}$, and one that comprises the entire network, i.e.\ $\mathcal{P} = \{\mathcal{C}_1\}$, $N_{\mathcal{C}} = 1$. 

{\CBf
\begin{example}
    We continue Ex.\ \ref{ex:subsystems} by showing a possible selection of the fundamental subsystems for that network. To this aim, we apply the algorithm for selecting fundamental subsystems defined in \cite{riccardi_GeneralPartitioningStrategy}, which iterates over network nodes allocating them to subsystems according to coupling strengths. The resulting definition of the subsystems is given in Fig.\ \ref{fig:Network_States_10_Inputs_3_atomic_control_agents}. This is one definition of the subsystems, and others are possible. 
\end{example}}
\begin{figure}[t]
	\centering
	\includegraphics[width=.9\linewidth]{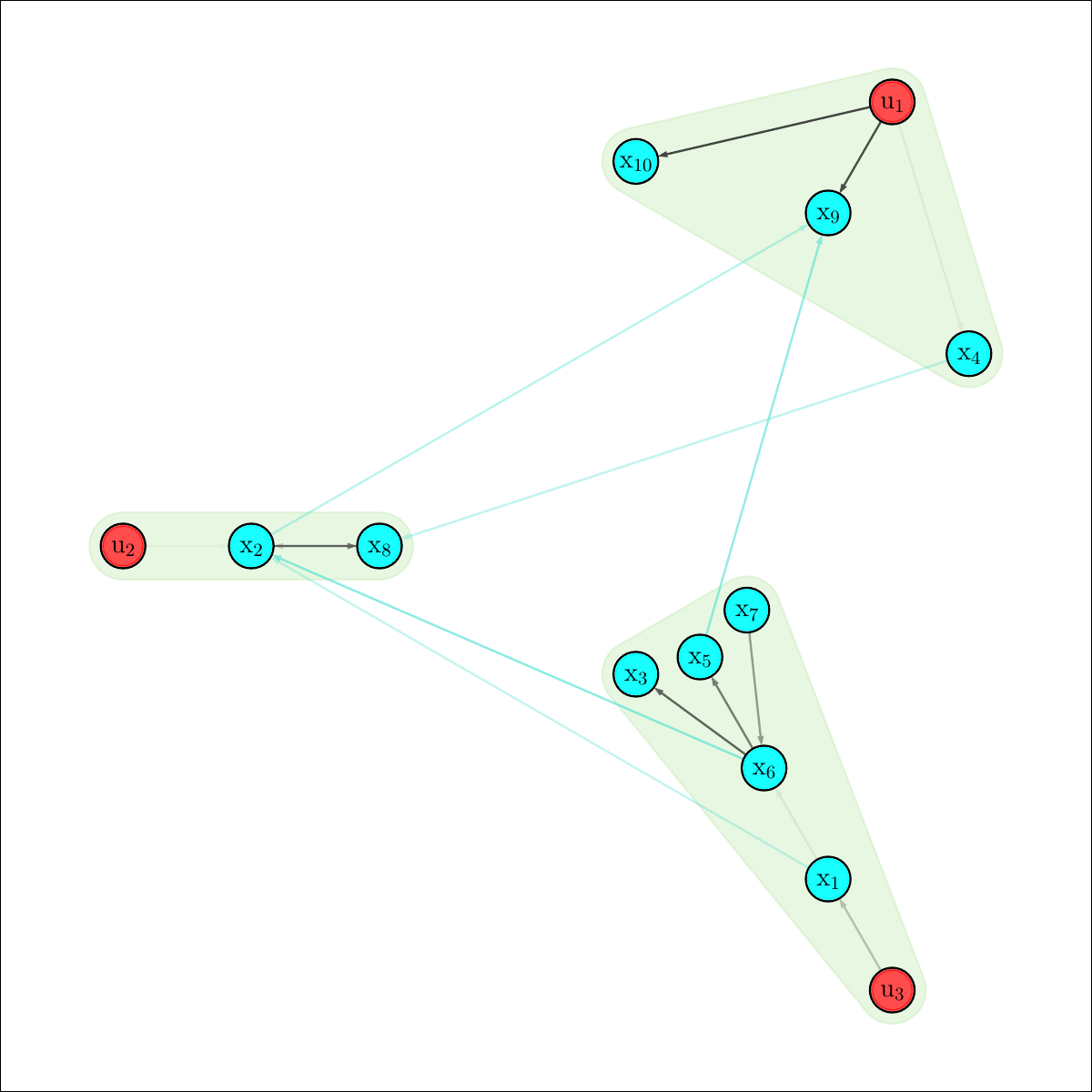}
	\caption{\CBf A possible selection of the subsystems for the network in Ex.\ \ref{ex:subsystems}. The green areas indicate the subsystems and comprehend several input and state variables. The arrows that go from one subsystem to the other can be interpreted as the dynamical coupling among control agents.}
	\label{fig:Network_States_10_Inputs_3_atomic_control_agents}
\end{figure}

\paragraph{Top-down and bottom-up approaches} 
From the discussion above, it is clear that the problem of partitioning a network can be approached from two different directions: a top-down and a bottom-up approach. In the top-down approach, a network $\mathcal{N}$ is considered a monolithic system (generally without any natural decomposition) that must be divided into smaller units. This approach is generally considered when complex nonlinear plants have to be decomposed for non-centralized control. In the bottom-up approach, instead, the problem is solved by aggregating fundamental subsystems that are given a priori, i.e.\ the network is assumed as $\mathcal{N} = \{\mathcal{S}_1,\ldots,\mathcal{S}_{N_{\mathcal{S}}}\}$. 
Both top-down and bottom-up strategies are generally valid approaches, and the preferred direction is usually dictated by the application considered.

\begin{remark} 
	When referring to a group of subsystems, we can also call it a set, cluster, or coalition. All these terms are necessarily used interchangeably throughout the survey because they all represent the same concept of a group of objects. There are subtle distinctions between the terms that will be remarked on in the specific sections. In general, the term cluster is used in the machine learning literature to indicate a group of objects that are strongly connected \cite{xu_SurveyClusteringAlgorithms_2005}, while coalition is a term used in cooperative game theory to denote a group of players \cite{apt_GenericApproachCoalition_2009}.
\end{remark}
}

\subsection{Metrics and evaluation methodology} \label{subsec:metrics}
{\CBd 
The fundamental question that each partitioning strategy in this survey tries to answer is: What is the best partition for non-centralized predictive control? This question only admits a posterior quantitative answer independently from the control strategy considered. A formal motivation for this fact is given in Sec.\ \ref{sucsec:optimal-partitioning}; instead, in this section, we focus on the metrics that allow us to assess the quality of a partition, and on the methodologies to do it. First of all, the best partition {\CBe for a selected evaluation criterion} must be defined for a specific non-centralized predictive control, i.e.\ Dec-MPC, DMPC, HMPC, or Coal-MPC, and w.r.t.\ CMPC. Throughout the section, we assume that the partition associated with CMPC is denoted by $\mathcal{P}^{\textrm{CMPC}}$ (this is the entire network), the one under evaluation by $\mathcal{P}^{\textrm{NMPC}}$, i.e.\ the partitioning for the application of a desired NCen-MPC strategy, and one generic partition by $\mathcal{P}^{\textrm{gen}}$. In this section, we first present the main metrics used to assess the quality of a partition, and then we briefly discuss the evaluation methodologies. 

\paragraph{Metrics} \label{subsubsec:metrics}
In the literature, four main key performance indicators are used to assess the quality of a partition: 1) the cumulative stage cost $J^{\textrm{stage}}$; 2) the computation time $J^{\textrm{time}}$; 3) the computational cost $J^{\textrm{comp.}}$; and 4) the communication cost $J^{\textrm{comm.}}$. To validate the partition, it is necessary to simulate the system using both CMPC and NCen-MPC using the desired partitioning strategy. Then, the key performance indicators are computed as follows.

\paragraph{Cumulative stage cost} 
Assume that the stage cost for CMPC at the time step $k$ is defined by the cost function $J(x(k),u(k-1))$. Moreover, assume a simulation horizon of $N_{\textrm{sim}}$ time steps. At time step $k$, the optimal control problem for CMPC is solved over a horizon $N$, yielding a solution control sequence $\tilde{u}^{*}_{\textrm{CMPC}}(k)$, of which the first element $u^{*}_{\textrm{CMPC}}(k)$ is applied to the system, providing the next step value for the state $x_{\textrm{CMPC}}(k)$. Consequently, the cumulative stage cost for CMPC is
\begin{equation} \label{eq:cumulative-stage}
	J^{\textrm{stage}}(\mathcal{P}^{\textrm{CMPC}}) = \sum_{k = 1}^{N_{\textrm{sim}}}  J(x_{\textrm{CMPC}}(k),u^{*}_{\textrm{CMPC}}(k-1))
\end{equation} 
The cumulative stage cost for the non-centralized strategy and a selected partitioning $\mathcal{P}^{\textrm{NMPC}}$ is obtained similarly. However, in this case, a number $N_{\mathcal{C}} = |\mathcal{P}^{\textrm{NMPC}}|$ of local problems is solved in parallel, providing local solutions $u^{*}_{\textrm{NMPC}, i}(k)$ for $i = 1,\ldots, N_{\mathcal{C}}$. Then, a global vector $u^{*}_{\textrm{NMPC}}(k)$ is obtained by grouping local solutions, and is applied to the plant to compute the global state transition $x_{\textrm{NMPC}}(k)$. This procedure allows the computation of the cumulative stage cost for NMPC, i.e.\ $J^{\textrm{stage}}(\mathcal{P}^{\textrm{NMPC}})$, as done in \eqref{eq:cumulative-stage} but using the non-centralized vectors. In general, for cost minimization it holds that $J^{\textrm{stage}}(\mathcal{P}^{\textrm{CMPC}}) \leq J^{\textrm{stage}}(\mathcal{P}^{\textrm{NMPC}})$. There are exceptions if the dynamics is nonlinear and the solution is obtained for a linearized version around an operating point, or if the network is subject to external uncertain signals. However, the centralized solution of the optimization problem is the reference to assess the optimality of the selected partition (and of the partitioning methodology) for a given non-centralized strategy. 

An approach to compare the cumulative stage cost of different architectures consists in normalizing these results such that, for a given partition $\mathcal{P}^{\textrm{gen}}$ under evaluation, the normalized cumulative stage cost is given by:
\begin{equation}
	J^{\textrm{stage}}_{\textrm{norm.}}(\mathcal{P}^{\textrm{gen}}) = \frac{J^{\textrm{stage}}(\mathcal{P}^{\textrm{gen}})}{J^{\textrm{stage}}(\mathcal{P}^{\textrm{CMPC}})}
\end{equation}
It holds that $J^{\textrm{stage}}_{\textrm{norm.}}(\mathcal{P}^{\textrm{CMPC}}) = 1$, and in general $J^{\textrm{stage}}_{\textrm{norm.}}(\mathcal{P}^{\textrm{gen}}) \geq 1$, so various partitions can be evaluated easily according to a metric that is valid across all possible strategies and applications. 

\paragraph{Computation time} 
This metric is straightforward to obtain. It is sufficient to measure the execution time in seconds necessary to execute the simulation over a horizon $N_{\textrm{sim}}$. For CMPC, one CPU core is used to execute this task\footnote{\CBd In some cases, parallel computing can also be used for CMPC. An example is when the network is constituted by hybrid systems. In this case, the MPC problem requires mixed-integer optimization, for which parallel execution algorithms are available. In such cases, instead of using one CPU for CMPC, it is possible to use any available number, given that each set of subsystems in the non-centralized strategy has such CPUs available at each time step.}, and the time in seconds to perform the simulation constitutes the computation time cost $J^{\textrm{time}}(\mathcal{P}^{\textrm{CMPC}})$. For NMPC, local optimization problems are solved in parallel at each time step, which requires $N_{\mathcal{C}}$ CPUs\footnote{\CBf The analysis of the computation time can be easily extended to the case in which the number of CPUs is time-varying, i.e.\ for $N_{\mathcal{C}}(k)$. This case occurs either when there is a time-varying partitioning $\mathcal{P}(k)$, or when the computational resources can change over time. Such extension also applies to other performance indicators.} The time required for this parallel execution constitutes the computation time cost for NMPC, i.e. \ $J^{\textrm{time}}(\mathcal{P}^{\textrm{NMPC}})$. For any well-designed non-centralized strategy and good choice of partition, it holds that $J^{\textrm{time}}(\mathcal{P}^{\textrm{NMPC}}) < J^{\textrm{time}}(\mathcal{P}^{\textrm{CMPC}})$. The gain in computation time is often one of the main reasons for deploying a non-centralized strategy. In fact, centralized computations may be prohibitive in several settings. For a partition $\mathcal{P}^{\textrm{gen}}$ under evaluation, the normalized version of the computation time is:
\begin{equation}
	J^{\textrm{time}}_{\textrm{norm.}}(\mathcal{P}^{\textrm{gen}}) = \frac{J^{\textrm{time}}(\mathcal{P}^{\textrm{gen}})}{J^{\textrm{time}}(\mathcal{P}^{\textrm{CMPC}})}
\end{equation}
where $J^{\textrm{time}}_{\textrm{norm.}}(\mathcal{P}^{\textrm{CMPC}}) = 1$, and for a well-designed non-centralized strategy and partition $J^{\textrm{time}}_{\textrm{norm.}}(\mathcal{P}^{\textrm{gen}}) < 1$.

\paragraph{Computation cost} 
We discussed how, to assess the computation time in non-centralized control, it is necessary to deploy the strategy in parallel, or alternatively, perform a simulation replicating such a situation. The computation cost is a metric that quantifies the cost associated with the usage of CPUs for these parallel operations, and was introduced in \cite{riccardi_GeneralPartitioningStrategy,riccardi_GeneralizedPartitioningStrategy_2024a} for the evaluation of different partitions of the same network in DMPC. The best way to do so is to look at the CPU usage time, which translates immediately into power and monetary requirements once a specific technology is selected. Consequently, the unit measure of the computation cost is [core $\cdot$ seconds], i.e.\ how much CPU time in parallel is required to perform the distributed computation. 
For a generic predictive control strategy, being it centralized or non-centralized, the computation cost is thus assessed by computing for the simulation horizon $N_{\textrm{sim}}$ the sum over the number of CPUs
of the active CPUs usage time for that time step, which for a CPU $i$ we denote by $\tau_{i}(k)$. If we assume that, in the non-centralized control strategy considered, one CPU is available for each agent in the partition $\mathcal{P}^{\textrm{NMPC}}$, then it holds that $N_{\textrm{CPU}}=N_{\mathcal{C}}$, and the computation cost can be written as:
\begin{equation}
	J^{\textrm{comp.}}(\mathcal{P}^{\textrm{NMPC}}) = \sum_{k = 1}^{N_{\textrm{sim}}} \sum_{i = 1}^{N_{\mathcal{C}}} \tau_{i}(k)
\end{equation}
It is possible to simplify this expression further if we assume that at each time step $N_{\textrm{sim}}$ all local controllers will wait and idle for the slowest controller to obtain its result without performing any operation. Then, the computation cost can be written as $J^{\textrm{comp.}}(\mathcal{P}^{\textrm{NMPC}}) =\sum_{k = 1}^{N_{\textrm{sim}}} N_{\mathcal{C}} \tau^{\textrm{slowest}}(k)$. For both definitions of $J^{\textrm{comp.}}$, the normalized version of the metric for a generic partition $\mathcal{P}^{\textrm{gen}}$ is given by:
\begin{equation}
	J^{\textrm{comp.}}_{\textrm{norm.}}(\mathcal{P}^{\textrm{gen}}) = \frac{J^{\textrm{comp.}}(\mathcal{P}^{\textrm{gen}})}{J^{\textrm{comp.}}(\mathcal{P}^{\textrm{CMPC}})}
\end{equation}
where $J^{\textrm{comp.}}_{\textrm{norm.}}(\mathcal{P}^{\textrm{CMPC}}) = 1$. In general $J^{\textrm{comp.}}_{\textrm{norm.}}(\mathcal{P}^{\textrm{gen}}) > 1$, but very efficient strategies can also achieve $J^{\textrm{comp.}}_{\textrm{norm.}}(\mathcal{P}^{\textrm{gen}}) < 1$.

\begin{remark}
	In literature, to the authors' best knowledge, the only a priori assessment of the computational cost associated with a specific non-centralized predictive control strategy has been performed in \cite{arastou_OptimizationbasedNetworkPartitioning_2025}. However, in that work, the determination is rather qualitative since it is performed through a Big-O analysis of the computational complexity of the algorithm for non-centralized predictive control. In practice, such an approach cannot always establish which is better among algorithms with the same Big-O complexity, as in iterative schemes.   
\end{remark}

%

\paragraph{Communication cost} 
The communication cost assesses the impact of information transmission in different non-centralized control architectures. In its original formulation, see e.g.\ \cite{fele_CoalitionalModelPredictive_2014,fele_CoalitionalControlCooperative_2017b,maestre_PageRankBasedCoalitional_2017a,masero_CoalitionalModelPredictive_2020} among others, the communication cost is a function of the information topology defining how coalitions in a network share information to achieve a coordinated control action. Therefore, to the non-centralized control architecture an information graph $\mathcal{G}_{\text{info}}^{\textrm{NMPC}}=\{\mathcal{V}_{\text{info}}^{\textrm{NMPC}},\mathcal{E}_{\text{info}}^{\textrm{NMPC}}\}$ is assigned, where the set of the nodes is constituted by the coalitions in the network, and the set of the edges by the active communication links. Then, to each link $\epsilon_{ij}\in\mathcal{E}_{\text{info}}^{\textrm{NMPC}}$ a cost is assigned, i.e.\ $\nu(\epsilon_{ij})$, and the communication cost is therefore computed as:
\begin{equation}
	J^{\textrm{comm}}(\mathcal{P}^{\textrm{NMPC}}) = \sum_{\epsilon_{ij}\in\mathcal{E}_{\text{info}}^{\textrm{NMPC}}} \nu(\epsilon_{ij})
\end{equation}
This formulation of the communication cost has been used consistently in deriving coalitional control strategies, leading to partitions of the network minimizing the information sharing. The communication cost of CMPC is obtained by considering the cost associated with each possible active link in the network. The value of the cost of communication can be quantified using distance-based criteria, or the operational costs of the lines. Additionally, we stress that this approach in defining the communication cost can be used to obtain a partition, i.e.\ it is available a priori since it is a pure topological metric, whereas the other costs introduced before are only available a posteriori after the simulation.

While this formulation of the communication cost is direct and straightforward,
it can be insufficient to establish the cost associated with iterative non-centralized control strategies. In fact, if the coordination protocol relies on the iterative sharing of information among agents to achieve an agreement about the control action to deploy, then a static topological metric can only be used to quantify the maximum amount of information shared once the maximum number of iterations of the coordination protocol is given. Posterior measurement of the true amount of information shared is, therefore, a more precise way to assess communication cost in this case. For example, assume that for an NMPC iterative strategy with information topology $\mathcal{G}_{\text{info}}^{\textrm{NMPC}}=\{\mathcal{V}_{\text{info}}^{\textrm{NMPC}},\mathcal{E}_{\text{info}}^{\textrm{NMPC}}\}$, at each time step $k$ several iterations $N_{\textrm{iter}}(k)$, and at every iteration a sequence of state-input predictions of length $N_{\textrm{seq}}$ is shared among the controllers. Then, for a simulation horizon $N_{\textrm{sim}}$, and assuming that each state and input variables vectors have an information transmission cost $\nu(x_i)$, $\nu(u_i)$, $i\in\mathcal{V}_{\text{info}}^{\textrm{NMPC}}$, then the communication cost can be defined as:
\begin{equation}
	J^{\textrm{comm}}(\mathcal{P}^{\textrm{NMPC}}) = \sum_{k = 1}^{N_{\textrm{sim}}} N_{\textrm{iter}}(k) \sum_{i \in \mathcal{V}_{\text{info}}^{\textrm{NMPC}}} \sum_{j\in\mathcal{N}_i} N_{\textrm{seq}} (\nu(x_i) + \nu(u_i))
\end{equation}
where the cost $\nu$ associated with the information transmission can then be directly translated into network operation or economic requirements. The CMPC strategy does not need any iteration; only variables at the current time step are shared. Therefore, its communication cost is:
\begin{equation}
	J^{\textrm{comm}}(\mathcal{P}^{\textrm{CMPC}}) = N_{\textrm{sim}} \sum_{i \in \mathcal{V}_{\text{info}}^{\textrm{CMPC}}} \nu(x_i) + \nu(u_i)
\end{equation}
For both formulations of the communication cost, a normalization assessment is possible. Therefore, for a given partition $\mathcal{P}^{\textrm{gen}}$ associated with a non-centralized MPC strategy, its normalized version is:
\begin{equation}
	J^{\textrm{comm}}_{\textrm{norm}}(\mathcal{P}^{\textrm{gen}}) = \frac{J^{\textrm{comm}}(\mathcal{P}^{\textrm{gen}})}{J^{\textrm{comm}}(\mathcal{P}^{\textrm{CMPC}})}
\end{equation}
where $J^{\textrm{comm}}_{\textrm{norm}}(\mathcal{P}^{\textrm{CMPC}}) = 1$, for decentralized MPC or non-iterative strategies usually holds $J^{\textrm{comm}}_{\textrm{norm}}(\mathcal{P}^{\textrm{gen}}) \leq 1$, while for iterative strategies $J^{\textrm{comm}}_{\textrm{norm}}(\mathcal{P}^{\textrm{gen}}) \geq 1$.

\paragraph{Evaluation methodology} 
From the above discussion about metrics, it is clear that assessing the quality of a partition is mainly a task performed after a simulation or experiment is completed. This fundamental fact, i.e.\ the impossibility of establishing the best partitioning prior to the deployment of the strategy, is one of the main limiting factors in developing partitioning strategies for non-centralized predictive control. In fact, once a partition is selected, computationally intensive simulations involving often large (in number or size) optimization problems have to be performed. Once the metrics of interest are selected for a specific problem and control strategy, the only effective way to determine the best partition is by complete enumeration, see e.g.\ \cite{atam_OptimalPartitioningMultithermal_2021}. However, enumerating and testing all possible partitioning quickly becomes intractable once the number of subsystems grows by more than a few units, due to a combinatorial explosion in the number of possible partitions. Therefore, most partitioning strategies have either developed paradigms for the topological a priori evaluation of partitions, or approached the problem by maximizing the immediate gain of a performance criterion by iterative exchange of agents. A definitive statement about what is the best approach cannot be formulated yet with the current literature, which leaves open many directions for future research. In practice, there might not even be a single partition minimizing simultaneously all four indicators $J^{\textrm{stage}}$, $J^{\textrm{time}}$, $J^{\textrm{comp.}}$, and $J^{\textrm{comm.}}$. Therefore, the desired partition should be selected according to control requirements among the most promising ones. 

}

\subsection{Optimal partition for performance maximization} \label{sucsec:optimal-partitioning}

{\CBd An agent $\mathcal{A}_i$ in the network is a structure with autonomy constituted by a group of subsystems $\mathcal{C}_i$, a local controller $\mathcal{K}_i$, and further devices allowing communication with other agents, or other digital features, such as the execution of algorithmic procedures. 

The problem of partitioning consists of finding an allocation of the agents of the network into groups such that a set of specifications is satisfied. Different criteria, including geographical distribution, communication and coordination effort, operational constraints, security and privacy guarantees, and design choices, can guide the selection of these groups. Often, these criteria are application-dependent and, in almost all cases, are related to the control strategy to deploy. Consequently, there is no common rationale underlying all the different partitioning approaches. However, when the partitioning problem is considered in the context of non-centralized predictive control, it assumes a more precise connotation, and an optimal version can be formulated. 

Assume to have a network with $N_{\mathcal{A}}$ agents, i.e.\ a collection $\mathcal{N} = \{\mathcal{A}_1,\ldots,\mathcal{A}_{N_{\mathcal{A}}}\}$. A set $\mathcal{C}_i$ of $N_{\mathcal{C}_i}$ agents is defined as $\mathcal{C}_i = \{\mathcal{A}_{i,1},\ldots,\mathcal{A}_{i,N_{\mathcal{C}_i}}\}$. We introduce a matrix of binary variables $\delta\in\mathbb{M}_{N_{\mathcal{A}}}(0,1)$\footnote{The class of square binary matrices of dimension $N_{\mathcal{A}}$.} s.t.\ $\delta_{ij} = 1 \Leftrightarrow \mathcal{A}_i \in \mathcal{C}_j$. In general, we can assume $\delta$ to be time-dependent, i.e. $\delta(k)$, but time-dependence is omitted in the following for simplicity, and only used when essentially required. For a given choice $\delta$, we denote a partition of network $\mathcal{N}$ into $N_{\mathcal{C}(\delta)}$ sets of agents by $\mathcal{P}(\delta) = \{\mathcal{C}_1,\ldots,\mathcal{C}_{N_{\mathcal{C}(\delta)}}\}$. Now we consider the control performance of the network that is measured through a cost function $J(x,u,\delta)$, where $x$ is the state of the network, $u$ is the applied control action, and $\delta$ is the selected partitioning, a set of binary decision variables. Once the non-centralized predictive control strategy is selected, the cost $J$ is minimized iteratively at each time step over a selected horizon $N$. For this, we use the vector notation $\tilde{x}_k = [x(1|k),\ldots,x(N|k)]$, $\tilde{u}_k = [u(0|k),\ldots,u(N-1|k)]$, $\tilde{\delta}_k = [\delta(0|k),\ldots,\delta(N-1|k)]$ to define state and input sequences over the horizon $N$. The global control problem is then defined as:
\begin{align} \label{eq:global-opt}
	\min_{\tilde{x}_k, \tilde{u}_k, \tilde{\delta}_k} J(\tilde{x}_k, \tilde{u}_k, \tilde{\delta}_k) = &  \sum_{i=1}^{N-1}J_{\textrm{s}}(x(i|k), u(i-1|k), \delta(i-1|k)) \\
	& + J_{\textrm{f}}(x(N|k), u(N-1|k), \delta(N-1|k)) \nonumber  \\
	\textrm{s.t.} \quad& x(k+1) = f(x(k),u(k)) \nonumber \\
	& x(0|k) = x(k) \nonumber \\
	& g(\tilde{x}_k, \tilde{u}_k, \tilde{\delta}_k) \leq 0 \nonumber 
\end{align}
where $J_{\textrm{s}}$ is the stage cost, $J_{\textrm{f}}$ the terminal cost, and $g$ a set of inequality constraints. This formulation of the optimal partitioning problem assumes that it is possible to simultaneously select the variables in matrix $\delta$, and perform the steps to deploy the non-centralized control strategy. Conceptually, this contemporaneous optimization is not always possible for non-centralized control, especially if communication and coordination protocols are involved, i.e.\ in all cases except for purely decentralized MPC. This limitation can be overcome with a nested reformulation of \eqref{eq:global-opt}. Specifically, the outer level is an integer optimization problem for the selection of $\delta$, and the inner level is associated with the solution of the non-centralized control problem: 
\begin{equation} \label{eq:local-opt}
	\begin{matrix}
		\displaystyle \min_{\tilde{\delta}_k} \hfill&  J^*(\tilde{\delta}_k) \hfill\\
		\textrm{s.t.} & g_{\text{out}}(\tilde{\delta}_k) \leq 0 \hfill \\
		&\hfill J^*(\tilde{\delta}_k) = & \displaystyle \min_{\tilde{x}_k, \tilde{u}_k} \sum_{i=1}^{N-1}J_{\textrm{s}}(x(i|k), u(i-1|k))\lvert_{\tilde{\delta}_k}   \hfill\\
		& &+ J_{\textrm{f}}(x(N|k), u(N-1|k))\lvert_{\tilde{\delta}_k} \hfill \hfill \\
		& \textrm{s.t.} & x(k+1) = f(x(k),u(k)) \hfill\\
		& & x(0|k) = x(k)\hfill \\
		& & g_{\text{in}}(\tilde{x}_k, \tilde{u}_k) \leq 0 \hfill
	\end{matrix}
\end{equation}
where, at the inner level, algorithmic procedures that ensure coordination among the agents might be present. In this formulation we assumed that the inequalities $g$ in \eqref{eq:global-opt} can be split in an outer $g_{\text{out}}$ and inner $g_{\text{in}}$ sets depending on variables $\tilde{\delta}_k$, and $(\tilde{x}_k, \tilde{u}_k)$ respectively. This assumption usually holds since once variables $\tilde{\delta}_k$ are fixed, they do not affect further the non-centralized control strategy. Moreover, the set of constraints $g_{\text{out}}$ can be used to impose desired properties on the partitioning. One common choice is to assume that sets $\mathcal{C}_i$ are non-overlapping, which can be codified with the constraints 
\begin{equation}
	\forall i \quad\sum_{j = 1}^{N_{\mathcal{A}}}\delta_{ij} = 1
\end{equation}
The complexity of the nested optimization problem \eqref{eq:local-opt} is NP-hard due to the outer mixed-integer layer. Moreover, from an implementation perspective, the time requirements to find the optimal partitioning and the optimal control action with this approach can quickly become prohibitive with a growing number of agents because, for each choice of $\tilde{\delta}$, the inner non-centralized predictive control strategy might be required to perform many iterative steps involving optimization.  

{\textit{Remark:} \textit{Optimal partitioning is intended for performance, but partitioning can be done according to other criteria, for which the optimal solution can be different. See Sec.\ \ref{subsubsec:metrics} for a list of common metrics that can be used. }}

\subsection{Solution methodologies} \label{subsec:solution-methodologies}
Partitioning approaches in current literature usually do not consider the level of complexity of the problem formulation \eqref{eq:local-opt}. Instead, simplified formulations, often application-oriented, are considered. These solution approaches can be broadly categorized into the following four methodologies:
\begin{itemize}
	\item \textit{Static partitioning}: this is the case in which the selection of $\delta$ is made prior to the deployment of the non-centralized strategy, and the partitioning $\mathcal{P}(\delta)$ is fixed at all instants. Most approaches follow this logic due to its simplicity and the fact that the partitioning can be computed offline. The disadvantage is that changes in the network's topology cannot be compensated for with this method, making it a viable option only for stationary networks. 
	\item \textit{Event driven partitioning}: it is the first extension of static partitioning. When a topological change is detected, a new network partitioning is deployed. This strategy is reactive since network alterations can be detected, but no assumptions or predictions about their future behavior are made. Suppose the number of possible different topologies of the network is known a priori. In that case, all the associated partitionings of the network can be computed offline and only deployed when necessary. In other cases, the new partitioning is computed as soon as the topological change is detected, implying that the partitioning method is fast enough to be executed between two distinct MPC computations. For large networks, this is not usually suitable through optimization-based approaches. Therefore, algorithmic solutions can be considered to perform local adjustments to partitioning in the neighborhood of the topological change. Also, tabular methods can be implemented to track the topology-partitioning couples, thus avoiding re-computations in known situations. 
	\item \textit{Fixed partitioning over the prediction horizon}: in this case, it is assumed that the topological changes that will occur over the network during the prediction horizon are known at the current time step, either accurately or with some uncertainty. Consequently, before the start of the optimization process in the MPC, a fixed sequence $\tilde{\delta}$ can be established, and the non-centralized MPC is deployed knowing all the changes in topology and partitioning during the prediction horizon. A limited number of techniques of this type are currently available in the literature.
	\item \textit{Time-varying partitioning}: this is the most complex case, where a potentially different network partitioning is allowed for each time step. In this way, all possible input-state-dependent topological changes that will occur in the network according to the available prediction model can be compensated, and uncertain topological changes might be accounted for using robustness arguments. This approach is also the only one that might guarantee the stability of the resulting non-centralized predictive control architecture under predictable topological changes. In current literature, no work is present in this category, and future research might consider addressing this problem.
\end{itemize}

Formally speaking, the last two approaches assume that a \textit{predictive partitioning} of the network can be implemented for the NCen-MPC strategy developed. Such partitioning can assume the network topology to be static, or to change according to known rules or dynamical models. In the first case, the predictive partitioning is performed purely to improve the NCen-MPC approach. For the other two cases, there is no known approach in the literature, making predictive partitioning using models of the network topology dynamics an open problem.

{\CBh We conclude this section by showing two examples of how to obtain the partition of two networks with different structures in Ex.\ \ref{ex:partitioning}, and of how to perform the posterior assessment of the performance of an NCen-MPC strategy applied to different partitionings of the same network in Ex.\ \ref{ex:simulation}.}

{\CBf
\begin{example} \label{ex:partitioning}
    We continue the examples started in Ex.\ \ref{ex:agents} by showing possible partitions of the modular and random networks. 
    
    We start by considering the modular network with 64 agents, and we apply the optimization-based partitioning technique developed in \cite{riccardi_GeneralPartitioningStrategy}. This methodology returns different optimal partitions according to a selected value for the granularity parameter, which balances coupling strengths with the size of the resulting sets of agents. Applying this partitioning methodology to the modular network returns four different partitions: the one constituted by individual agents, two partitions aggregating groups of four agents according to their modules, and the grand coalition accounting for all the agents. The examples of the two intermediate partitions are shown in Fig.\ \ref{fig:modular_partitioning}. 

    We also show the application of partitioning procedures defined in \cite{riccardi_GeneralPartitioningStrategy} to the random network with 50 agents. The use of an algorithmic approach here is advised because the previously deployed optimization-based strategy has a slow convergence rate, which is a consequence of the NP-hard nature of the problem. The algorithmic approach is instead known to have a computational complexity of at most $O(n^4)$, where $n$ is the number of nodes of the graph, after which improvements in the partitioning quality are usually marginal, and it can be potentially optimized and parallelized as commonly done in clustering procedures \cite{xu_SurveyClusteringAlgorithms_2005}. However, which method provides the best partitions cannot be established a priori, and the results should be validated through control experiments, which we show in the next example.  Two different network partitions, one obtained through the optimization-based approach, the other through the algorithmic approach, are shown in Fig.\ \ref{fig:random_partitioning}.    
\end{example}}

\begin{figure}[t]
    \centering
    \begin{subfigure}[b]{0.24\textwidth}
        \centering
        \includegraphics[width=\textwidth]{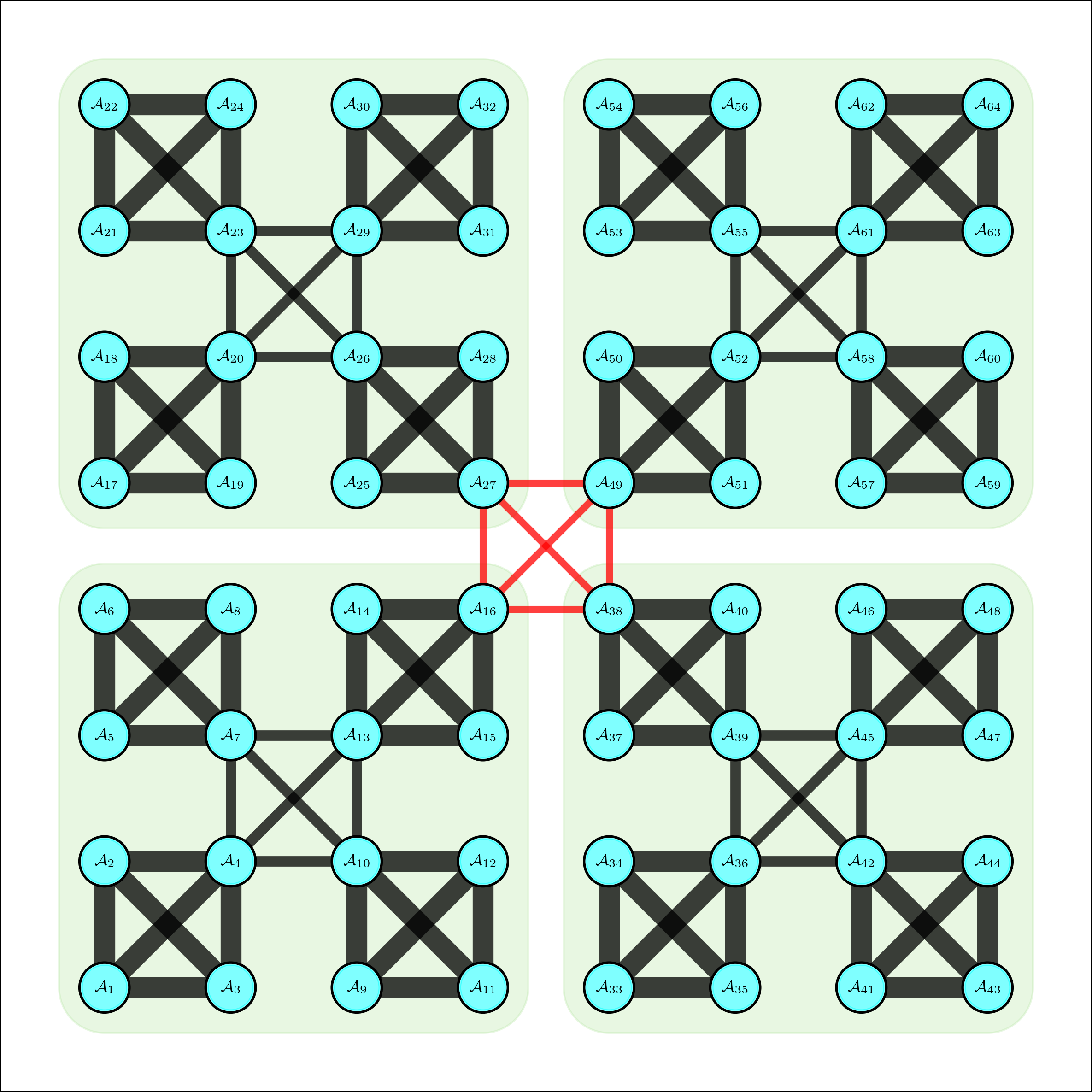}
        \caption{}
        \label{fig:first}
    \end{subfigure}%
    \hfill
    \begin{subfigure}[b]{0.24\textwidth}
        \centering
        \includegraphics[width=\textwidth]{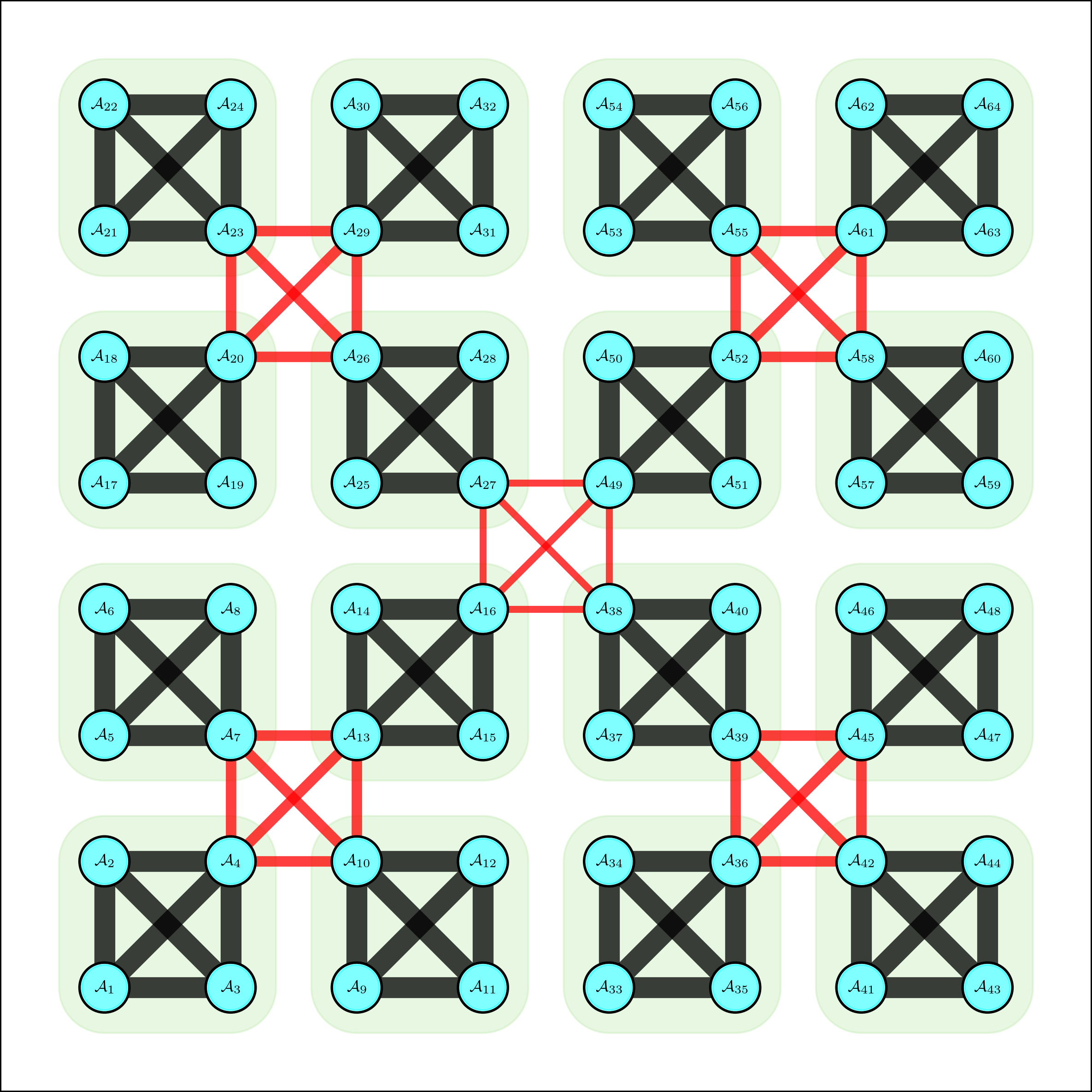}
        \caption{}
        \label{fig:second}
    \end{subfigure}
    \caption{\CBf Two different partitions for the modular network. The green areas represent the control agents, the black links are the interactions inside the same control agent, while the links in red represent the interactions among the control agents. }
    \label{fig:modular_partitioning}
\end{figure}

\begin{figure}[t]
    \centering
    \begin{subfigure}[b]{0.24\textwidth}
        \centering
        \includegraphics[width=\textwidth]{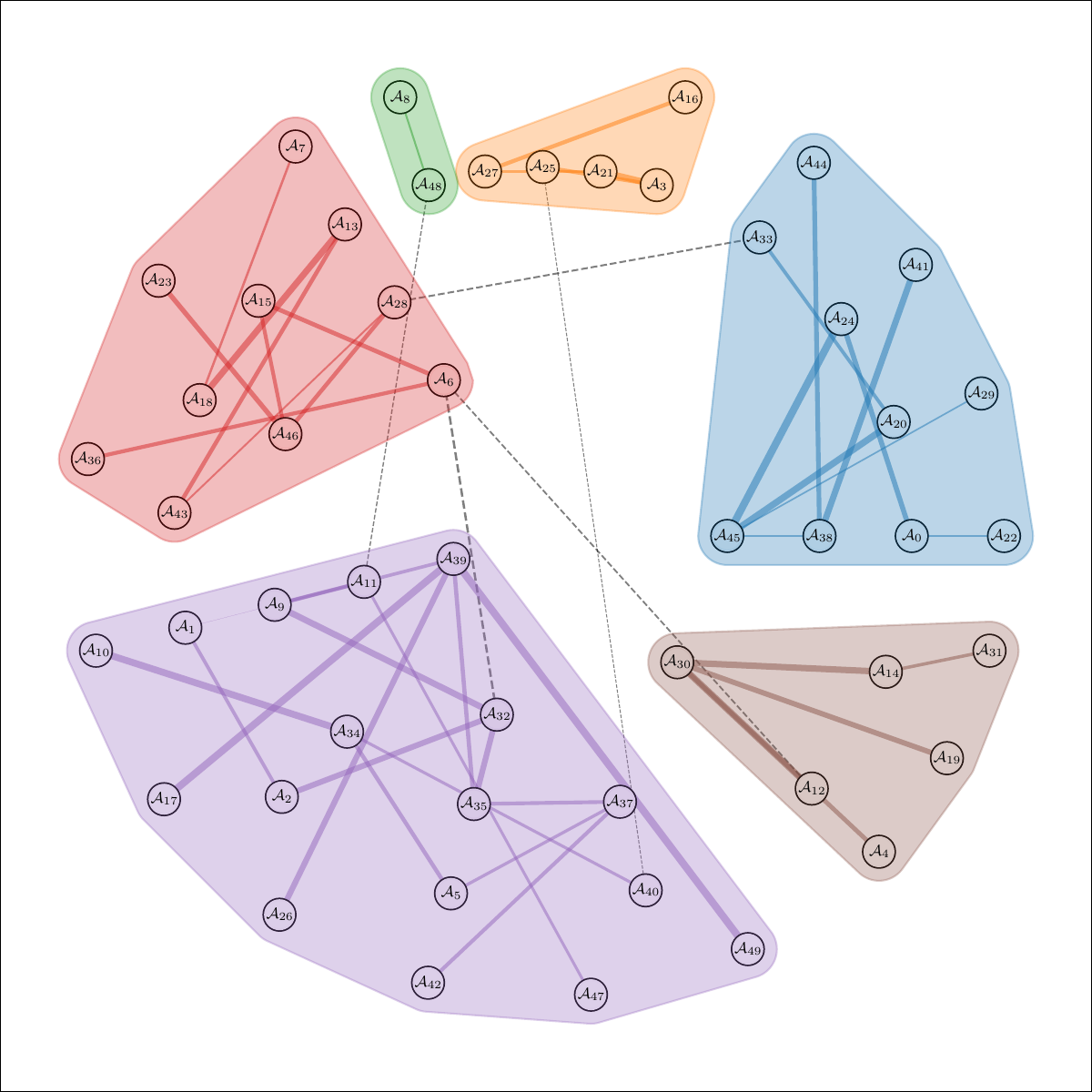}
        \caption{Optimization-based partition.}
        \label{fig:first}
    \end{subfigure}%
    \hfill
    \begin{subfigure}[b]{0.24\textwidth}
        \centering
        \includegraphics[width=\textwidth]{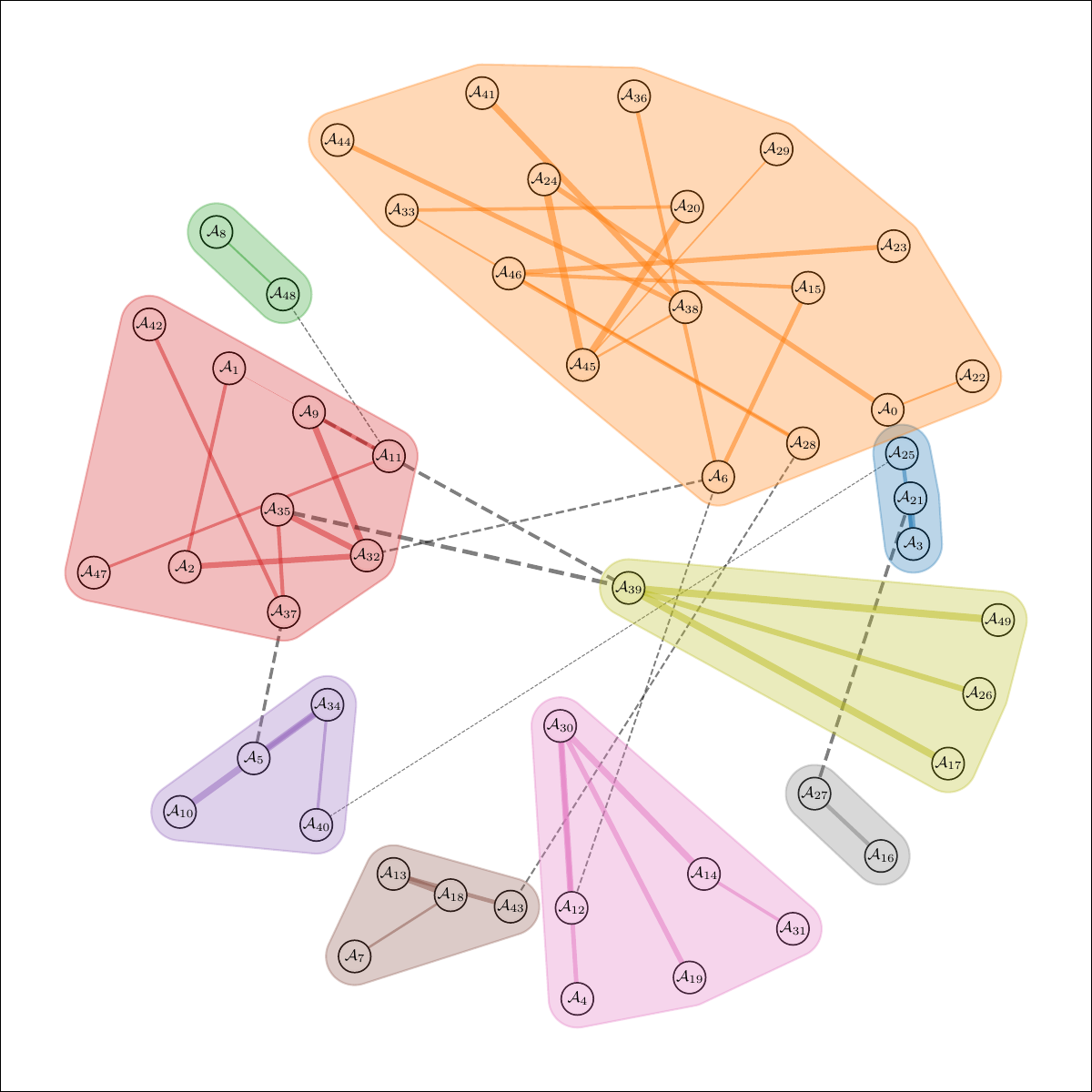}
        \caption{Algorithmic partition.}
        \label{fig:second}
    \end{subfigure}
    \caption{\CBf Two possible different partitions of the random network for selected levels of the granularity parameter obtained with different strategies. These partitions are obtained with the scope of minimizing the strength of the interaction among control agents in different sets, while maximizing the interaction among control agents in the same set. While these two partitions appear to be very similar, the effects they have on network control can be quite different, as shown in Ex.\ \ref{ex:simulation}.}
    \label{fig:random_partitioning}
\end{figure}
\begin{table*}[t]
\ra{1.3}
\CBf
	\centering
	\caption{ Comparison of DMPC-ADMM performance applied to a random network of hybrid systems for different partitioning strategies}
    \resizebox{\textwidth}{!}{%
	\begin{tabular}{c | c |c| c |c | c | c | c }
		\toprule[1.5pt]
		\textbf{Partition}  &   \textbf{Cores}  & \textbf{Cost fun.\ value\ } & \textbf{Opt.\ loss $\%$} & \textbf{Comp.\ time [s]}  & \textbf{Comp.\ time  ratio} &  \textbf{Core seconds [s]} & \textbf{Core seconds ratio}\\
		\midrule[1.5pt]
		$\mathcal{P}^{\text{CMPC}}$& 1 & 6899.9750 &0.00 &2628.04 & 26.4870 & 2628.04 & 1.3736 \\ 
		$\mathcal{P}^{\text{ADMM}}$& 50 & 7749.2102 & 12.31 &99.22 & 1.0000 & 4960.99& 2.5930 \\
		$\mathcal{P}^{\text{Opt}}$&  6& 6916.7114 & 0.24 &436.29 &  4.3972 & 2617.71 & 1.3682 \\ 
		$\mathcal{P}^{\text{Alg}}_1$& 11 & 6982.5798 & 1.20 &173.93 & 1.7530 &1913.28 & 1.0000 \\
		$\mathcal{P}^{\text{Alg}}_2$& 9& 6975.5149 & 1.09 &353.81 &  3.5660 & 3184.27& 1.6643 \\ 
		$\mathcal{P}^{\text{Alg}}_3$&  5& 6911.0475 & 0.16 &2818.69 &  28.4085 & 14093.47 & 7.3661 \\ 
		\bottomrule[1.5pt]
	\end{tabular}
	\label{tab:results-random}
    }
\end{table*}
{\CBf
\begin{example} \label{ex:simulation}
    For this example, we consider again the random network with 50 agents. We further assume that each agent $\mathcal{A}_i$ controls a subsystem with hybrid dynamics, defined as:
    \begin{equation*} \label{eq:hybrid}
    \begin{matrix}
        \displaystyle x_i(k+1) = 0.5 x_i(k) + u_i(k) + \sum_{j\in \mathcal{N}_i}  w_{i,j}x_j(k) \quad \text{if} \quad  x_i(k) \geq 0 \\
        \displaystyle x_i(k+1) = -0.5 x_i(k) + u_i(k) + \sum_{j\in \mathcal{N}_i}  w_{i,j}x_j(k) \quad \text{if} \quad  x_i(k) < 0
    \end{matrix}
    \end{equation*}
    Thus, subsystem $\mathcal{S}_i$ is coupled through state interactions to its neighboring subsystems $\mathcal{S}_j$ with $j\in\mathcal{N}_i$, and is subject to local constraints $u_i \in [-0.5;0.5]$, $x_j \in [-0.9;0.9]$ $\forall i,j$, but not to coupling constraints or objectives. The dynamical coupling occurs through the weights $w_{i,j}$, which define the topology of the network and are reported in Tab.\ \ref{tab:random_topology}. We want to deploy a DMPC strategy based on the alternating-direction method of multipliers (ADMM). We use hybrid dynamics because these are nonlinear systems, for which the effect of partitioning on pure network control performance is evident. Additional technical details about the case study are in \cite{riccardi_GeneralPartitioningStrategy}. Here we focus on the results of control simulations to show how the metrics and the evaluation methodology developed in Sec.\ \ref{subsec:metrics} can be used to assess the quality of the partitions, and to select the most appropriate partitioning strategy for the considered application. To this end, we compare CMPC and the respective coalition denoted by $\mathcal{P}^{\text{CMPC}}$, which is made by all agents; conventional DMPC, where each agent acts independently, denoted by $\mathcal{P}^{\text{ADMM}}$; one of the partitions obtained using the optimization-based method $\mathcal{P}^{\text{Opt}}$ and reported in Fig.\ \ref{fig:random_partitioning}; and three partitions $\mathcal{P}^{\text{Alg}}_i$ obtained with the algorithmic partitioning procedure. We propose only one optimization-based partition because they produce control simulations that are generally similar w.r.t.\ the algorithmic approaches that have more interesting aspects to show. The results of the control simulations are reported in Tab.\ \ref{tab:results-random}. The CMPC approach has the best control performance, and is used as a reference in this category, while conventional ADMM presents a noticeable gap in performance, above the $12\%$. However, it is the fastest control approach, more than 26 times faster than CMPC, which can be the determining factor for selecting a specific partition in many applications. On the other hand, the computational cost in terms of core seconds w.r.t.\ CMPC is approximately double. The optimization-based and algorithmic control approaches provide a trade-off regarding performance gain, computation time, and cost. The strategy based on $\mathcal{P}^{\text{Opt}}$ has a negligible loss in terms of optimality, while being 6 times faster than CMPC and having approximately the same computational cost. Therefore, if these are a priority over speed, $\mathcal{P}^{\text{Opt}}$ is preferable w.r.t.\ conventional DMPC. Algorithmic partitioning approaches have mixed results. The strategy based on $\mathcal{P}^{\text{Alg}}_3$ will give the best results in terms of optimality gap, but it is also slower and more computationally expensive than CMPC; therefore, it is undoubtedly an option to discard. The approach that uses $\mathcal{P}^{\text{Alg}}_1$ has a relatively small loss in optimality, but it is also the least expensive in terms of computational cost, while retaining a good computation time. It is thus a good alternative to $\mathcal{P}^{\text{Opt}}$. The partition $\mathcal{P}^{\text{Alg}}_2$, which is the one reported in Fig.\ \ref{fig:random_partitioning}, offers similar results, and can also be considered. In the end, the most appropriate partition to use will depend on the requirements for the specific application, and can be selected among the listed options with a clear indication of the gains and tradeoffs. A possible way to visualize computational time and costs for different partitions, which can help guide such decisions, is reported in Fig.\ \ref{fig:cost_comparison}.

    This illustrative example shows how posterior evaluation of operational performance for different partitions is fundamental in NCen-MPC. In particular, for the same partitioning strategy, variations in the parameters to perform the partition can lead to very different control results. This fact motivates using a solid methodological assessment of control performance under different partitions. 
\end{example}}

\begin{figure}[t]
    \centering
    \begin{subfigure}[b]{0.49\textwidth}
        \centering
        \includegraphics[width=\textwidth]{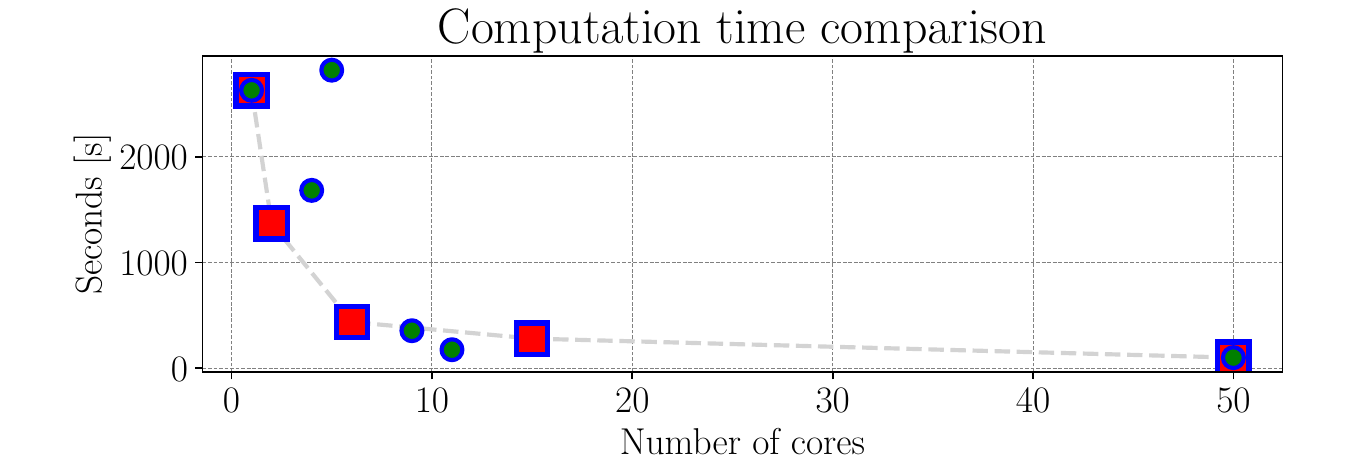}
        \caption{}
        \label{fig:first}
    \end{subfigure}
    \hfill
    \begin{subfigure}[b]{0.49\textwidth}
        \centering
        \includegraphics[width=\textwidth]{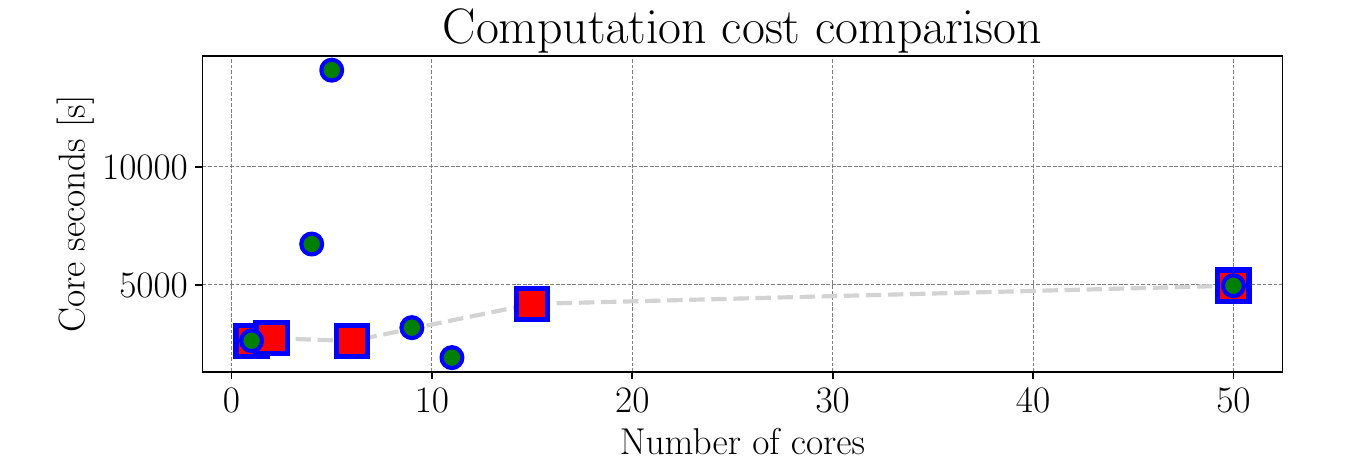}
        \caption{}
        \label{fig:second}
    \end{subfigure}
    \caption{\CBf Computation times and costs for solving the same NCen-MPC problem using different partitions. The data points represented with squares refer to partitions obtained through an optimization-based technique, and they approximately follow well-defined exponential or linear patterns, represented by the dashed lines. The data points using circles refer instead to the results of simulations using partitions obtained through algorithmic approaches. We see that the latter have a less clear evolution across the number of sets, which might be related to the suboptimality of algorithmic approach.}
    \label{fig:cost_comparison}
\end{figure}

{
\section{Analysis and Classification of the Partitioning Techniques for Non-Centralized Predictive Control}
\label{sec:analysis}

In this section, we analyze and classify the partitioning techniques for NCen-MPC that we found in the literature. The analysis we perform here is oriented toward the definition of the main characteristic of each methodology, highlighting their strengths and limitations, which generally apply to all techniques belonging to that category. For a detailed technical discussion both about the general methodologies and the specific papers presented, we developed instead the sections from Sec.\ \ref{sec:optimization-based} to Sec.\ \ref{sec:heuristic}.

Regarding the classification of the partitioning techniques, we propose and discuss in the following three different perspectives:
\begin{enumerate}
	\item A categorization according to the general partitioning class, i.e.\ optimization-based, algorithmic, community-detection-based, game-theory-based, and heuristic.
	\item A categorization in subclasses of the partitioning methods based on specific structures in the problem, or objectives to achieve through its deployment.
	\item A classification according to the NCen-MPC control architecture used in the strategy.
\end{enumerate}
The classification tables for the techniques in this survey are provided in Tab.\ \ref{tab:classification} and Tab.\ \ref{tab:classification-control}. In the first table, we collocate the works found in the literature according to class and subclass. In the second table, we classify them according to the control approach used.

\subsection{Classification according to the partitioning class} \label{subsec:classification-main-branch}
{\begin{table*}
	\centering
	\ra{1.3}
	\resizebox{\textwidth}{!}{%
		\begin{tabular}{p{.20\linewidth}|p{.16\linewidth}p{.12\linewidth}p{.17\linewidth}p{.16\linewidth}p{.08\linewidth}} 
			\toprule[1.5pt] 
			\multirow{2}{*}{\textbf{Partitioning subclass}}& \multicolumn{5}{|c}{\textbf{Partitioning class}} \\
			\cmidrule{2-6}
			& Optimization-based & Algorithmic& Community detection & Game-theory-based & Heuristics \\
			\midrule[1.5pt]
			Unique techniques & 
			\cite{nunez_TimevaryingSchemeNoncentralized_2015,xie_GABasedDecomposition_2016b,barreiro-gomez_TimevaryingPartitioningPredictive_2019} &
			\cite{ocampo-martinez_PartitioningApproachOriented_2011,ocampo-martinez_HierarchicalDecentralisedModel_2012,kamelian_NovelGraphbasedPartitioning_2015,zheng_CouplingDegreeClusteringbased_2018b,rocha_PartitioningDistributedModel_2018,arastou_OptimizationbasedNetworkPartitioning_2025}; $k$-means: \cite{changqing_FrequencyRegulationControl_2022,labella_SupervisedModelPredictive_2022,zhao_OptimalSchedulingStrategy_2023,lin_HierarchicalClusteringbasedOptimization_2020,zhang_EnhancingCooperativeDistributed_2019,chanfreut_ClusteringbasedModelPredictive_2023}&
			\cite{jogwar_CommunitybasedSynthesisDistributed_2017b,pourkargar_ImpactDecompositionDistributed_2017d,jogwar_DistributedControlArchitecture_2019}&
			\cite{muros_GameTheoreticalRandomized_2018b,baldiviesomonasterios_CoalitionalPredictiveControl_2021,sanchez-amores_CoalitionalModelPredictive_2023}& 
			\cite{pourkargar_ImpactDecompositionDistributed_2017d,jain_DistributedWideareaControl_2018,huanca_DesignDistributedSwitching_2023a} \\ \midrule
			Hierarchical & 
			&
			\cite{chen_CooperativeDistributedModel_2020b,wang_HierarchicalClusteringConstrained_2023}&
			\cite{guo_DynamicIdentificationUrban_2019}&
			& 
			\\ \midrule
			Time-varying & 
			&
			\cite{kamelian_NovelGraphbasedPartitioning_2015,rocha_PartitioningDistributedModel_2018,wei_EventtriggeredDistributedModel_2020}&
			\cite{arastou_OptimizationbasedNetworkPartitioning_2025,wang_MPCbasedDecentralizedVoltage_2022}&
			\cite{fele_CoalitionalControlCooperative_2017b,fele_CoalitionalControlSelforganizing_2018,maestre_PageRankBasedCoalitional_2017a}& 
			\cite{ananduta_EventtriggeredPartitioningNoncentralized_2021,liu_DistributedModelPredictive_2019a}\\ \midrule
			Hierarchical time-varying & 
			\cite{riccardi_GeneralPartitioningStrategy} &
			\cite{chanfreut_ClusteringbasedModelPredictive_2023}&
			\cite{riccardi_GeneralPartitioningStrategy}&
			\cite{chanfreut_DistributedModelPredictive_2022,masero_MarketbasedClusteringModel_2022b,masero_FastImplementationCoalitional_2023,masero_RobustCoalitionalModel_2021,masero_LightClusteringModel_2021,sanchez-amores_CoalitionalModelPredictive_2023,chanfreut_CoalitionalModelPredictive_2021a}& 
			\cite{ye_HierarchicalModelPredictive_2019}\\ \midrule
			Problem decomposition & 
			\cite{kersbergen_DistributedModelPredictive_2016} &
			&
			\cite{tang_OptimalDecompositionDistributed_2018,segovia_DistributedModelPredictive_2021}&
			& 
			\\ \midrule
			Input coupling & 
			\cite{chanfreut_FastClusteringMultiagent_2022} &
			\cite{wei_EventtriggeredDistributedModel_2020,wang_HierarchicalClusteringConstrained_2023}&
			&
			\cite{masero_CoalitionalModelPredictive_2020a,sanchez-amores_CoalitionalModelPredictive_2022,sanchez-amores_RobustCoalitionalModel_2023,masero_RobustCoalitionalModel_2023}& 
			\\ \midrule
			Frequency-based & 
			&
			\cite{tang_RelativeTimeaveragedGain_2018}&
			\cite{wang_DistributedModelPredictive_2023}&
			& 
			\\ \midrule
			Applications & 
			\cite{siniscalchi-minna_NoncentralizedPredictiveControl_2020,atam_OptimalPartitioningMultithermal_2021} &
			&
			\cite{moharir_DistributedModelPredictive_2018,pourkargar_DistributedEstimationNonlinear_2019,he_EnhancingTopologicalInformation_2023,tang_AutomaticDecompositionLargescale_2023,guo_DynamicIdentificationUrban_2019,wang_MPCbasedDecentralizedVoltage_2022}&
			\cite{fele_CoalitionalModelPredictive_2014,chanfreut_CoalitionalModelPredictive_2021a,maxim_CoalitionalDistributedModel_2021,maxim_CoalitionalDistributedModel_2022,maxim_CoalitionalDistributedModel_2024,maxim_DistributedModelPredictive_2023}& 
			\\ \midrule[1.5pt]
		\end{tabular}%
	}
	\caption{Categorization of the partitioning techniques according to class and subclass.}
	\label{tab:classification}
\end{table*}}

\paragraph{Optimization-based partitioning} As introduced in Sec.\ \ref{sec:partitioning-predictive-control}, the problem of partitioning can be seen, in an abstract way, as the problem of assigning a set of objects to several given sets. This type of problem can be naturally formulated as an MIP, see e.g.\ Sec.\ \ref{sucsec:optimal-partitioning}, whose solution will provide the optimal network partitioning according to the selected metric. At the basis of this formulation, there is a binary decision variable $\delta_{ij}$ that equals $1$ if the object $i$ belongs to the set $j$. All partitioning methodologies based on this descriptive approach using binary variables fall into the category of optimization-based partitioning and are discussed in Sec.\ \ref{sec:optimization-based}. When considering an optimization-based partitioning technique, it is essential to consider the fact that the associated MIP is NP-hard \cite{karp_ReducibilityCombinatorialProblems_1972,sandholm_CoalitionStructureGeneration_1999,brandes_MaximizingModularityHard_2006}. Consequently, their scalability is limited, and optimization-based partitioning is suitable only for relatively small problems and static network topologies. This also means that online re-partitioning of a network using optimization-based approaches is prohibitive. Approximate solutions of mixed-integer problems can be found using, e.g.\, the genetic algorithm \cite{goldberg_GeneticAlgorithmsSearch_1989,srinivas_GeneticAlgorithmsSurvey_1994}, which does not guarantee global optimality, and still suffers from considerable computational complexity.

\paragraph{Algorithmic partitioning} Partitioning approaches based on algorithmic procedures are a faster and computationally less intensive alternative to optimization-based ones. The trade-off for these gains is that, unless extensive search is performed, their results are suboptimal w.r.t.\ the alternative optimization-based strategies, which constitutes their main disadvantage. However, for large problems or in time-varying settings, algorithmic partitioning approaches result to be the only viable option thanks to their scalability. Additionally, through algorithmic procedures, it is possible to obtain partitions according to more sophisticated requirements, such as the satisfaction of control properties, more directly and straightforwardly than through optimization-based strategies. All the approaches discussed in Sec.\ \ref{sec:algorithmic} fall in this category of algorithmic partitioning. However, we also stress that the works based on the community detection method reported in Sec.\ \ref{sec:modularity-based} are algorithmic procedures. Despite this fact, we decided to discuss community detection methods separately because: 1) it represents by itself a branch of graph and network methods, in this case applied to partitioning for NCen-MPC control; 2) a rich body of studies and approaches has been developed in partitioning for NCen-MPC control exclusively through this method; 3) in this survey, almost every community detection methodology is based on a metric called modularity. Considering these characteristics, we dedicate Sec.\ \ref{sec:algorithmic} to all the algorithmic methods in the literature that do not belong to the community detection approaches, and are not based on the modularity metric or its extensions. {\CBf A similar consideration holds for the game-theoretic oriented partitioning approaches of Sec. \ref{sec:coalitional}. In fact, these approaches are also mainly based on algorithms; however, the fundamental presence of game-theoretic arguments in the selection of the partitions, as well as the extensive development of the coalition control methodology rooted in this technique, deserves a separate discussion in a dedicated section.  }

\paragraph{Community-detection-based partitioning} As mentioned in the discussion for algorithmic approaches, community detection methodologies have been developed in graph and network theory for the identification of strongly connected components of a graph for various applications \cite{fortunato_CommunityDetectionNetworks_2016}. Among all the techniques, great attention has been devoted to community-detection-based partitioning to methods based on the maximization of the \textit{modularity} metric \cite{newman_ModularityCommunityStructure_2006b}. Most of the techniques in this section are conceptually based on this approach. {\CBf The maximization of modularity can be either sought through the solution of an optimization problem, an NP-hard problem, or with a heuristic or greedy algorithm, where the latter approach will, in general, provide a suboptimal result. All the techniques presented here are based on the aforementioned algorithmic approaches, thus allowing for scalability and real-time applicability for time-varying partitioning. The maximization of modularity and other derived metrics will provide groups of agents that exhibit weak inter-group coupling strengths, and, potentially, strong intra-group coupling. The unproven paradigm at the basis of modularity maximization for control problems is that a partition maximizing modularity will also provide optimal NCen-MPC performance. While this statement has not been proven true or false yet, a large body of studies, presented in Sec.\ \ref{sec:modularity-based}, has shown that partitions maximizing modularity will, in general, improve control performance w.r.t.\ heuristic, expert, or random partitions.}

\paragraph{Game-theory-based partitioning} The partitioning approaches based on game-theoretic methodologies find their roots in the theory of coalition formation \cite{ray_GameTheoreticPerspectiveCoalition_2007}. Agents in such networks participate in a game in which they seek their best allocation in a coalition to maximize the collective outcome, which, in this context, corresponds to the global cost function of the MPC problem. Most of the partitioning strategies developed in this field are based on algorithmic procedures; however, the prominent presence of game-theoretic techniques, and the fact that a whole body of literature has been developed about the resulting control strategy, i.e.\ Coal-MPC, motivate the treatment of these methodologies in a dedicated section. {\CBf Game-theoretic partitioning methodologies are, in general, more complex to develop w.r.t.\ other algorithmic approaches, and require the clear definition of cooperative games and the associated cost functions. However, these approaches also allow for obtaining interpretable performance gains in the deployment of the Coal-MPC strategy, a point often missing in most algorithmic approaches.}

\paragraph{Heuristic partitioning} In this class, we include all the partitioning strategies for NCen-MPC that we found in the literature based on heuristic methodologies, which have not been developed originally to be extended to other applications. While the scope and generalizability of these strategies may appear limited, they can still be highly effective in specific contexts, and may offer inspiration for developing more broadly applicable methods.  



\subsection{Classification according to the partitioning subclass} \label{subsec:classification-sub-class}

As it is possible to see in Tab.\ \ref{tab:classification}, there are common features shared among partitioning techniques across different general partitioning strategies. 

First, we can identify hierarchical strategies, in which we collocate approaches that either have multiple aggregation levels for the resulting partition, or are developed using a partitioning layer distinguished from the control layer. {\CBf All purely hierarchical approaches presented in Tab.\ \ref{tab:classification} belong to the first category. Among these, we find works that use a hierarchy to introduce a sequential decision-making ordering into the NCen-MPC strategy, or works with multi-level partitioning approaches, generally used for partition refinement. The former approaches allow obtaining coordinated actions prioritizing the performance of the controllers at the highest level of the hierarchy, and sacrificing the others; the latter generally use purely topological metrics, thus not being directly oriented toward performance optimization. }

Time-varying approaches include the techniques that allow for a reconfiguration of the network, either online during the execution of the control strategy or offline through the derivation of look-up tables. These methods are developed to react to topological changes in the network with the objective of maximizing the global operation cost. {\CBf While real-time adaptability of the partition is advisable (when possible) to improve performance, the computational complexity of the partitioning problem can make it prohibitive if the network has fast dynamics. On the other hand, the offline computation of pre-defined partitions will surely allow for fast online reaction to topological changes, but on the other hand, it assumes either that it is possible to compute all these desired partitions, or there is a trade-off between performance and quality of the partition according to heuristics. }

Hierarchical time-varying strategies are obtained by combining the two previous concepts. The most common setting is the following: a partitioning layer generally operates at a higher hierarchy level and a slower time scale w.r.t.\ a control layer. This approach has been extensively explored because the execution of a partitioning strategy cannot generally be performed in real time according to the control sampling time. Therefore, a slower time scale is used for the partitioning layer, allowing either periodic or event-driven network reconfiguration. {\CBf Hierarchical time-varying strategies allow to obtain enhanced control performance, generally adapting the partitioning (reactively) w.r.t.\ network performance; however, two main aspects deserve some attention: 1) these are complex strategies, and therefore they require a higher level of coordination and communication w.r.t.\ more direct approaches 2) operating at different time scales allows for online re-partitioning, but assumes that the performance degradation during the partitioning intervals is acceptable, and eventual topological changes between re-partitioning intervals will not harm network operation. 
}

Partitioning for input-coupled dynamics has been addressed separately because the underlying dynamics lead to strategies that present unique features, {\CBf such as the definition of private and public control actions and related negotiation strategies,} which are usually not considered when the dynamics present coupling through state interactions. {\CBf In theory, most of the techniques defined for dynamical coupling among network subsystems can be extended to input-coupled dynamics with the necessary care. The most critical aspect for these systems is their limited applicability to real-world problems, which is also reflected in the limited amount of related studies. 
} 

Frequency-based approaches are defined based on the network's transfer functions that link input-output channels. {\CBf These approaches find their roots in the MIMO decoupling approaches \cite{skogestad_MultivariableFeedbackControl_2001} for selecting control channels. Frequency-based approaches are generally developed for linear or linearized systems, and instead of using a direct performance assessment for partitioning, they use frequency-based performance metrics. 
}

A range of approaches in the literature can be seen as applicative work of previously developed strategies, or as prototype techniques that have been extended later. These works can be used to develop comparative case studies for future developments. 

Finally, a range of techniques has been uniquely defined in each partitioning methodology. These works do not share their direct scope with others; thus, we have placed them in a separate category. However, their features can potentially be extended to other techniques, and direct comparisons might be possible. 

{
\subsection{\CBg Classification according to the partitioning methodology} \label{sucsec:classification-methodology}
\CBg
A further classification of the partitioning techniques can be provided in terms of the methodology they are designed for. Specifically, a partitioning strategy can be either developed to operate on a given structure, or to address a specific problem. This classification is provided in Fig.\ \ref{fig:classification-partitioning} as a coloring scheme to distinguish the methodology to which all the subclass entries belong, where mixed approaches indicate that both methodologies have been used in the same subclass. In the following, we discuss their characteristics.  

Structure-based partitioning strategies leverage the presence of a structure in the topology of the network or optimization problem to obtain the partition. Generally speaking, these approaches only require information about the network connections, and can use well-known tools from network and graph theory, such as spectral clustering or $k$-means. One reason to use such approaches is that for some applications, knowing the dynamics of the network is not essential for the specific partitioning problem, and other factors, such as achieving a particular decomposition for ease of operation, accessibility, or maintenance of the network, must be taken into account. Additionally, structure-based approaches do not generally need any information about the dynamics of the subsystems in the network, which can be advantageous in settings where security and privacy are of main concern. In this context, pairing structure-based partitioning with Dec-MPC approaches can be advisable. In such settings, there will be no requirement for real-time data or a communication infrastructure, and the approach can work well in situations where the network does not change over time, or changes slowly and predictably. The main trade-off in such implementations will be the loss in control performance, and the adaptability of the control structure. However, it is important to stress that structure-based approaches should not be limited to static networks, because they can also be developed for time-varying networks and be used with communication-based NCen-MPC approaches. Their main drawback in this sense is that they do not generally account directly for the dynamics of the subsystems; therefore, their actual impact on the performance should be quantified a posteriori.

Goal-oriented partitioning strategies are, in a sense, oriented toward the opposite direction compared to structure-based ones. In fact, they are developed to achieve a given goal without explicitly accounting for the structure of the problem. Usually, this is a control goal, and often, performance optimization. To this, goal-oriented partitioning must have access to some form of information that can relate to the predictable behaviors of the network, such as subsystem dynamics, time-series predictions from local controllers, or the operation cost of the local optimization problems. Additionally, communication and coordination structures are required to leverage and process such information, which increases development costs and complexity; but also affects the privacy of agents and security of network operation. Additionally, goal-oriented partitioning is naturally suited to work with time-varying networks, because it already requires real-time data about the current operation. It also follows that goal-oriented partitioning can be paired effectively with communication-based NCen-MPC, such as DMPC, HMPC, and Coal-MPC. The advantage here is generally sought in performance optimization, or to achieve particular configurations of agents for specific tasks. 

From this discussion, it is clear that both partitioning methodologies are fundamental in the literature, and research in the field of MPC should keep addressing both themes. 
}

\subsection{Classification according to the control strategy} \label{sucsec:classification-control}

In Tab.\ \ref{tab:classification-control}, we categorize the works in partitioning according to the control architecture to which they have been applied. Other than the more conventional Dec-MPC, DMPC, and HMPC strategies, we report that extensive work has been performed on the Coal-MPC methodology. Instead, few studies involve nonlinear MPC strategies. Finally, we mention the presence of a few mixed control strategies that allow for switching between control architectures according to control necessities. {\CBf In the following, we briefly discuss each strategy, but for a detailed discussion, we refer the reader to \cite{maestre_DistributedModelPredictive_2014,scattolini_ArchitecturesDistributedHierarchical_2009, fele_CoalitionalControlCooperative_2017b}. 

Starting from the simplest form of NCen-MPC, we have Dec-MPC in which local controllers do not share any information with their neighbors and compute the local control actions either independently, or using some approximated or estimated information about the strength of the incoming dynamical coupling. Robustness arguments are used to ensure the stability of the network under uncoordinated operation. The biggest strength of Dec-MPC, other than the non-centralized computation of the control action, lies in the ability to preserve the privacy of local subsystems during network operation, since there is no information sharing. The main drawback is the loss of performance w.r.t.\ CMPC, given the conservative nature of local actions.  

In the DMPC approach, information about the current state of the local subsystem, the current control action, or even the predicted state-input sequence is shared among local controllers. This communication is supported by a coordination protocol, which allows local controllers to refine the local actions to achieve superior global performance for the network. The communication and coordination strategy can be structured according to different criteria, thus producing different DMPC approaches. In linear settings, DMPC strategies can converge to near CMPC performance, which is the main advantage of DMPC. However, DMPC also has drawbacks: more expensive hardware requirements w.r.t.\ Dec-MPC, due to the communication infrastructure and the necessity of more advanced abilities for local controllers; complex coordination algorithms, which can also be iterative and must operate within the limits of real-time control; information sharing, which is not always guaranteed to be possible or real-time.  

HMPC includes any control strategy having local controllers and a coordination layer in the form of a centralized decision maker. Such approaches are usually designed to achieve performance advantages, while allowing to overcome other technical challenges, such as model complexity reduction, multi-scale network operation, privacy preservation, or optimization of global coordination. Given the flexibility of HMPC approaches, the specific drawbacks of each technique depend on its implementation, but all approaches unquestionably come at the cost of an increased technical complexity and increased hardware requirements w.r.t.\ simpler NCen-MPC approaches. 

The Coal-MPC strategy was born to fuse MPC with game theory in a non-centralized control setting. The result is a control strategy in which local control agents can merge into coalitions according to game-theoretic strategies to achieve superior control performance. Therefore, the Coal-MPC strategy can also be interpreted by itself as a game-theoretic-oriented partitioning strategy for distributed MPC, with dynamic allocation of local controllers into time-varying coalitions. In this regard, the Coal-MPC problem inherits the computational complexity of the general partitioning problem, or coalition formation problem, i.e.\ it requires the online solution of an NP-hard problem. This main drawback has been solved through different algorithmic procedures, which has led to the development of a large body of literature also discussed in this survey. The main theoretical advantage of Coal-MPC is that it allows for online dynamic partitioning with the objective of global performance optimization in a game-theoretic sense.  

Regarding NLin-MPC, the above considerations have to be extended in a setting where the MPC model is nonlinear. This approach can allow for superior operational performance, but has several drawbacks, mainly: the complexity of defining an appropriate nonlinear model, the computational complexity related to nonlinear optimization, the eventual presence of local minima in the cost function, and the difficulty in ensuring stability of operation. 

Mixed strategies for NCen-MPC use any combination of the previous techniques, trying to balance their strengths and limitations with online reconfiguration of the controllers' settings and (sometimes) partitions. This fact necessarily implies that such strategies have a high implementation complexity, and a combinatorial number of possible approaches at each time step, which is usually addressed through the use of heuristics. 
}

\begin{table*}[t]
	\centering
	\ra{1.3}
		\resizebox{\textwidth}{!}{%
		\begin{tabular}{p{.20\linewidth}|p{.16\linewidth}p{.10\linewidth}p{.17\linewidth}p{.16\linewidth}p{.08\linewidth}}
			\toprule[1.5pt]
			\multirow{2}{*}{\textbf{Control approach}}& \multicolumn{5}{|c}{\textbf{Partitioning class}} \\
			\cmidrule{2-6}
			& Optimization-based & Algorithmic& Community detection & Game-theory-based & Heuristics \\
			\midrule[1.5pt]
			Decentralized MPC & 
			\cite{atam_OptimalPartitioningMultithermal_2021,nunez_TimevaryingSchemeNoncentralized_2015} &
			\cite{ocampo-martinez_PartitioningApproachOriented_2011,ocampo-martinez_HierarchicalDecentralisedModel_2012,kamelian_NovelGraphbasedPartitioning_2015,wang_HierarchicalClusteringConstrained_2023} &
			\cite{arastou_OptimizationbasedNetworkPartitioning_2025,wang_MPCbasedDecentralizedVoltage_2022}&
			\cite{baldiviesomonasterios_CoalitionalPredictiveControl_2021}& 
			\cite{jain_DistributedWideareaControl_2018}\\ \midrule
			Distributed MPC &
			\cite{riccardi_GeneralPartitioningStrategy,nunez_TimevaryingSchemeNoncentralized_2015,xie_GABasedDecomposition_2016b, barreiro-gomez_TimevaryingPartitioningPredictive_2019,kersbergen_DistributedModelPredictive_2016} &
			\cite{zheng_CouplingDegreeClusteringbased_2018b,rocha_PartitioningDistributedModel_2018,labella_SupervisedModelPredictive_2022,tang_RelativeTimeaveragedGain_2018,zhang_EnhancingCooperativeDistributed_2019,wei_EventtriggeredDistributedModel_2020,arastou_OptimizationbasedNetworkPartitioning_2025} &
			\cite{pourkargar_ImpactDecompositionDistributed_2017d,jogwar_DistributedControlArchitecture_2019,tang_OptimalDecompositionDistributed_2018,segovia_DistributedModelPredictive_2021,wang_DistributedModelPredictive_2023,arastou_OptimizationbasedNetworkPartitioning_2025,riccardi_GeneralPartitioningStrategy,moharir_DistributedModelPredictive_2018,pourkargar_DistributedEstimationNonlinear_2019,tang_AutomaticDecompositionLargescale_2023,guo_DynamicIdentificationUrban_2019}&
			\cite{maxim_CoalitionalDistributedModel_2021,maxim_DistributedModelPredictive_2023}& 
			\cite{pourkargar_ImpactDecompositionDistributed_2017d,liu_DistributedModelPredictive_2019a,huanca_DesignDistributedSwitching_2023a}\\ \midrule
			Hierarchical MPC &
			\cite{nunez_TimevaryingSchemeNoncentralized_2015,siniscalchi-minna_NoncentralizedPredictiveControl_2020} &
			\cite{ocampo-martinez_HierarchicalDecentralisedModel_2012,changqing_FrequencyRegulationControl_2022,zhao_OptimalSchedulingStrategy_2023,lin_HierarchicalClusteringbasedOptimization_2020,chen_CooperativeDistributedModel_2020b,chanfreut_ClusteringbasedModelPredictive_2023} &
			\cite{he_EnhancingTopologicalInformation_2023}&
			& 
			\cite{ye_HierarchicalModelPredictive_2019}\\ \midrule
			Coalitional MPC &
			\cite{chanfreut_FastClusteringMultiagent_2022} &
			&
			&
			\cite{fele_CoalitionalControlCooperative_2017b,fele_CoalitionalControlSelforganizing_2018,maestre_PageRankBasedCoalitional_2017a,muros_GameTheoreticalRandomized_2018b,masero_MarketbasedClusteringModel_2022b,masero_FastImplementationCoalitional_2023,masero_RobustCoalitionalModel_2021,masero_LightClusteringModel_2021,sanchez-amores_CoalitionalModelPredictive_2023,masero_CoalitionalModelPredictive_2020a,sanchez-amores_RobustCoalitionalModel_2023,masero_RobustCoalitionalModel_2023,fele_CoalitionalModelPredictive_2014,chanfreut_CoalitionalModelPredictive_2021a,maxim_CoalitionalDistributedModel_2021,maxim_CoalitionalDistributedModel_2022,maxim_CoalitionalDistributedModel_2024,maxim_DistributedModelPredictive_2023}& 
			\\ \midrule
			Nonlinear MPC &
			&
			\cite{kamelian_NovelGraphbasedPartitioning_2015,rocha_PartitioningDistributedModel_2018}&
			\cite{tang_OptimalDecompositionDistributed_2018}&
			& \\ \midrule
			Mixed strategies &
			\cite{nunez_TimevaryingSchemeNoncentralized_2015}&
			&
			&
			\cite{chanfreut_DistributedModelPredictive_2022,maxim_CoalitionalDistributedModel_2021,maxim_DistributedModelPredictive_2023}& 
			\cite{ananduta_EventtriggeredPartitioningNoncentralized_2021}\\ \midrule[1.5pt]
		\end{tabular}%
			}
	\caption{Categorization of the partitioning techniques according to the control strategy deployed.}
	\label{tab:classification-control}
\end{table*}

}
{\CBa
\section{Optimization-Based Partitioning} \label{sec:optimization-based}

\subsection{General techniques}
An advanced optimization-based partitioning strategy for implementing NCen-MPC techniques has been presented in \cite{nunez_TimevaryingSchemeNoncentralized_2015}. The objective is to design a strategy that can work for every predictive control structure, being decentralized, distributed, or hierarchical. To this aim, a variable communication topology is considered. In particular, a communication graph is introduced, linking together systems that directly interact on a dynamics level. Then, an integer variable $\delta_{ij}$ is introduced for each edge $\epsilon_{ij}$ of the information graph, assuming the following values: $\delta_{ij} = 0$ if controller $\mathcal{K}_i$ does not share any information with controller $\mathcal{K}_j$; $\delta_{ij} = 1$ if the control sequence $\tilde{u}_{ij}(k-1)$ at the previous time step is shared from $\mathcal{K}_i$ to $\mathcal{K}_j$; and $\delta_{ij} = 2$ if control sequence $\tilde{u}_{ij}(k)$ at the current time step is shared from $\mathcal{K}_i$ to $\mathcal{K}_j$. The use of these variables for the communication topology in any number of their possible combinations realizes a number of $3^{N_{\mathcal{S}}}$ possible decentralized, distributed, or hierarchical communication topologies for a number $N_{\mathcal{S}}$ of agents in the network, i.e.\ a combinatorial explosion. In \cite{nunez_TimevaryingSchemeNoncentralized_2015}, the number of these configurations is drastically reduced by introducing heuristics. The selection of the optimal communication topology is then achieved by introducing a term into the global cost function of the problem that accounts for the cost of communication. Accordingly, the framework adapts a non-centralized control strategy depending on how information is shared. The methodology is validated on a 16 water tanks system \cite{maestre_AssessmentCoalitionalControl_2015}, showing how the optimal partitioning is affected by the change in operating conditions.  

An original approach to optimization-based network partitioning for DMPC can be found in \cite{xie_GABasedDecomposition_2016b}, where an input-output decomposition of large-scale linear systems is sought. In particular, to overcome the limitations of previous techniques such as \cite{ocampo-martinez_PartitioningApproachOriented_2011}, or system decompositions based on interaction analysis approaches (e.g.\ RGA \cite{skogestad_MultivariableFeedbackControl_2001}), the paper proposes a two-stage procedure based first on input clustering decomposition (ICD), and then on input-output pairing decomposition (IOPD). For the former, ICD consists of finding a matrix $G=(g_{ij})\in\mathbb{R}^{m\times M}$, where $m$ is the number of inputs and $M\leq m$ the number of subsystems, such that $g_{ij} = 1$ if input $i$ belongs to subsystem $j$ and zero otherwise. Then, a transformation matrix $T(G)\in \mathbb(R)^{m\times m}$ is built from $G$, allowing to find the input decoposition of the original vector $u$ into the form $\bar{u} = [\bar{u}_1^\intercal, \ldots, \bar{u}_M^\intercal]^\intercal = T(G)u$, and $T(G)$ is an orthogonal matrix allowing easy back transformation. After the ICD, the IOPD is found by minimizing the coupling effect between the subsystems. This is achieved by leveraging the condensed formulation of the MPC problem as a quadratic program \cite{machowski_PredictiveControlConstraints_2002}: for a prediction horizon $N$, and a given initial condition $x(0|k) = x_{0, k}$ at a time step $k$, by defining $\tilde{u}_k$ as the input sequence over $N$ time steps, the cost of MPC is $J_k = \tilde{u}_k^\intercal H \tilde{u}_k + 2 x_{0, k}^\intercal F^\intercal \tilde{u}_k$, with matrices $H$, $F$ constructed using the system model and the weighting matrices (this is a standard derivation, for further details see \cite{machowski_PredictiveControlConstraints_2002}). Then, applying the ICD transformation for the input decomposition, the cost $J$ is rewritten in terms of the input vector $\bar{u}$, through a new matrix $O$ that is a function of $T(G), H$, and $F$. The coupling effect among the subsystems is then quantified as:
\begin{eqnarray}
	J^{\textrm{coupling}} = \frac{\|O - \textrm{diag}(O_{11},\ldots,O_{MM})\|_{\textrm{F}}}{\|O\|_\textrm{F}}
\end{eqnarray}
The minimization of this cost is achieved by approximately selecting the entries of $G$, which provides the input clustering sought. Given the binary nature of $G$, this is a nonlinear integer programming problem. Therefore, the authors approach it using the Genetic Algorithm (GA) \cite{srinivas_GeneticAlgorithmsSurvey_1994}. The partitioning approach is further developed for partial observability (for details see \cite{xie_GABasedDecomposition_2016b}). Application of the partitioning methodology is performed on a chemical plant with five operation units known in the literature as the Tennessee Eastman problem \cite{downs_PlantwideIndustrialProcess_1993, lyman_PlantwideControlTennessee_1995}. Partitioning of the latter is executed on a linearized version of the plant around operating points generated through a stabilizing control action \cite{mcavov_BaseControlTennessee_1994}. No control validation of the proposed DMPC architecture is performed. 

\subsection{Multi-objective optimization in partitioning}
Multi-objective optimization is at the basis of the partitioning strategy proposed in \cite{barreiro-gomez_TimevaryingPartitioningPredictive_2019}. Specifically, an optimization problem of the following form is considered at each time step $k$:
\begin{align} \label{eq:multi-obj-opt}
	\min_{\mathcal{P}(k)} & \,\, \sum_{i = 1}^{4} \varphi_i\sigma_i(\mathcal{P}(k)) \\
	\textrm{s.t.} & \,\, \bigcup_i \mathcal{C}_{i}(k) = \mathcal{P}(k) \nonumber\\
	& \,\,  \bigcap_i \mathcal{C}_{i}(k) = \emptyset\nonumber
\end{align}
where the constraints require that the sets $\mathcal{C}_{i}(k)$ constitute a nonovelapping partitioning, i.e.\ $\mathcal{P}(k) = \{\mathcal{C}_1(k),\ldots, \mathcal{C}_{N_{\mathcal{C}}}(k)\}$; $\varphi_i$ are weights; and the four indicators $\sigma_i$ defined for the sets $\mathcal{C}_{i}$ account for: the number of links connecting the sets, the difference in the number of nodes between the sets, the distance between the sets, and the relevance of the information shared between the sets. To solve the proposed optimization-based problem, an algorithmic distributed approach based on the Kernighan-Lin algorithm \cite{gupta_FastEffectiveAlgorithms_1996} for graph partitioning is proposed. In this algorithm, each node is a decision maker, which selects the set to move it to based on a utility linked to the local cost in \eqref{eq:multi-obj-opt}. This partitioning approach is applied to the Barcelona drinking water transport network using a DMPC approach based on density-dependent population games \cite{sandholm_PopulationGamesEvolutionary_2010}.

\subsection{For optimization problem decomposition}
A partitioning technique developed for railway traffic management is presented in \cite{kersbergen_DistributedModelPredictive_2016}, where a switching max-plus-linear model is used to describe the Dutch railway network \cite{kersbergen_RailwayTrafficManagement_2016}. Given the hybrid nature of the model, optimization of such a system over a prediction horizon provides an MILP, and the aim of \cite{kersbergen_DistributedModelPredictive_2016} is to develop a DMPC strategy to solve it. This strategy is a cooperative iterative approach with a consensus threshold. The centralized optimization problem over a given horizon $N$ is stated as:
\begin{align}\label{eq:raylway-opt}
	& \min_{z(k)} c^\intercal(k) z(k) \\
	& \textrm{s.t.} \,\,\, A(k)z(k) \leq b(k) \nonumber
\end{align}
where the state and input sequences for the network are in $z(k) = [\tilde{x}^\intercal(k) \,\, \tilde{u}^\intercal(k)]^\intercal$. The partitioning of this global optimization problem is performed according to a set of desired properties for the local optimization problems. Specifically, the partitioning has to provide non-overlapping subproblems, such that constraints are decoupled, and the size and number of variables of the problems are approximately the same. This is achieved through an MIQP setup for a selected number $n_{\textrm{sub}}$ of subsystems. Specifically, for each constraint of problem \eqref{eq:raylway-opt} (in total $n_{\textrm{T}_2}$) a number $N_{\mathrm{C}}$ of binary variables $\delta$ is introduced. A set of $N_{\mathcal{C}}$ continuous variables $S$ representing the number of constraints of each problem is also introduced, and an additional variable $M_{\textrm{MAX}} \geq M_i - M_j$, for $i,j = 1,\ldots, N_{\mathrm{C}}$ represents the maximum difference in the number of constraints. Then, the cost of the partitioning problem is:
\begin{equation}
	J = \rho M_{\textrm{MAX}} - \sum_{j = 1}^{n_{\textrm{T}_2}} \sum_{k = 1}^{n_{\textrm{T}_2}}  \sum_{i = 1}^{N_{\mathcal{C}}} \delta_{ji} Q_{jk}\delta_{ki}
\end{equation}
where $\rho$ is a tuning parameter, and the weighting matrix $Q$ is constructed assigning to each element $Q_{ij}$ the number of constraints connecting the constraints set $i$ to $j$. Constraints are added to the MIQP to enforce the properties listed above. Solving this problem provides a decomposition of \eqref{eq:raylway-opt} into $N_{\mathrm{C}}$ subproblems. The approach is validated on the model of the Dutch railway network \cite{kersbergen_RailwayTrafficManagement_2016} against a CMPC implementation. The results show that the distributed implementation is up to $90\%$ faster in computing the predictive control action w.r.t.\ CMPC with only marginal performance losses.

\subsection{Ad-hoc performance indicators}
Optimization-based partitioning using as a performance indicator the wake-effect\footnote{The wake-effect refers to the wind reduction and increased turbulence that downstream turbines experience due to the extraction of wind power from upstream turbines.} coupling in wind turbines is used in \cite{siniscalchi-minna_NoncentralizedPredictiveControl_2020}. This work addresses non-centralized hierarchical control of wind farms, where MPC is used for reference point setting at the control partition level, while conventional controllers are used for individual turbines. The definition of the groups in the partition is performed based on the coupling among the turbines induced by the wake effect \cite{katic_SimpleModelCluster_1986}, whose intensity is derived using first-principles modeling of the system \cite{bianchi_WindTurbineControl_2007}. Accordingly, a weighted directed graph is constructed using the intensity of the wakes as labeling \cite{annoni_EfficientOptimizationLarge_2018}. A multi-objective integer optimization program is constructed, where a binary variable $\delta_{ij}$ is 1 if turbine $\mathcal{S}_i$ belongs to the group $\mathcal{C}_j$ and 0 otherwise. Three concurrent costs are considered: the first requires that the wake effect among turbines is maximized, the second involves the minimization of the distance among the turbines, and the third balances the size of the resulting groups, requiring that the difference of elements among them is minimized.
The constraints ensure nonoverlapping partitioning. Due to the formulation of the costs funciton, this is a nonlinear optimization program, which is transformed into a linear form by introducing auxiliary variables \cite{bemporad_ControlSystemsIntegrating_1999}. 
To overcome the computational complexity of the proposed strategy, the operating conditions of the wind turbines are discretized, and 12 wind directions are considered. Additionally, the strategy is only suited for offline construction due to the computational cost, but a look-up table can be constructed for different operational settings. The approach is validated on a wind farm with 42 NREL-5MW wind turbines \cite{jonkman_Definition5MWReference_2009a}, and modeled using SimWindFarm \cite{grunnet_AeolusToolboxDynamics_2010}. 
The non-centralized strategy is compared with its centralized counterpart, showing a significant reduction in computation times while ensuring a good level of performance.

\subsection{Robust and stochastic optimization}
Robust and stochastic methodologies for partitioning a system have been developed in \cite{atam_OptimalPartitioningMultithermal_2021} for Dec-MPC of the thermal zones of a building. A graph representing the thermal interactions among the zones is constructed and weighted by a representative metric depending on the temperature difference between areas, and external disturbances. This representation allows the formulation of a mixed-integer optimization problem for partitioning, where specifications about the resulting clusters are imposed through constraints \cite{boulle_CompactMathematicalFormulation_2004}. In addition, in the stochastic formulation the definition of the thermal interactions is replaced by their expectations in the cost function; while in the robust formulation their maximum values are used to account for the worst-case scenario. The former approach assumes that the probability of the occurrence of the disturbances is known, while the former approach is more conservative. The efficacy of the resulting partitioning is assessed through ad-hoc performance indicators for the specific application. The approach is extensively validated for the Dec-MPC control of a 5 and a 20 zones case study, and compared with the partitioning approach of \cite{chandan_OptimalPartitioningDecentralized_2013}.  

\subsection{Input-coupled dynamics} \label{subsec:opt-input-coupling}
A binary programming approach for partitioning a network of linear systems coupled through input interactions has been proposed in \cite{chanfreut_FastClusteringMultiagent_2022}, i.e.\ a dynamics of the following form is considered:
\begin{align}\label{eq:subsystem-linear-CHANGE}
	& x_i(k+1) = A_{ii}x_i(k) + B_{ii}u_i(k) + w_i(k) \\
	& w_i(k) = \sum_{j\in\mathcal{N}_i}B_{ij}u_j(k) . \nonumber
\end{align}
Using the condensed formulation of the MPC problem \cite{machowski_PredictiveControlConstraints_2002}, and the structure of the input-coupled dynamics, the gradient of the global cost function $J$ for the selected prediction horizon can be approximated \cite{deoliveira_MultiagentModelPredictive_2010} for a selected topology $\Lambda$ by the sum of the gradients for the local contributions $g^{\textrm{local}}$, and the one for the coupling inputs, thus providing:
\begin{equation}
	g^{\Lambda} = \nabla J(\tilde{x},\tilde{u})\approx g^{\textrm{local}} + \sum_{ij \in \Lambda} \Delta g^{\Lambda}_{ij}
\end{equation}
where $\Delta g^{\Lambda}_{ij}$ are coupling contributions for $\Lambda$. Then, a vector of $|\mathcal{E}|$ binary variables $\delta$ is defined to establish if a link $(i,j)\in\mathcal{E}$ belongs to a certain topology. Accordingly, the partitioning of the network is retried by solving a binary quadratic program with cost function $\delta^T Q \delta + R \delta$, 
where the weighting matrices $Q$ and $R$ are functions of the gradient approximation $g^{\Lambda}$. Further details about the problem derivation, and a theorem bounding the performance degradation in the computation of the cost function based on the topology selection are given in \cite{chanfreut_FastClusteringMultiagent_2022}. An analysis of the scalability of the approach is also performed, comparing it against genetic algorithm optimization, and a greedy algorithm. The control approach is then validated on an urban transportation network \cite{deoliveira_MultiagentModelPredictive_2010} with eight intersections, and performance is compared w.r.t.\ CMPC. Simulations show how this strategy can reduce the number of active communication links more than the $40\%$ while retaining good levels of performance.

\subsection{Hierarchical approaches for time-varying graphs}
\label{sec:opt-hier-time}
In \cite{riccardi_GeneralPartitioningStrategy,riccardi_CodePublicationGeneral_2025}, a binary quadratic programming (BQP) approach is used to obtain the partitioning of a network at different levels of aggregation through a granularity parameter $\alpha$, for a potentially time-varying topology. The approach is based on constructing an equivalent graph representation of the network using partial derivatives of the dynamics. Specifically, for a network characterized by the nonlinear difference equation \eqref{eq:nonlinear-system}, 
a graph $\mathcal{G}= (\mathcal{V},\mathcal{E})$ is constructed by defining the set of input and state nodes, respectively $\mathcal{V}_x, \mathcal{V}_u$, according to the variables in $x$ and $u$. The weighting, and therefore the definition, of the edges in $\mathcal{E}$ is given by \eqref{eq:weights-edges}.
This definition provides a time-varying graph for the general nonlinear case, for which re-partitioning should be performed at each time step. 
For this equivalent graph, an intermediate algorithmic step, detailed in Sec.\ \ref{sucsec:alg-hier-time}, is used to select the fundamental system units (FSUs), denoted by $\mathcal{S}_i$, to be used for the construction of the agents (but these can also be given a priori). Once a number $N_{\textrm{FSU}}$ of FUSs is constructed,
a global quadratic metric is used into a BQP to select collections $C_i$  of FSUs that have a strong interaction inside the set, and a weak interaction among the sets. This is done introducing a number $N_{\textrm{FSU}}^2$ of binary variables $\delta_{ij}$ with $\delta_{ij} = 1 \Leftrightarrow \mathcal{S}_i \in\mathcal{C}_j$, and zero otherwise,
and defining the intra- and inter-collection weights as:
\begin{align} \label{eq:IQP_metric_1}
	W^{\text{inter}}(\delta) &= \sum_{m = 1}^{N_{\textrm{FSU}}} \sum_{i = 1}^{N_{\textrm{FSU}}}\sum_{\substack{j = 1\\ j\neq i}}^{N_{\textrm{FSU}}}\sum_{\substack{l = 1\\ l\neq m}}^{N_{\textrm{FSU}}}\delta_{i,m}\delta_{j,l}\left(|w(i,j)| + |w(j,i)|\right) \\
	W^{\text{intra}}(\delta) &= \sum_{m = 1}^{N_{\textrm{FSU}}} \sum_{i = 1}^{N_{\textrm{FSU}}}\sum_{j = 1}^{N_{\textrm{FSU}}}\delta_{i,m}\delta_{j,m}\left(|w(i,i)|+|w(i,j)|+\right.\\
	& \,\,\,\qquad\quad\qquad\left.+|w(j,i)|+  |w(j,j)|\right) \nonumber
\end{align}
Additionally, a weighting term to minimize the size of the resulting sets is defined as:
\begin{equation}\label{eq:IQP_metric_3}
	W^{\text{size}}(\delta) = \sum_{m = 1}^{N_{\textrm{FSU}}} \left(\sum_{i = 1}^{N_{\textrm{FSU}}} \delta_{i,m}\right)^2  
\end{equation}
These three weights are used as the cost function in the following BQP that allows finding the optimal partitioning maximizing the interaction strength in the collections for a given value of the parameter $\alpha$ that influences the level of granularity:
\begin{flalign}\label{eq:IQP}
	\min_\delta \quad &  W^{\text{inter}}(\delta) - W^{\text{intra}}(\delta) +\alpha  W^{\text{size}}(\delta) \nonumber \\
	\textrm{s.t.} \quad  &\sum_{j= 1}^{N_{\textrm{FSU}}}\delta_{ij} = 1  \qquad \forall i \\ 
	& \delta_{ij} \in \{0,1\} \hfill \nonumber
\end{flalign}
The constraints ensure that nonoverlapping sets constitute the resulting partitioning, and varying $\alpha$ allows to obtain collections of different sizes. 
The partitioning approach thus defined in \cite{riccardi_GeneralPartitioningStrategy} is applied for partitioning a modular network with 64 agents, and a random network of hybrid systems with 50 agents. The first case shows how varying the granularity $\alpha$ allows to retrieve modules at different aggregation levels, allowing a hierarchical clustering. In the second case, an ADMM-based DMPC approach \cite{summers_DistributedModelPredictive_2012} is deployed for network control. Different simulations are performed: one for CMPC, one for the conventional DMPC-ADMM with 50 agents, and three for partitionings obtained with varying levels of $\alpha$. The simulation results show how the optimization-based partitioning DMPC controllers have a loss of performance w.r.t.\ CMPC below $0.3\%$, while the conventional DMPC-AMM approach with 50 agents has a loss of more than $12\%$. This performance advantage is paid in computation time, which is generally higher for partitioned system w.r.t.\ the conventional ADMM formulation. The approaches are also compared in terms of computational cost by calculating the core seconds for the simulations, i.e., \ the number of seconds necessary to compute the control action in parallel times the number of agents working in parallel. In this regard, optimization-based partitioning allows a computational cost in line with CMPC, while conventional DMPC-ADMM is at least $2.59$ times more expensive. It is important to stress that, even if the framework allows the partitioning for time-varying topologies, the computational cost of a BQP is prohibitive for the online re-partitioning of large systems. This problem is solved if the system only transitions among a given number of known topologies, which allows an offline computation of all the optimal partitionings. Otherwise, an algorithmic modification of the strategy in \cite{riccardi_GeneralPartitioningStrategy} is proposed in Sec.\ \ref{sucsec:alg-hier-time} to overcome the computational complexity associated with this class of problems.

}
{\CBa
\section{Algorithmic Partitioning} \label{sec:algorithmic}
\subsection{Applied to equivalent graph-based representations}
One of the first contributions to graph-based partitioning for the application of non-centralized predictive control is found in \cite{ocampo-martinez_PartitioningApproachOriented_2011}. The starting point of the partitioning strategy is a graph-based representation, proposed as a control-oriented representation described by an incidence matrix \cite{zecevic_ControlComplexSystems_2010, bondy_GraphTheory_2008}.
The graph is divided into non-overlapping subgraphs according to an algorithm developed starting from the graph-partitioning-based ordering algorithm (GPB) \cite{gupta_FastEffectiveAlgorithms_1996}, with various modifications and heuristics to adapt it for control of a complex system. One of the core components of the algorithm is the cut size, i.e.\ the number of links that belong to different subgraphs, which is an indirect measure of the desired subgraph size. 
The partitioning approach is applied to the case study representing the Barcelona drinking water transport network \cite{ocampo-martinez_ImprovingWaterManagement_2009}, using the Dec-MPC technique previously developed in \cite{ocampo-martinez_ModelPredictiveControl_2010}, obtaining six groups of systems distributed to three level of control hierarchy. Simulation results show a reduction of about $50\%$ in computation time w.r.t.\ CMPC, while a reduction in the electricity usage is achieved at the expense of a higher water volume required. The overall loss of performance is contained to the $15\%$. This version of the algorithm is developed to work offline, before the control strategy is deployed.

Partitioning based on nested $\epsilon$-decomposition \cite{sezer_NestedEdecompositionsClustering_1986} is proposed in \cite{ocampo-martinez_HierarchicalDecentralisedModel_2012} for decentralized predictive control. For a linear causal system, the $\epsilon$-decomposition works as follows. Construct a matrix $M$ using all system variables as nodes of a graph, i.e.\ build the weighted adjacency matrix:
\begin{eqnarray}
	M = \begin{bmatrix}
		A & B & 0 \\ 0 & 0 & 0  \\  C & 0 & 0
	\end{bmatrix}
\end{eqnarray}
Then, for a given threshold $\epsilon$, compute the permutation matrix $P$ that provides a new block decomposed matrix $\tilde{M} = P^\intercal M P $ consisting of $N$ block such that, for the off-diagonal terms, it holds that $\tilde{M}_{ij} \leq \epsilon$. This decomposition transforms the network into $N$ connected subgraphs where interconnections are defined by the off-diagonal terms of $\tilde{M}$ and their strength constrained by the choice of $\epsilon$. A maximum number of $|M|$ nested $\epsilon$-decompositions is possible for any given $M$. Further details and stability analysis of this decomposition are presented in \cite{sezer_NestedEdecompositionsClustering_1986}. In \cite{ocampo-martinez_HierarchicalDecentralisedModel_2012}, the $\epsilon$-decomposition is applied to the Barcelona drinking water network \cite{ocampo-martinez_ImprovingWaterManagement_2009} incorporating a heuristic selection of $\epsilon$, and a hierarchical Dec-MPC strategy is applied to the resulting three-subsystem network. The architecture is validated against a CMPC controller implementation showing an overall performance loss always smaller than $2\%$, with a reduction of computation times up to $35\%$.  

An algorithmic approach for nonlinear systems is devised in \cite{kamelian_NovelGraphbasedPartitioning_2015}. This approach is also based on the control-oriented representation and the derivation of the incidence matrix \cite{bondy_GraphTheory_2008}; however, in this case the matrix is constructed accounting for relations among system variables, where each input state and output is considered as a distinct node. A general nonlinear dynamics of the form:
\begin{eqnarray}
	\mathcal{S} : \left\{\begin{matrix}
		x_{k+1} = f(x_k,u_k,w_k) \\ y_k = g(x_k,u_k,w_k) \hfill
	\end{matrix}\right.
\end{eqnarray}
is used to construct the graph; however, this dynamics is linearized around an operating point to derive a weighting of the associated graph; specifically, the matrices $(A,B,C,D)$ resulting from the linearization are used. The algorithmic approach starts by the centers of the clusters as the input variables.
Then, a sorting procedure is used to order the state and output vertices according to their degree.
A merging phase groups subgraphs based on their number of edges.
The procedure is regulated by the cut size, according to \cite{jamoom_OptimalDistributedActuator_1998}, but also considering the number of resulting groups. 
The resulting partitioning is used in the deployment of a Dec-NLin-MPC strategy over an industrial chemical plant constituted by two continuous stirred-tank reactors in cascade \cite{venkat_DistributedModelPredictive_2006, bakule_DecentralizedControlOverview_2008a}. The results show that the decentralized approach proposed in \cite{kamelian_NovelGraphbasedPartitioning_2015} has a performance comparable to C-NLin-MPC, and superior performance w.r.t.\ the Dec-NLin-MPC approach proposed in \cite{venkat_DistributedModelPredictive_2006}.

A partitioning approach based on the strength of interaction among subsystems is proposed in \cite{zheng_CouplingDegreeClusteringbased_2018b}. The underlying DMPC strategy in this work is the dual-mode DMPC proposed in \cite{li_ImpactedregionOptimizationDistributed_2015}. Specifically, for a network composed of subsystems interacting through a dynamical coupling, the approach in \cite{zheng_CouplingDegreeClusteringbased_2018b} requires grouping these subsystems into larger virtual middle-scale subsystem, also called M-subsystems. Then, a cooperative DMPC strategy is deployed, for which, at each time step, iterative optimization is performed within the systems in the M-subsystem, and communication occurs only once between different M-subsystems. The clustering approach requires the determination of these M-subsystems to minimize the coupling strength. Further details about the definition of weakly coupled M-subsystems are detailed in the paper.
Once the weakly-coupling condition is established, for a selected threshold $\delta$ the variable adjacency matrix ${A}(\delta) = (a)_{ij}$ defining the M-subsystems is obtained as $a_{ij} = 1$ if $\|A_{ij}\| \geq \delta$, and  $a_{ij} = 1$ otherwise. The algorithmic clustering approach consists of finding such $\delta$ and a permutation matrix $T$ such that $T^\intercal\tilde{A}T$ is block-diagonal, and the overall system is weakly coupled.
The clustering algorithm consists of gradually reducing $\delta$ from a given initial value $\delta_0$ until the decomposition into weakly coupled M-subsystems is achieved, where conditions are validated at each time step. Algorithms details are in  \cite{zheng_CouplingDegreeClusteringbased_2018b}. The overall approach is validated on a case study for building temperature regulation against CMPC. The system comprehends eight rooms that should keep the temperature variation at zero despite external influences. The DMPC approach can stabilize the network, as CMPC, but only qualitative results are provided, and some performance degradation is present. The paper also provides theorems for the stability and recursive feasibility of the DMPC strategy.

A framework for algorithmic partitioning of nonlinear systems based on the equivalent graph representation of linearized dynamics around an operating point is proposed in \cite{rocha_PartitioningDistributedModel_2018}. In this approach, each time a re-linearization of the nonlinear dynamics is performed, the system might be re-partitioned. The re-partitioning trigger is not specified, but it is reasonable to assume that it coincides with a change in the topology of the associated graph. The partitioning algorithm proposed is based on the iterative grouping of input-state-output variables, followed by a controllability check. The algorithm does not guarantee the terminability, or that controllable groups are achievable. The other contributions of the paper are the derivation of two DMPC techniques, cooperative and non-cooperative, working on linearized versions of the models. The viability of the approach is demonstrated for the reactor-separator process \cite{stewart_CooperativeDistributedModel_2010a}, with two reactors in series and a separator.

\subsection{Applied to flow graph representations}
Algorithmic partitioning for power networks using a flow-graph representation is considered in \cite{labella_SupervisedModelPredictive_2022}. The approach is developed in multiple conceptual steps. First, the power network is divided into sources for generators (a set of nodes $\mathcal{V}^{\textrm{source}}$), and sinks for the loads (a set of nodes $\mathcal{V}^{\textrm{sink}}$), thus constructing a flow graph. Then, the optimal power flow problem \cite{frank_IntroductionOptimalPower_2016} for the best and worst case scenarios is solved, i.e.\ treating separately maximum demands and generations to obtain two optimal solutions for transactions of flows between sources and sinks.
The average of these two solutions allows defining the average transaction $x_{ij}^{*}(k)$ between sources $i\in\mathcal{V}^{\textrm{source}}$, and sinks $j\in\mathcal{V}^{\textrm{sink}}$. Then, for each $i\in\mathcal{V}^{\textrm{source}}$ and $j\in\mathcal{V}^{\textrm{sink}}$ the shortest path $\mathcal{L}_{ij}$ is defined \cite{dijkstra_NoteTwoProblems_1959}, and the value $x_{ij}^{*}(k)$ is assigned to all edges in $\epsilon\in\mathcal{L}_{ij}$. Finally, the weight of each edge in the network
is computed by summing all the values $x_{ij}^{*}(k)$ of the shortest paths passing by that edge. A weighted flow graph is thus constructed. A partitioning of this graph for a given number of clusters is obtained using the $k$-way partitioning method minimizing the edge cut 
using the METIS algorithm \cite{karypis_METISSoftwarePackage_1997}. This procedure is performed at the time scale of the clustering procedure, slower than the time scale of the control process. A HMPC approach with local Dec-MPC is then deployed to control local power clusters independently. Local requests of energy activate a supervisory layer if clusters cannot satisfy the demand. Further features, such as using energy storage systems, multiple time scales, and ADMM distributed computations in the supervisory layer, are detailed in \cite{labella_SupervisedModelPredictive_2022}. The approach is implemented on the IEEE 118-bus, showing the online clustering capabilities of the approach.

\subsection{Using frequency-based performance indicators}
The use of the Relative Time-averaged Gain Array (RTGA) as a metric to perform partitioning for DMPC is explored in \cite{tang_RelativeTimeaveragedGain_2018}. The traditional Relative Gain Array (RGA) \cite{bristol_NewMeasureInteraction_1966} has been extensively used in control of multiple-input multiple-output (MIMO) systems \cite{skogestad_MultivariableFeedbackControl_2001}. It is defined by considering the MIMO transfer function of a system $G(s)$, and computing the RGA matrix $\Lambda = G(0) \cdot G^\intercal(0)$. Then, if the element $\lambda_{ij}$ of $\Lambda$ is close to 1, the I/O loop of the MIMO system from input $u_j$ to output $y_i$ is well decoupled from other loops in the systems, and the variables should be paired for control design. This feature has been used for optimal pairing using integer programming in \cite{kookos_AlgorithmicMethodControl_1998}. 
The authors in \cite{tang_RelativeTimeaveragedGain_2018} propose a new metric called Relative Time-Avaraged Gain Array (RTAGA), based on the step response of the system averaged by an exponential distribution function $f(t,\tau) = (1/\tau)\cdot e^{-t/\tau}$, for a parameter $\tau$ characterizing the decay of the exponential. Then, for matrix $G$ of transfer functions, the element $g_{ij}(1/\tau)$ is the intensity of the response $y_i$ for a step input $u_i$ weighted by the distribution $f(t,\tau)$ decaying at time scale $\tau$. Accordingly, the RTAGA matrix is defined as $\Lambda(1/\tau) = G(1/\tau)\cdot G^\intercal(1/\tau)$. The authors proposed the RTGA because it provides a better framework for system decomposition w.r.t.\ RGA. Further details are in \cite{tang_RelativeTimeaveragedGain_2018}. For partitioning the system using the RTGA, the authors of \cite{tang_RelativeTimeaveragedGain_2018} rely on the input-output bipartite graph, and the weighting of each edge $(u_j,y_i)$ representing input-output loops is given by the scalar $w_{ij}$ defined according to the entries of $\Lambda(1/\tau)$ as:
\begin{align}
	w_{ij} = \left\{\begin{matrix} 
		\lambda_{ij} \hfill& 0\leq \lambda_{ij} \leq 1  \hfill\\
		1/ \lambda_{ij} \hfill& \lambda_{ij} > 1 \hfill\\
		0 \hfill& \lambda_{ij} < 0\hfill
	\end{matrix}\right.
\end{align}
Then the modularity $Q$ of the bipartite weighted graph is defined according to \cite{barber_ModularityCommunityDetection_2007}, and modularity maximization is achieved through the Louvain fast unfolding algorithm \cite{blondel_FastUnfoldingCommunities_2008}. Further, heuristics on the choice of the parameter $\tau$ are given. 
The approach is applied for deploying a noncooperative and iterative DMPC control scheme \cite{liu_SequentialIterativeArchitectures_2010} over a reactor-separator process with two continuously stirred tank reactors in series \cite{pourkargar_ImpactDecompositionDistributed_2017d}. Different decompositions of the networks are achieved, and results are compared against CMPC through a quality index normalizing the performance-computation-time product w.r.t.\ CMPC. This quality index is used to determine the best partitioning of the network.

\subsection{Using k-means}
One of the most used algorithms for clustering is k-means \cite{xu_SurveyClusteringAlgorithms_2005}. At the core of the algorithm there is the problem of organizing $N$ objects, e.g.\ vectors $x\in\mathbb{R}^d$, into $K$ subsets. This is achieved by using the definition of Euclidean distance, and an algorithm is developed to minimize the squared error between each object and the center of the clusters. The algorithm starts with an initialization of the centers of the $K$ clusters (either random or informed). Then, each object is assigned to the nearest cluster. Accordingly, the prototype matrix, i.e.\ the matrix containing centroids or the means of the clustering, is updated with the given assignment. The last two steps are iterated until there is no further change in the clusters. The computational cost of the algorithm is $O(N\, K\, d)$. The k-means clustering is well developed, and parallel implementations are available \cite{stoffel_ParallelHmeansClustering_1999} to improve computation times. The interested reader can refer to the survey \cite{xu_SurveyClusteringAlgorithms_2005} for further information. 

The clustering of a wind farm using k-means has been performed in \cite{changqing_FrequencyRegulationControl_2022}. The article focuses on the frequency regulation of a double-fed induction generator, which is affected by both the operating conditions of the plant, and the wind orientation and strength. To improve the frequency regulation of the system, a multi-layer control approach is proposed: MPC \cite{afram_ArtificialNeuralNetwork_2017a} is used for frequency regulation and power output maximization, whereas k-means clustering \cite{vallee_OptimalWindClustering_2011} based on wake-effect interaction is used to spatially cluster the wind turbines. The clustering thus performed allows the division of wind turbines into minimally coupled clusters. The approach is applied to a 25-turbine farm (1.5 MW), showing its effects on frequency regulation and power output w.r.t.\ more traditional control approaches. 

An improved version of $k$-means, i.e. crow search \cite{lakshmi_DataClusteringUsing_2018}, is used in \cite{zhao_OptimalSchedulingStrategy_2023} to cluster a wind farm. Crow search is used in this approach for its improved clustering accuracy and cluster stability, allowing the authors of \cite{zhao_OptimalSchedulingStrategy_2023} to achieve superior cluster quality w.r.t.\ traditional $k$-means. The wind farm is partitioned according to four key performance indicators, which are the power characteristic of the turbine, the smooth coefficient, the generation potential coefficient, and the anomaly coefficient \cite{howlader_IntegratedPowerSmoothing_2015, yin_AbnormalDataCleaning_2022}. 
Given this dataset, the algorithmic partitioning is performed for a given number of clusters. The resulting partitioning is used in a HMPC scheme, and the performance w.r.t.\ CMPC are qualitatively compared in a 12-turbine wind farm case study.

An approach for clustering wind farms based on an approximate linear model of their power tracking \cite{jha_WindTurbineTechnology_2010, chen_ClosedloopActivePower_2019} is proposed in \cite{lin_HierarchicalClusteringbasedOptimization_2020}. Once an estimate of this transfer function is available for each turbine in the farm, \cite{lin_HierarchicalClusteringbasedOptimization_2020} proposes to apply a global fuzzy $c$-means algorithm for clustering the network \cite{siringoringo_InitializingFuzzyCmeans_2019, heo_ExtensionGlobalFuzzy_2010}. The approach is deployed on a farm with 20 NREL 5-MW wind turbines \cite{jonkman_Definition5MWReference_2009a}, modeled using SimWindFarm \cite{grunnet_AeolusToolboxDynamics_2010}, and obtaining four clusters. The control approach is hierarchical and employs a proportional controller in the lower layer and an MPC in the upper layer, where in the latter, all the clusters are aggregated into a single performance index. Simulation results show how the proposed strategy outperforms both conventional PD control and CMPC, while reducing computation times.

\subsection{Data-driven decomposition}
Partitioning in a data-driven application is discussed in \cite{zhang_EnhancingCooperativeDistributed_2019}, where water distribution networks are used as examples of large-scale nonlinear systems with coupling dynamics. The scope of a data-driven approach is to capture the nonlinear dynamics that might not figure in purely model-based approaches as \cite{ocampo-martinez_HierarchicalDecentralisedModel_2012}. Once time series data about inputs $\mathbb{U} = \{\tilde{u}_\ell\}_{\ell=1}^{n_u}$ (outlet pressure at the water pumps), states $\mathbb{X} = \{\tilde{x}_\ell\}_{\ell=1}^{n_x}$ (level of water reservoirs), and outputs inputs $\mathbb{Y} = \{\tilde{y}_\ell\}_{\ell=1}^{n_y}$ (pressure at users' nodes) are collected, a system model is defined as $\mathcal{S}(\mathbb{U},\mathbb{X},\mathbb{Y})$. The partitioning problem is then formulated s.t.\ the WDN is divided into $k$ subsystems $\mathcal{S}_i$, where $k$ is a number defined by inspection depending on the shape of the time-series data in matrix $\mathbb{Y}$. The underlying partitioning procedure is then provided by the $k$-shape clustering algorithm for time series sequences \cite{paparrizos_KShapeEfficientAccurate_2015}, and canonical correlation analysis to establish the strength of interaction among the groups of variables, which allows to define strong and weakly coupled neighbors according to heuristic thresholds. The algorithmic procedure for partitioning allows to retrieve $k$ groups of strongly coupled non-overlapping subsystems with approximately the same number of variables. The approach is applied to the water distribution network of Shanghai, using 800 samples captured every 10 minutes from 44 sensors in the network. Different partitionings are obtained by varying the parameters of the algorithm, but the one providing 6 groups is selected since it gives the minimum variance. 
Simulations are performed to compare the proposed enhancing DMPC strategy with the Dec-MPC approach defined in \cite{ocampo-martinez_HierarchicalDecentralisedModel_2012, ocampo-martinez_PartitioningApproachOriented_2011}. Overall, the strategy proposed by \cite{zhang_EnhancingCooperativeDistributed_2019} allows to achieve a reduction in the water pressure of the network, while ensuring stability and robustness, thus reducing leakages and energy requirements.  

\subsection{Hierarchical clustering}
The study \cite{chen_CooperativeDistributedModel_2020b} introduces a cooperative DMPC framework based on topological hierarchy decomposition, aiming to optimize communication efficiency while maintaining global system performance. The theory at the basis of the approach is interpretive structural modeling \cite{attri_InterpretiveStructuralModelling_2013}, which allows to hierarchically structure subsystems based on their coupling strength, ensuring that strongly coupled subsystems are grouped within the same layer, while weakly coupled ones are placed in lower layers. Moreover, it is assumed, not without loss of generality, that only the upper layer influences the lower layer in a sequential cascade. This hierarchical order prioritizes the resolution of the local MPC problems, and their coordination, in the upper-layer subsystems, propagating their optimal control inputs downward, and iterating the process over the fixed down-streamed variables in the lower layer. The update of the input trajectories in the cooperative DMPC is performed through the Gauss-Jacobi distributed optimization method \cite{bertsekas_ParallelDistributedComputation_2015}. Proofs of feasibility and stability of the overall architecture are provided. The approach is tested over a walking beam reheating furnace system and a six-area power system, and validated against the DMPC formulation of \cite{venkat_StabilityOptimalityDistributed_2005a}. In the tests, the hierarchical approach of \cite{chen_CooperativeDistributedModel_2020b} shows the ability to reduce the communication burden, avoiding the transmission of unnecessary information while ensuring system performance. 

The study \cite{chanfreut_ClusteringbasedModelPredictive_2023} introduces a hierarchical clustering-based MPC strategy for optimizing heat transfer fluid flow rates in solar parabolic trough plants \cite{boukelia_ParabolicTroughSolar_2013}. In such systems, parabolic installations focus solar radiation on a pipe transporting a fluid, which will then be used for diverse applications. The challenge in these systems is maintaining the heat transfer homogeneous across the loops the pipe performs in the plant, regardless of meteorological conditions, in this case related to the presence of clouds. The architecture proposed has a two-layer structure. The bottom layer consists of local MPC agents controlling coalitions of loops, while the top layer dynamically clusters loops. For this, the $k$-means clustering algorithm groups loops with similar dynamics, determined by lumped parameters, and recursive least squares estimation \cite{shaferman_ContinuoustimeLeastsquaresForgetting_2021} adapts system parameters in real-time. Moreover, the top layer accounts for variable weather conditions to assign MPC constraints to local agents so that the flow in the pipe can be restricted or increased in specific loops depending on the presence of clouds. The method allows scalability of the MPC architecture, but is sensitive to parameter estimation errors and relies on fixed cluster numbers. Simulations performed on 10-loop and 80-loop plants show significant effectiveness of the technique and minimal performance loss.

\subsection{Input-coupled systems} \label{subsec:alg-input-coupling}
An algorithmic partitioning approach for input-coupled systems is proposed in \cite{wei_EventtriggeredDistributedModel_2020}, where the objective is to derive a novel iterative DMPC strategy with a dynamic communication topology to improve the communication burden of conventional cooperative DMPC with a static communication topology. The network is assumed to be composed by a number $n$ of coupled linear dynamics of the form 
\begin{equation}
	x_i(k+1) = A_{ii}x_i(k) + B_{ii}u_i(k) + \sum_{j\in\mathcal{N}_i}\left[A_{ij}x_j(k) + B_{ij}u_j(k)\right].
\end{equation}
By using the Kalman canonical form, the state coupling can be avoided with an appropriate selection of the new subsystem states \cite{stewart_CooperativeDistributedModel_2010a}, providing new input-coupled local dynamics\footnote{note that this state transformation can be already considered a partitioning of the state of the network.} $\bar{x}_i(k+1) = \bar{A}_{ii}\bar{x}_i(k) + \bar{B}_{ii}\bar{u}_i(k) + \sum_{j\in\mathcal{N}_i} \bar{B}_{ij}\bar{u}_j(k)$. A sensitivity analysis is performed to establish the effect of the coupling variables on the optimization problem. On this basis, a threshold triggering communication between local controllers is derived. Accordingly, an algorithmic procedure determines the entries of a communication matrix at each time step, thus obtaining an event-triggered topology change for the communication networks defining the local controllers. The proposed DMPC strategy is validated for the four-tank water system \cite{gross_DistributedModelPredictive_2015} against cooperative DMPC with static topology, effectively reducing the communication burden.

\subsection{Hierarchical clustering for input-coupled systems} \label{subsec:alg-hierarchical-input-coupling}
Hierarchical clustering for input-coupled systems is proposed in \cite{wang_HierarchicalClusteringConstrained_2023}, where a distance function induced over minimal robust positively invariant sets is used as an underlying metric for the clustering algorithm. Specifically, the hierarchical clustering of \cite{xu_SurveyClusteringAlgorithms_2005} is used to to design a robust Dec-MPC, as the one of \cite{trodden_DistributedPredictiveControl_2017}. 
The approach is iterative and defined for a given number of hierarchy levels, starting from the network considering each agent as an individual cluster. A tuning parameter $\alpha > 0$ is defined to perform the clustering. At each step, the minimum distance $d^{\textrm{min}} = \min_{ij}d_{ij}$ is computed. Then, the procedure aggregates together the agents for which $d_{ij} > (1+\alpha) d^{\textrm{min}}$. Some refinements are performed on the resulting clusters to ensure consistency. Then, the procedure is iterated for the next hierarchy level until one single agent representing the entire network is obtained. The partitioning approach proposed in \cite{wang_HierarchicalClusteringConstrained_2023} is validated by computing the size of the resulting minimal robust positively invariant sets for different clustering procedures, showing how it outperforms other strategies in maximizing the sizes of the sets. The case study is a 43 agents flow system \cite{koeln_StabilityDecentralizedModel_2017}. However, the impact of the proposed partitioning on the performance of the robust Dec-MPC strategy has not been explored in the work.

\subsection{Computational complexity and controllability}
An algorithmic partitioning approach oriented at the minimization of the computational complexity of the resulting DMPC architecture while ensuring the controllability of the resulting subsystems is developed in \cite{arastou_OptimizationbasedNetworkPartitioning_2025}. This approach is motivated by previous studies oriented at the minimization of the communication cost in DMPC approaches, such as \cite{barreiro-gomez_TimevaryingPartitioningPredictive_2019}. To this aim, in \cite{arastou_OptimizationbasedNetworkPartitioning_2025}, the authors develop an algorithm for the reduction of the number of iterations $\bar{r}$ required to retrieve an (approximate) solution of a distributed optimization problem with a desired accuracy $\epsilon$. The idea behind this approach is that by finding the partitioning that minimizes the number of iterations of the DMPC, the amount of information shared among the agents will also be minimized. However, while it is correct that the communication burden grows with the number of iterations of the DMPC strategy, it is not exact to say that minimizing the number of iterations automatically minimizes the communication burden, or computation time. In \cite{arastou_OptimizationbasedNetworkPartitioning_2025}, the desired partitioning is obtained through the minimization of the cost function $F$ dependent by the selected partitioning $\mathcal{P}^j$ is defined as:
\begin{equation}\label{eq:global_computaiton}
	F(\mathcal{P}^j)= \left(\log_{\beta(\mathcal{P}^j)}\frac{\epsilon}{J(\boldsymbol{x}_{(0|k)},\boldsymbol{u}_{(0|k)}^0)}-1\right)\sum_{i = 1}^{N_{\mathcal{P}_a}} g(n_i,m_i,N,n_c)
\end{equation}
where $f(\mathcal{C}_i) = g(n_i,m_i,N,n_c)$ is a function of the number of states and inputs of the collection $\mathcal{C}_k$, the prediction horizon $N$, and of the number of constraints $n_c$; and $J(\boldsymbol{x}_{(0|k)},\boldsymbol{u}_{(0|k)}^0)$ is the cost function for the first prediction step, evaluated with the first iteration of the control action. The minimization of \eqref{eq:global_computaiton} is sought using the Kernighan-Lin algorithm \cite{kernighan_EfficientHeuristicProcedure_1970}, also at the basis of the approach \cite{ananduta_OnlinePartitioningMethod_2019}, and based on iterative node exchange.
The approach is applied to the control of a simplified version of the Richmond water distribution network, Yorkshire, UK \cite{vanzyl_OperationalOptimizationWater_2004}, using a flow-based graph representation. The simulations show how the DMPC strategy applied to different network partitionings always ensures a negligible loss in performance, while showing computation times that gradually decrease with a higher number of sets in the partition.

}
{\CBa
\section{Community-Detection-Based Partitioning}
\label{sec:modularity-based}

\subsection{Fundamentals and modularity metric}
Community detection is a fundamental branch of modern network theory, and its scope is the identification of groups of elements in the network that have a higher probability of being strictly connected to each other w.r.t.\ other member in the network \cite{fortunato_CommunityDetectionNetworks_2016}. Among the methodologies for community detection, we find optimization-based, algorithmic, dynamics-based, and consensus-based approaches, as well as methods based on statistical inference, and spectral or hierarchical clustering: an extended discussion about these topics can be found in \cite{fortunato_CommunityDetectionGraphs_2010a, fortunato_CommunityDetectionNetworks_2016}. Partitioning approaches based on the quality function called \textit{modularity} belong to the broader class of methods for community-detection in graphs \cite{fortunato_CommunityDetectionGraphs_2010a}, i.e.\, they are clustering methodologies, often algorithmic. In this context, modularity is a metric that has been consistently used to quantify the quality of the resulting clusters, not only in network theory, but also for control systems. Several studies in the field of partitioning for predictive control use modularity as a fundamental metric. Therefore, we treat this topic separately from other partitioning approaches.

In control theory, modularity has been applied to compute the partitioning of the graph associated with a dynamical system. The method to derive this graph has been discussed in Sec.\ \ref{subsec:equivalent-graph}. However, modularity-based partitioning can also be deployed over agent-based representations of the form Sec.\ \ref{subsec:agent-graph}, which is a conceptually different use case. In general, for a network with a given adjacency matrix $\mathcal{A}$, and a partition into $N$ communities $\mathcal{P} = \{\mathcal{C}_1, \ldots, \mathcal{C}_N\}$, the modularity $Q$ index is constructed as:
\begin{equation}
	Q = \frac{1}{2m} \sum_{ij}\left(\mathcal{A}_{(i,j)} - \frac{k_i^{\textrm{in}}k_j^{\textrm{out}}}{2m}\right)\delta_{ij}
\end{equation}
where $\mathcal{A}_{(i,j)}$ is the $ij$-th element of the adjacency matrix, $k_i^{\textrm{in}}$ and $k_i^{\textrm{out}}$ are respectively the in- and out-degree of node $i$ in the network, $m$ is the total number of edges, and the binary variable $\delta_{ij}$ is equal to 1 if nodes $i$ and $j$ are in the same community, and zero otherwise. Modularity-based partitioning approaches all focus on finding the partitioning $\mathcal{P}$ that maximizes the modularity $Q$ (usually, for a given number $N$ of communities). In the remainder of this section, we will discuss how modularity-based partitioning has been used in predictive control, and provide different examples.

\subsection{Maximization of modularity by iterative bipartition of the network}
The most used methodology for modularity maximization in control has been presented in \cite{jogwar_CommunitybasedSynthesisDistributed_2017b}. The approach is based on the construction of the modularity matrix $\mathcal{B}$, whose entries are defined as:
\begin{equation}
	\mathcal{B}_{(i,j)} = \mathcal{A}_{(i,j)} - \frac{k_i^{\textrm{in}}k_j^{\textrm{out}}}{m}
\end{equation}
Then, the partitioning approach iteratively splits the network into two communities. To this aim, a vector $\boldsymbol{s}$ with a size equal to the number of nodes in the network is defined as follows. When a split is performed, the network $\mathcal{N}$ is divided into two communities: $\mathcal{C}_a$ and $\mathcal{C}_b$. Accordingly, the $i$-th entry of $\boldsymbol{s}$ is defined to be equal to 1 if $i\in \mathcal{C}_a$, and $-1$ if $i\in \mathcal{C}_b$. The modularity associated with this new partition of the network is then:
\begin{equation}
	Q = \frac{1}{4m} \boldsymbol{s}^\intercal \left(\mathcal{B} + \mathcal{B}^\intercal\right) \boldsymbol{s}
\end{equation}
The specific partitioning algorithm used to perform the modularity maximization of the basis of the iterative division is \cite{leicht_CommunityStructureDirected_2008}, which successively divides the network into communities using approximate spectral optimization for the divisions. Fine-tuning by node shifting \cite{newman_ModularityCommunityStructure_2006b} is performed at each step to improve the partitioning quality. The limitation of this approach is that it neglects any measure of the strength of interaction among nodes and only accounts for topological information. Also, there is no guarantee that the resulting partitions will be non-autonomous systems because no distinction is made between node variables. Consequently, the algorithm may result in partitions containing only input or state nodes, which has limited applicability from the control perspective. Another aspect to consider is the potential controllability of the resulting partitions, for which structural controllability can be tested by verifying the input reachability and no dilation condition \cite{siljak_DecentralizedControlComplex_1991}, but this is not explicitly performed in \cite{jogwar_CommunitybasedSynthesisDistributed_2017b}.

The paper \cite{pourkargar_ImpactDecompositionDistributed_2017d} investigates the impact of system decomposition on the performance and computational efficiency of DMPC applied to nonlinear process networks \cite{liu_SequentialIterativeArchitectures_2010}. The study compares the partitioning of a network obtained through community detection 
with intuitive partitioning given by expert subsystem selection according to energy or technical considerations. The metrics used for comparison are the closed-loop control performance and computational burden. The analysis is conducted on a reactor-separator process, where sequential and iterative DMPC formulations \cite{christofides_DistributedModelPredictive_2013b} are compared against CMPC.
The empirical results obtained by applying the different predictive control strategies show that modularity-based community detection methods perform close to CMPC with significantly reduced computation time. 
However, the study has some limitations. The optimality of the decomposition, as always for modularity-based approaches, is not explicitly guaranteed, as modularity maximization does not necessarily imply the maximization of control performance. Additionally, the method does not consider adaptive decomposition, meaning the partitioning remains static even under changes in system conditions. 

Extension of the partitioning methodology \cite{jogwar_CommunitybasedSynthesisDistributed_2017b} to weighted graphs using the module of the partial derivatives of the dynamics around the operating points for nonlinear systems is proposed in \cite{jogwar_DistributedControlArchitecture_2019}. 
The partitioning procedure relies on a modified version of the multiway spectral community detection algorithm \cite{zhang_MultiwaySpectralCommunity_2015} developed for unweighted graphs.  

\subsection{For optimization problem decomposition}
An algorithmic partitioning approach for the optimization problem decomposition using community detection has been proposed in \cite{tang_OptimalDecompositionDistributed_2018}. The optimization problem related to DMPC considered in this work is assumed to be in a ``separable'' form:
\begin{align}
	\min_{v} & \,\,f_1(v_1) + \ldots + f_n(v_n) \\
	\textrm{s.t.} & \,\, c_j(v_1,\ldots v_n) = 0, \,\, j = 1,\ldots,m \\
	& \,\, v_i \in\mathcal{V}_i, \,\, i = 1,\ldots,n
\end{align}
where the scalar variables in $v$ belong to decoupled intervals, the objective function is separable, and the coupling in the problem only arises in the equality constraints. This setting is common in MPC, where in a network of agents, each has its own individual objective, and is subject to local constraints in the state and input spaces, but they are dynamically coupled with neighbors. To decompose the problem, the authors of \cite{tang_OptimalDecompositionDistributed_2018} use two different graph representations. In the first, they use a bipartite graph, where variables are linked to constraints according to the existence of their partial derivatives, thus capturing their functional interaction. In the second graph, they use a unipartite representation using the variables as nodes, and the number of coupling constraints as arcs. From these two graphs, it is possible to obtain adjacency matrices, and accordingly find the partitioning of these graphs that minimizes the modularity, both for unipartite \cite{newman_FindingEvaluatingCommunity_2004}, and bipartite \cite{barber_ModularityCommunityDetection_2007} representations. Modularity optimization is achieved using the Louvain fast unfold algorithm \cite{blondel_FastUnfoldingCommunities_2008}. The approach is validated for control of a reactor-separator process \cite{liu_DistributedModelPredictive_2009, stewart_CooperativeDistributedModel_2010a}, with two reactors in series and a separator. The approach deployed is an ADMM-based DMPC \cite{bertsekas_NonlinearProgramming_1999}, and is validated against nonlinear CMPC. The results show how the DMPC implementation can outperform CMPC for this nonlinear setting while reducing computation time by more than $50\%$.

Optimization problem decomposition based on modularity optimization is proposed in \cite{segovia_DistributedModelPredictive_2021} through the use of optimality condition decomposition (OCD) \cite{conejo_DecompositionTechniquesMathematical_2006}, to overcome the assumption that the cost function of the optimization problem must be separable to decompose it. For a given non-completely-coupled optimization problem:
\begin{align}
	\min_{\boldsymbol{z}} & \,\,f(\boldsymbol{z}) \\
	\textrm{s.t.} & \,\, b(\boldsymbol{z}) \leq 0
\end{align}
the OCD allows the ploblem to be decomposed into $N$ subploems, for which a relaxed formulation \cite{bertsekas_ConstrainedOptimizationLagrange_1996} takes the form
\begin{align} \label{eq:OCD-Lagrange}
	\min_{\{z^{(i)}\}_{i=1}^N} & \,\,\sum_{i=1}^{N}f^{(i)}(z^{(i)}) + \lambda^{(i)}h^{(i)}\left(z^{(1)},\ldots,z^{(N)}\right) \\
	\textrm{s.t.} & \,\,h^{(i)}\left(z^{(1)},\ldots,z^{(N)}\right) \leq 0 \qquad i \in \{1,\ldots,N\} \nonumber \\
	& \,\,g^{(i)}\left(z^{(i)}\right) \leq 0 \qquad i \in \{1,\ldots,N\} \nonumber
\end{align}
where $z^{(i)}$ is the variable of the $i$-th subproblem, $\boldsymbol{h}$ is a set of complicating constraints without which the subproblems would be independent, $\boldsymbol{g}$ are the constraints resulting from the conversion of $b(\boldsymbol{z}) \leq 0$, and $\lambda$ are the Lagrange multipliers. To the problem \eqref{eq:OCD-Lagrange} can be associated the matrix of first-order Karush-Kuhn-Tucker condition \cite{boyd_ConvexOptimization_2004}. This matrix can naturally be interpreted as a graph $\mathcal{G} = (\mathcal{V}, \mathcal{E})$, for which modularity-based community detection can be applied. Specifically, in the work \cite{segovia_DistributedModelPredictive_2021}, modularity maximization is achieved through the fast unfold algorithm \cite{blondel_FastUnfoldingCommunities_2008}, thus providing a decomposition of the optimization problem and consequently a partitioning of the system. The resulting OCD-DMPC approach is qualitatively validated on the quadruple-tank benchmark system \cite{alvarado_ComparativeAnalysisDistributed_2011} against other MPC strategies, and on the Barcelona drinking water transport network.

\subsection{Frequency-based graph weighting}
The use of a frequency-based index to perform partitioning through community detection is explored in \cite{wang_DistributedModelPredictive_2023}. In this paper, the network is represented through an input-output bipartite graph, as in Sec.\ \ref{subsec:io-graph}. The edges connecting I/O variables are weighted through the linearized frequency response between each pair of variables. Specifically, the integral of the magnitude of the transfer function between two variables $(u_i,y_j)$ for a given range of frequencies $[\omega_1, \omega_2]$ is computed as:
\begin{equation}
	\beta_{ij} = \int_{\omega_1}^{\omega_2} \frac{|G_{ij}(j\omega)|}{\sqrt{1+|G_{ij}(j\omega)|^2}} d(\omega)
\end{equation}
and then a normalization is used to obtain the weights $w_{ij}=1-e^{-\beta_{ij}}$. This allows to retrieve a monotonically increasing weighting in the range $[0,1]$ for all the edges. The computation of the partitioning based on this weighting is performed through a modified version of Barber's algorithm \cite{barber_ModularityCommunityDetection_2007}. Moreover, the gap metric \cite{zames_UncertaintyUnstableSystems_1981} is introduced as a way to quantify the stability of the I/O functions, and incorporated into the partitioning algorithm as a quantity to be minimized along with modularity maximization. The resulting partitioning algorithm is used to deploy DMPC over two different case studies, and compared with CMPC and DMPC with partitioning computed using the conventional modularity maximization. The first experiment involves a reactor separator process consisting of two continuously stirred tank reactors and a flash separator \cite{liu_TwotierControlArchitecture_2010}; the second is an air separation process. The empirical results show how different decompositions of the network impact the performance of the DMPC, showing that also frequency-based modularity maximization is not always the best choice, which is in line with the concept that modularity maximization does not provide by itself the best partitioning in terms of performance. Additionally, the technique proposed only works with linear systems. 

\subsection{Time-varying graph representations}
Exploration of a partitioning algorithm for time-varying systems is proposed in \cite{arastou_OptimizationbasedNetworkPartitioning_2025} where nonlinear dynamics of the following form are considered:
\begin{align}\label{eq:nonlinear_continuous}
	\dot{x}(t) & = f(x(t)) + g(x(t),u(t)) \\
	y(t) & = h(x(t)). \nonumber
\end{align}
For this class of systems, an associated graph representation is constructed using as weighting for the edges the partial derivatives of the dynamics w.r.t.\ the variables. Specifically, denoting with an arrow an edge between variables, the corresponding weights are defined as in \cite{kravaris_GeometricMethodsNonlinear_1990}:
\begin{equation} \label{eq:weights_nonlinear}
	u_i \rightarrow x_j :  \left|\frac{\partial g_j}{u_i}\right|; \quad x_i \rightarrow x_j :  \left|\frac{\partial f_j}{x_i}\right|; \quad x_i \rightarrow y_j :  \left|\frac{\partial h_j}{x_i}\right|
\end{equation}
Once all the weights are defined, the corresponding adjacency matrix $A^{\textrm{adj}}$ is constructed, and accordingly, the modularity metric $Q$ can be used for graph partitioning. The algorithm used in this case is the spectral community detection detailed in \cite{zhang_MultiwaySpectralCommunity_2015}. It is important to notice that if a change in the structure of the dynamics \eqref{eq:nonlinear_continuous} occurs such that the weights in \eqref{eq:weights_nonlinear} are altered, then the graph associated with the network changes. This aspect is explored in the work \cite{arastou_OptimizationbasedNetworkPartitioning_2025} by considering the same plant in two different operating points. The case study is the benzene alkylation process using four continuous stirred tank reactors and a flash separator controlled through the DMPC strategy developed in \cite{pourkargar_DistributedEstimationNonlinear_2019}, which also involves the partitioning of the process using community detection. The strategy developed in \cite{arastou_OptimizationbasedNetworkPartitioning_2025} shows an improvement in the performance up to $26.9\%$ w.r.t.\ the one in \cite{pourkargar_DistributedEstimationNonlinear_2019}. 
	
\subsection{Hierarchical approach for time-varying graphs}
\label{sucsec:alg-hier-time}
A hierarchical algorithmic approach for time-varying topologies is presented in \cite{riccardi_GeneralPartitioningStrategy,riccardi_CodePublicationGeneral_2025}. Starting from the graph representation already introduced in Sec.\ \ref{sec:opt-hier-time}, the partitioning problem is divided into two parts: first, a selection of fundamental and indivisible systems dynamics, called FSUs, is performed algorithmically; then the FSUs are aggregated into collections, called composite system units (CSUs), for which a controller is designed. The algorithmic procedure for the selection of FSUs is motivated by the fact that the subsystems in the sense of Sec.\ \ref{subsec:terminology} might not be given a priori. In this case, a selection of the subsystems is necessary to transform a network into a multi-agent representation. In \cite{riccardi_GeneralPartitioningStrategy}, an algorithmic procedure addressing this problem is proposed. 
Application of the algorithmic selection of FSUs allows to obtain a network structure $\mathcal{N} = \{\mathcal{S}_1,\ldots,\mathcal{S}_{N_{\textrm{FSU}}}\}$ from any given dynamics of the form \eqref{eq:nonlinear-system}. The second part of the partitioning strategy is an aggregative procedure for merging FSUs into CSUs, the collections constituting the partitioning. To this aim, a modularity-inspired metric is designed to capture the strength of the interaction intra- and inter-CSUs, while balancing their size. These individual components of the metric are:
\begin{align}
	& W_{\mathcal{C}_i}^{\text{intra}} = \sum_{s,t\in\mathcal{V}_{i}} |w_{i}(s,t)| \label{eq:intra-agent-nonlinear} \\
	& W_{\mathcal{C}_i}^{\text{inter}} = \sum_{s\in\mathcal{F}_{\mathcal{C}_i}} \sum_{j\in\mathcal{N}_{\mathcal{C}_i}} \sum_{t\in\mathcal{N}_{s}\cap\mathcal{V}_j} |w_{i}(s,t)| + |w_{j}(t,s)|  \label{eq:inter} \\
	& W_{\mathcal{C}_i}^{\text{size}} = |\mathcal{C}_i|^2
\end{align}
where $\mathcal{V}_i$ is the set of the nodes in the set $\mathcal{C}_i$, and $\mathcal{F}_i$ its frontier. Using these terms, the global metric for partitioning, named partition index, is defined as:
\begin{equation} \label{eq:metric_greedy}
	p^{\textrm{idx}}(\mathcal{P}) = \frac{\displaystyle  \sum_{i=1}^{m}W_{\mathcal{S}_i}^{\text{Intra}}}{\displaystyle 1 + \sum_{i=1}^{m}W_{\mathcal{S}_i}^{\text{Inter}}} + \frac{\alpha}{\displaystyle 1 + \sum_{i=1}^{m} W_{\mathcal{S}_i}^{\text{size}}}
\end{equation}
where $\alpha$ is the parameter affecting the granularity, thus allowing balancing the effect of the size of the collections in the partitioning. A greedy algorithmic procedure is used to iteratively assign the subsystems $\mathcal{S}_i$ to the collections $\mathcal{C}_i$ such that at each assignment the variation $\Delta p^{\textrm{idx}} = p^{\textrm{idx}}(\mathcal{P^{\textrm{new}}}) - p^{\textrm{idx}}(\mathcal{P^{\textrm{old}}})$ is maximized. 
The approach is validated on the same random network of hybrid dynamical systems described in Sec.\ \ref{sec:opt-hier-time} using the same DMPC strategy. The simulation results show how the loss in performance for agents designed with algorithmic partitioning is of an additional $1\%$ w.r.t.\ the ones obtained with optimization-based partitioning. However, the computation times are comparable to the ones of conventional DMPC-ADMM with 50 agents (1.75 times slower), but having the smallest computational cost among all the approaches in terms of core seconds. Therefore, the algorithmic partitioning strategy proved to be an effective alternative to the optimization-based approach in \cite{riccardi_GeneralPartitioningStrategy} and detailed in Sec.\ \ref{sec:opt-hier-time}. The interesting aspect is that the algorithmic approach has a maximum computational complexity of $O(n_{\textrm{FSU}}^4)$, while the optimization-based is an NP-Hard problem. Accordingly, depending on the size complexity, time constant, and desired quality of the partitioning, the algorithmic approach can be suitable for online re-partition in case of time-varying topologies. We also note that the partition index defined in \eqref{eq:metric_greedy} can be used in global search optimization (genetic algorithm) to obtain a non-overlapping partitioning, as similarly proposed in \cite{riccardi_GeneralizedPartitioningStrategy_2024a,riccardi_CodeUnderlyingPublication_2024c}.	
	
}

\subsection{Applications and case studies}
Application of the modularity-based partitioning methodology derived in \cite{jogwar_CommunitybasedSynthesisDistributed_2017b} is performed in \cite{moharir_DistributedModelPredictive_2018} for iterative DMPC of an Amine gas sweetening plant. The decomposition of the relatively small plant shows how modularity maximization is achieved when two communities are obtained, and further partitioning the system into three communities does not improve the modularity. Modularity maximization also accounts for the structural information about the plant, ensuring the existence of well-posed subsystems (i.e.\ subsystems for which a controller can be defined, having at least one input and one output of the original plant). No further division of the plant is proposed. The DMPC architecture is compared against CMPC, Dec-MPC, and DMPC for a different partitioning (sub-optimal in terms of modularity). The modularity-based DMPC is the best-performing non-centralized strategy, approaching CMPC results while reducing computation times. Given the reduced size of the plant, all possible modularity-based partitions of the systems providing well-posed subsystems can be evaluated in this case; however, the procedure still relies on expert knowledge, heuristics, and inspection to be performed accurately.

The approach of \cite{jogwar_CommunitybasedSynthesisDistributed_2017b} is deployed in \cite{pourkargar_DistributedEstimationNonlinear_2019} for both distributed control and estimation of a benzene alkylation process consisting of four continuous stirred-tank reactors, and a flash tank separator. 
Deploying the DMPC architecture for the selected partition provides a good approximation of CMPC results with a reduced computation burden.

A modularity-based partitioning technique has been used in \cite{guo_DynamicIdentificationUrban_2019} to deploy a DMPC strategy for perimeter control of urban traffic. The approach is structured to divide urban networks into regions for which traffic control methods based on the macroscopic fundamental diagram \cite{geroliminis_ExistenceUrbanscaleMacroscopic_2008} can be implemented \cite{an_NetworkPartitioningAlgorithmic_2018}. To this, a two-layer partitioning method is proposed in \cite{guo_DynamicIdentificationUrban_2019}. In the upper layer, congested regions are selected using the dynamic modularity metric for urban traffic introduced in \cite{guo_DynamicIdentificationUrban_2019}. These regions are compact, and a macroscopic fundamental diagram can be identified for them. However, the regions do not cover the entirety of the urban network, i.e.\ non-congested regions are present at their interconnection, defining a boundary. At the lower layer of the partitioning strategy, the boundary region is divided into multiple areas based on spatial proximity using the Euclidean distance, so that a boundary region exists between each two congested areas. Validation of the partitioning approach is performed by applying the DMPC strategy \cite{kim_DistributedModelPredictive_2019} on the case study of the road network in downtown Jinan, China. The proposed approach is validated against a fixed signal control rate, and the boundary-feedback control strategy \cite{zhu_StudyDiscreteBoundaryfeedbackcontrol_2019}, demonstrating how the proposed strategy is the most effective in reducing the total time spent on the road by the drivers, and the total accumulated delaye of the vehicles.  

Modularity optimization has been used in \cite{wang_MPCbasedDecentralizedVoltage_2022} to partition a power network in the presence of photovoltaic inverters and electric vehicles, with the objective of using the charging/discharging capabilities of the latter to mitigate the curtailment of the former. In \cite{wang_MPCbasedDecentralizedVoltage_2022}, a two-step Dec-MPC strategy is developed: in the first phase a modified modularity index is used for partitioning, and in the second step local MPC actions are computed in parallel. The modularity metric is modified to incorporate two ad-hoc performance indicators for power networks. The first is voltage sensitivity, which describes how voltage magnitude changes in nodes after voltage injection in other nodes.
The second is the voltage regulation capacity used for reactive power compensation. The modularity is maximized through the Louvain algorithm \cite{girvan_CommunityStructureSocial_2002}. The resulting approach is qualitatively validated on the IEEE 123 node test feeder, showing the viability of the strategy.  

The paper \cite{he_EnhancingTopologicalInformation_2023} presents a graph-based hierarchical Lyapunov-based DMPC \cite{liu_SequentialIterativeArchitectures_2010} framework. The control framework is based on the selection of communities performed through the multiway spectral community detection algorithm \cite{zhang_MultiwaySpectralCommunity_2015}. This community detection algorithm approaches the modularity maximization problem using spectral methods through a heuristic approach that can work with any number of desired communities. The approach has the same computational complexity of $k$-means clustering; therefore, it is attractive for its scalability. The method partitions subsystems into a relative leader-follower hierarchy by integrating community detection algorithms. The work is posed as an extension of \cite{chen_CooperativeDistributedModel_2020b} to nonlinear systems. However, no formal guarantees are given, and the use of the interpretive structural modeling, as well as the communication strategy, are not entirely clear, contrary to its reference strategy. The proposed architecture minimizes all-to-all communication, requiring only a single inter-layer exchange per sample, reducing the computational burden. The approach is validated over a reactor-separator integrated system developed in  \cite{pourkargar_ImpactDecompositionDistributed_2017d}.

Modularity-based 
algorithmic partitioning using iterative bisection \cite{newman_ModularityCommunityStructure_2006b} is also at the basis of the automatic decomposition approach used in the Shell-Yokogawa platform for advanced Control and estimation \cite{tang_AutomaticDecompositionLargescale_2023}. In this advanced process control technology, partitioning is performed using an equivalent graph representation of the network, with the usual definition of nodes as variables and arcs as relations. 
Iterative bisection is performed according to the algorithm of \cite{newman_ModularityCommunityStructure_2006b}, and the resolution parameter \cite{reichardt_StatisticalMechanicsCommunity_2006} is used to limit the size of the resulting clusters. Two post-processing procedures are used to ensure the connectedness of the resulting components, and to re-balance the sets according to their sizes. Heuristics are used to define the number of clusters, and resolution. The partitioning algorithm is applied to three case studies: a crude distillation process for a refinery, a gas-to-liquid process, and a hydrocracking process, all plants with hundreds of nodes. The resulting partitions are used for the application of DMPC
showing how the distributed computation of the control action can improve the time required for online optimization up to 5 times. However, the impact on the control performance of this approach w.r.t.\ CMPC is not assessed.
{
\section{Partitioning Based on Game-Theoretical Coalition Formation} \label{sec:coalitional}

Coalitional predictive control is among the most recent formulations of non-centralized predictive control \cite{maestre_CoalitionalControlScheme_2014b}. It consists of a combination of optimization-based control and game theory in which dynamical groups of agents cooperate to achieve a coordinated action to optimize some given performance criteria. At the basis of this strategy, there is the concept of \textit{coalition formation}, explained in detail in \cite{ray_GameTheoreticPerspectiveCoalition_2007}, according to which agents in a network group themselves into coalitions to improve their collective outcome. In coalitional control this concept is used to obtain a distributed control strategy.

In this section, the main partitioning strategy used in coalitional predictive control will be introduced first, and then details about fundamental alternatives will be given. After, the theoretical properties of coalitional predictive control and their relation to partitioning are discussed. Various extensions and applications are presented in the remainder of the section.

\subsection{The concept of coalitional control: predictive control and game theory}

Consider a network $\mathcal{N}$ constituted by $N_{\mathcal{A}}$ agents, i.e.\ a collection $\mathcal{N} = \{\mathcal{A}_1,\ldots,\mathcal{A}_{N_{\mathcal{A}}}\}$. A \textit{coalition} $\mathcal{C}$ is any subset $\mathcal{C} \subseteq \mathcal{N}$ where agents in $\mathcal{C}$ cooperate. To each coalition it is assigned a \textit{characteristic function} $v(\mathcal{C})$, mapping the coalitions into real numbers, i.e.\ $v:2^{N_{\mathcal{A}}} \rightarrow \mathbb{R}$, $v(\mathcal{C}) \geq 0$. A \textit{coalitional structure} $\mathcal{P}$ is a collection of disjoint coalitions covering the entire network, in other words a non-overlapping partitioning of the network $\mathcal{P} = \{\mathcal{C}_1,\ldots,\mathcal{C}_{N_{\mathcal{C}}}\}$. The value of the coalitional structure is the sum of the individual contributions of each coalition:
\begin{equation}
	V(\mathcal{P}) = \sum_{\mathcal{C} \in \mathcal{P}} v(\mathcal{C})
\end{equation}
The objective of the \textit{characteristic function game} (CFG) \cite{sandholm_CoalitionStructureGeneration_1999} played by the agents, and that is considered in coalitional control, is to find the coalitional structure that maximizes the total welfare:
\begin{equation}
	\mathcal{P}^* = \arg \max_{\mathcal{P} \in \mathcal{M}} V(\mathcal{P})
\end{equation}
where $\mathcal{M}$ is the set of all possible disjoint partitions of $\mathcal{N}$. {\CBb Various methodologies can be deployed to solve this problem, as it will be presentend in the remainder of the section.}

The framework of the CFG is well suited for developing distributed predictive control strategies since it is, at its core, a distributed optimization approach. One of the first works that formalizes the coalitional predictive control strategy is \cite{fele_CoalitionalControlCooperative_2017b}, where a large-scale system is assumed to be composed of subsystems of the form:
\begin{align}\label{eq:subsystem-coal}
	& x_i(k+1) = f(x_i(k),u_i(k)) + w_i(k) \\
	& w_i(k) = \sum_{j\in\mathcal{N}_i}f(x_j(k),u_j(k)). \nonumber
\end{align}
Each of these subsystems is an agent $\mathcal{A}_i$, and it can participate in a coalition $\mathcal{C}_\ell$, such that $\bigcup_{\ell=1}^{N_{\mathcal{C}}} \mathcal{C}_\ell = \mathcal{N}$, $\bigcap_{\ell=1}^{N_{\mathcal{C}}} \mathcal{C}_\ell = \emptyset$, with $N_{\mathcal{C}}$ the number of coalitions. Each subsystem is associated with a local optimization problem:
\begin{align}\label{eq:opt-coal}
	\min_{\tilde{x}_{i, k}, \tilde{u}_{i, k}} J_i = &  \sum_{j=1}^{N-1}J_{\textrm{s}}(x_i(j|k), u_i(j-1|k)) \\ & + J_{\textrm{f}}(x_i(N|k), u_i(N-1|k)) \nonumber \\
	\textrm{s.t.} \quad& x_i(k+1) = f(x_i(k),u_i(k)) + \hat{w}_i(k) \nonumber \\ 
	& x_i(0|k) = x_i(k) \nonumber \\
	& g_i(\tilde{x}_{i, k}, \tilde{u}_{i, k}) \leq 0 \nonumber 
\end{align} 
where $\hat{w}_i$ is an estimate of the dynamical coupling of $x_i$ with its neighboring subsystems, and $\tilde{x}_{k}$, $\tilde{u}_{k}$ are the state and input sequences defined over the prediction horizon $N$ for a time step $k$. A coalition $\mathcal{C}_\ell$ is formed only if the value of the cost associated with the coalition, i.e.\ $J_\ell$, is lower than the sum of the costs of the individual subsystems. 
Thus, the coalition formation condition is:
\begin{equation}
	J_\ell^* < \sum_{i\in\mathcal{C}_{\ell}} J_i^*
\end{equation}
In the framework of CFG, the simplest characteristic function associated with a coalition $\mathcal{C}_\ell$ is $v(\mathcal{C}_\ell) = J_\ell^*$. In this case the coalition formation problem consists in finding the optimal coalitional structure $\mathcal{P}^* = \arg \max \sum_{\mathcal{C}_\ell \in \mathcal{P}}v(\mathcal{C}_\ell) = \sum_{\ell = 1}^{N_{\mathcal{C}}}J_\ell^*$, with a number $N_{\mathcal{C}}$ of coalitions. This problem is known to be NP-Complete \cite{sandholm_CoalitionStructureGeneration_1999}, inheriting the same complexity of the general partitioning problem. 

The underlying principle of coalition formation described above is shared among all coalitional control strategies, and variations are present in the definition of the characteristic function, the individual payoffs, the implementation of the local MPC controllers, the computation of ordering maps sorting agents costs, and the aggregation algorithm. In the remainder of the section, we report variations, extensions, and applications of the partitioning approach found in coalitional control literature. 

\subsection{Foundational works}

Coalitional predictive control is effectively formalized in the seminal work \cite{fele_CoalitionalControlCooperative_2017b}, where it is applied to energy management in smart grids, specifically to optimize local energy trade among consumer nodes with distributed generation and storage capabilities. In \cite{fele_CoalitionalControlCooperative_2017b}, prosumers (producers-consumers) \cite{larsen_PowerSupplyDemand_2014} cooperate to reduce power dependence from the main grid while minimizing energy exchange costs and transmission losses among them.
To overcome the computation complexity associated with the general coalition formation approach described in the previous section, the partitioning problem is addressed by looking at the coalitional structure $\mathcal{P}$ where the participation preference of each agent $\mathcal{A}$ is sorted according to their Pareto ordering. This is achieved by first using the Shapley value \cite{shapley_ValueNpersonGames_1953} to compute the individual payoffs of each agent $\mathcal{A}$ in each possible subset of agents $\mathcal{S}\subseteq\mathcal{N}$, that for agent $\mathcal{A}_i\in\mathcal{S}$ is defined as:
\begin{equation}
	\displaystyle
	\phi^{\mathcal{S}}_{\mathcal{A}_i} = \sum_{\mathcal{C}\subseteq\mathcal{S}\setminus\mathcal{A}_i} \frac{|\mathcal{C}|!(|\mathcal{S}|-|\mathcal{C}|-1)!}{|\mathcal{S}|!}\left[v(\mathcal{C}\cup\mathcal{A}_i)-v(\mathcal{C})\right]
\end{equation}
Using the Shapley value it is possible to build a mapping $\Phi:\mathcal{N}\times2^{\mathcal{N}}\times\mathbb{Z}\rightarrow\mathbb{R}$ for each agent in each possible coalition, i.e.\ at each time step $k$ a function $\Phi(\mathcal{A}_i, \mathcal{C}_j, k)$ is available. The function $\Phi$ provides for each agent their preferred participation order into coalitions. Accordingly, agents can autonomously organize into the coalitional structure\footnote{The partitioning $\mathcal{P}^{\Phi}$ does not necessarily coincide with the optimal partitioning $\mathcal{P}^*$ in terms of global minimization of the value of the cost function $J(\tilde{x}_k, \tilde{u}_k, \tilde{\delta}_k)$ in \eqref{eq:global-opt}.} $\mathcal{P}^{\Phi}$.
The dynamic coalition formation is guided by an individual payoff $\Phi$ coinciding with the energy exchange with the main grid. Simulation results illustrate how coalitional structures evolve over time, showing how coalitional trade reduces overall costs compared to grid-dependent strategies, as prosumers can access more favorable internal energy prices.

A first extension of \cite{fele_CoalitionalControlCooperative_2017b} is found in \cite{fele_CoalitionalControlSelforganizing_2018}, which proposes a coalitional predictive control strategy with self-organizing agents. The coalition formation strategy is based on a negotiation protocol allowing agents to autonomously form coalitions based on expected performance improvements and cooperation costs. In particular, the coalition formation problem is framed as a transferable utility game \cite{sandholm_CoalitionStructureGeneration_1999, stearns_ConvergentTransferSchemes_1968, cesco_ConvergentTransferScheme_1998}, where agents decide to merge or separate dynamically using a bargaining protocol. Specifically, the coalitional benefit is considered under the assumption of individual rationality, described in the following. Consider two coalitions $\mathcal{C}_1$, $\mathcal{C}_2$, and the value of their individual and aggregated characteristic functions, i.e.\ $v(\mathcal{C}_1)$, $v(\mathcal{C}_2)$, and $v(\mathcal{C}_1\cup\mathcal{C}_2)$. Also, consider the value associated with each of the players in the coalition, denoted by $v(\mathcal{C}_1\cup\mathcal{C}_2)|_{(i)}$ for $i=1,2$, and defined such that $v(\mathcal{C}_1\cup\mathcal{C}_2)|_{(1)} + v(\mathcal{C}_1\cup\mathcal{C}_2)|_{(2)} = v(\mathcal{C}_1\cup\mathcal{C}_2)$. Then the merger of  $v(\mathcal{C}_1)$, $v(\mathcal{C}_2)$ occurs if and only if the the condition $v(\mathcal{C}_1\cup\mathcal{C}_2)|_{(i)} \leq v(\mathcal{C}_i)$ holds for both $i=1,2$, which is known as individual rationality. The value associated with a player $v(\mathcal{C})$ is then considered as an economic index, a utility that can be transferred. Consequently, a bargaining procedure is designed to merge the coalitions considering that, when aggregating two coalitions, the value $v(\mathcal{C}_1) + v(\mathcal{C}_2) - v(\mathcal{C}_1\cup\mathcal{C}_2)$ is a surplus that can be reallocated between the remaining agents. Further details about the strategy and the stability of the coalitions are given in \cite{fele_CoalitionalControlSelforganizing_2018}. This strategy is applied for wide-area control of power networks \cite{chakrabortty_IntroductionWideareaControl_2013}, showing the ability of the architecture to adapt to topological changes that may arise with faults or network extensions. 

{\CBb 
Another bottom-up aggregative procedure for coalitions has been devised in \cite{maestre_PageRankBasedCoalitional_2017a} where a PageRank \cite{brin_AnatomyLargescaleHypertextual_1998, ishii_PageRankProblemMultiagent_2014a} approach is used as the metric to guide local node exchanges among coalitions. For a graph $\mathcal{G} = (\mathcal{V}, \mathcal{E})$, the PageRank associated with each node $i\in\mathcal{V}$ is a scalar $p_i\in[0,1]$, s.t.\ $\sum_{i\in\mathcal{V}}p_i = 1$. Given the neighborhood $\mathcal{N}_i$ of node $i$, its PageRank value is computed as $p_i = \sum_{j\in\mathcal{N}_i}p_j/n_j$, where $p_j$ is the value associated with node $j$, and $n_j$ its number of edges.Once the values $p$ are known for all the nodes, a weighting of the links is performed assigning to each $\epsilon_{ij}$ a weight $w_{ij} = p_i/n_i$ (undirected arcs are handled summing the weights in both directions). Once the weighting of the graph is available, an algorithm to aggregate nodes into coalition is set up using iterative aid requests, and closed-loop performance evaluations w.r.t.\ a given threshold to handle the merging. The distributed computation of the PageRank is performed using the algorithm \cite{ishii_DistributedRandomizedAlgorithms_2010}. This coalitional predictive control strategy is deployed over the 16 water tanks system \cite{maestre_AssessmentCoalitionalControl_2015}, and compared against CMPC, Dec-MPC, and the DMPC scheme \cite{nunez_TimevaryingSchemeNoncentralized_2015}. The strategy proposed in  \cite{maestre_PageRankBasedCoalitional_2017a} is the best performer in terms of optimality gap w.r.t.\ CMPC, after parameters calibration.

A combination of the methodologies \cite{fele_CoalitionalControlCooperative_2017b} and \cite{maestre_PageRankBasedCoalitional_2017a} is found in \cite{muros_GameTheoreticalRandomized_2018b} where a randomized method for the estimation of the Shapley value is applied. Specifically, the Shapley value defined as the vector $\phi(\mathcal{N},v)$ $\forall i \in \mathcal{N}$, for the game induced over the set of agents $\mathcal{N}$ and for a characteristic function $v$ (coinciding with the stage cost of the local MPC), is used to introduce a weighting of the links among agents, which is defined for the undirected link $ij\in\mathcal{E}$ as $w_{ij} = \phi_i(\mathcal{N},v)/|\mathcal{E}_i| + \phi_j(\mathcal{N},v)/|\mathcal{E}_j|$. To address the problem of the combinatorial explosion associated with the computation of the Shapley value associated with all possible coalitions, randomized methods \cite{castro_PolynomialCalculationShapley_2009,ishii_DistributedRandomizedAlgorithms_2010} are proposed to estimate it. In particular, using the modified definition of the Shapley value given in \cite{castro_PolynomialCalculationShapley_2009}, an estimation of its value is given for a set of $q$ samples of all possible coalitions, giving an approximation of the value, whose efficient estimate is distributed as $\tilde{\phi}_i(\mathcal{N},v)\sim N(\phi_i,\sigma^2_{\phi_i}/q)$, with bounded error. Further details about the algorithmic partitioning approach are given in \cite{muros_GameTheoreticalRandomized_2018b}. The partitioning methodology is validated over the Barcelona drinking water transport network by applying coalitional predictive control and comparing it against CMPC, showing how it can outperform Dec-MPC and other decentralized control architectures.

}

\subsection{Technical extensions: feasibility, stability, robustness}

Theorems for the stability and recursive feasibility \cite{mayne_ConstrainedModelPredictive_2000} of a coalitional predictive control formulation have been proposed in \cite{baldiviesomonasterios_CoalitionalPredictiveControl_2021}. In \cite{baldiviesomonasterios_CoalitionalPredictiveControl_2021}, a DMPC technique \cite{mayne_RobustModelPredictive_2005a} relying on tube-based MPC \cite{limon_RobustTubebasedMPC_2010} is considered as the underlying control strategy for each coalition. Then, the aggregation of coalitions is achieved through a consensus procedure, where for each coalition $\mathcal{C}_i$ in a given state $x$ a consensus optimization problem is defined as:
\begin{align}
	\min_{\mathcal{C}_i\in\mathcal{M}} & J_i(\mathcal{C}_i,\mathcal{C}_{-i},x) = J_i^{\textrm{consensus}}(\mathcal{C}_i,\mathcal{C}_{-i},x) + \rho J_i^{\textrm{power}}(\mathcal{C}_i,x)
\end{align}
and $\mathcal{C}_{-i} \triangleq \{\mathcal{C}_{j}\}_{j\in\mathcal{N}_i}$ is the set of possible neighboring coalitions. In this optimization problem, the term $J_i^{\textrm{consensus}} = 0$ if coalitions $\mathcal{C}_i$ and its neighbors agree on the current arrangement into coalitions, and the term $J_i^{\textrm{power}}$ weighted by the scalar $\rho$ represents the effect of coalition $\mathcal{C}_i$ on neighbors opinions \cite{muros_BanzhafValueDesign_2017}. The consensus optimization is achieved through an algorithm that leverages the theory of finite exact potential games \cite{monderer_PotentialGames_1996}.
The approach is successfully validated against CMPC over a four-agent mass-spring-damper planar chain, showing that the coalitional control scheme proposed can reach states that are otherwise not feasible for CMPC. 

Another extension is found in \cite{chanfreut_DistributedModelPredictive_2022}, where tracking of target sets is achieved. The technique is developed by deploying a tube-based MPC formulation for each coalition, thus obtaining a robust formulation of local controllers. In \cite{chanfreut_DistributedModelPredictive_2022}, coalitional control is used in combination with Dec-MPC. Coalitions are formed to enlarge the domain of attraction of MPC, but when sufficient, the decentralized formulation is used. The underlying partitioning strategy is hierarchical, where partitioning is executed at a slower time scale over a heuristic selection of possible communication topologies. The underlying coalitional scheme is defined by \cite{maestre_CoalitionalControlScheme_2014b}. In particular, given the set $\mathcal{M}$  of all possible communication topologies, and for a partitioning $\mathcal{P}\in\mathcal{M}$, the characteristic is defined as:
\begin{equation}
	V(\mathcal{P},x_{\mathcal{P}}) = (x_{\mathcal{P}} - x_{\Gamma})^{\intercal} P_{\mathcal{P}} (x_{\mathcal{P}} - x_{\Gamma}) + c|\mathcal{E}_{\mathcal{P}}|
\end{equation} 
where $x_{\mathcal{P}}$ is the aggregated state of all the coalitions at time step $k$, $x_{\Gamma}$ is the Chebyshev center of the target set, $|\mathcal{E}_{\mathcal{P}}|$ is the number of communication links enabled in the partitioning $\mathcal{P}$, $c>0$ is a scalar, and $P_{\mathcal{P}}$ is a positive definite matrix. Further details about the methodology are given in \cite{maestre_CoalitionalControlScheme_2014b}.
The approach proposed is validated over a 12-tracks system connected through springs and dumpers; an example also used in \cite{riverso_TubebasedDistributedControl_2012,trodden_DistributedPredictiveControl_2017}, showing a good performance retention w.r.t.\ centralized control with significant reductions in communication costs. 

\subsection{Market-based partitioning}

A market-based coalition formation approach applied to coalitional predictive control is introduced in \cite{masero_MarketbasedClusteringModel_2022b}. The approach optimizes the heat transfer fluid (HTF) flow in a parabolic-trough solar collector field. The strategy is inspired by other market-based approaches proposed for energy networks \cite{son_ShorttermElectricityMarket_2004}, and results in a hierarchical coalitional control strategy. A parabolic-trough solar collector field is a system composed of many loops of parabolic collectors focusing heat on a trough flowed by the HTF. This fluid is thus heated and can be used for electrical energy generation. The objective of a control strategy applied to this system is to maximize the thermal power output by regulating the flow $q$ of the HTF across the loops, where the dynamics of the plant is nonlinear and subject to disturbances caused, e.g.\, by the variability in atmospheric conditions. The objective function $J$ of the plant is a quadratic sum of the power output and $q$ contributions for each loop, where $q$ is the control variable. The market-based coalitional strategy is implemented by defining for each agent $i\in\mathcal{N}$ in the plant, i.e.\ the individual loop, a utility value $U_i(\cdot) = - J_i(\cdot)$ that the agent $i$ can supply or demand to purchase or sell a quantum of flow $\Delta q$. Accordingly, the set of agents is split into two disjoint subsets $\mathcal{L}_\textrm{s}$, $\mathcal{L}_\textrm{d}$ of supply and demand loops, with respective utilities. Then, the utility is computed and classified according to the two groups for each agent or coalition. This way, the requests can be sorted in descending and ascending order for demand and supply, and trades are performed according to this matching. The utility gain of each agent participating in a specific trade is equal to the difference between the demand and supply utilities divided by two. The hierarchical coalition formation procedure is then implemented starting from the coalition formed by individual agents, and runs periodically according to a fixed time step bigger than the control step. Heuristics ensure the terminability of the algorithm.
The overall control strategy is validated on the model of the real-world collector field ACUREX, located in Plataforma Solar de Almería \cite{galvez-carrillo_NonlinearPredictiveControl_2009, gallego_AdaptativeStatespaceModel_2012} composed by 10 loops, and its scaling to 100 loops. Comparison strategies include PI control, two different CMPC strategies,
and the control strategy based on loop-pair clustering devised in \cite{masero_LightClusteringModel_2021}. According to the simulation results, the market-based coalitional predictive control is the best-performing strategy with a gain of $12.51\%$ w.r.t.\ PI control, outperforming also the CMPC implementation with $0.37\%$. Additionally, an analysis of the computational burden is performed. In practice, the CMPC strategy is not deployable because its computation time exceeds the operating time step of the plant. On the contrary, market-based coalitional control is fast enough to be potentially scaled up to a plant of about 300 loops while maintaining the same performance gains.

Feedforward Neural Networks (NNs) \cite{fine_FeedforwardNeuralNetwork_2006} are used in \cite{masero_FastImplementationCoalitional_2023} to reduce the computational complexity of the coalition control algorithm. The market-based hierarchical formulation introduced in \cite{masero_MarketbasedClusteringModel_2022b} is considered as reference strategy, and the same application to parabolic-trough solar collector field is considered. Specifically, in \cite{masero_FastImplementationCoalitional_2023} sets of NNs are used with two different scopes in cascade. The first set of NNs uses information about states and disturbances to approximate the values of the utilities of supply and demand agents. These are used to implement the market-based coalition formation. Then, a second set of NNs, using the same information and considering the coalition obtained, approximate the value of the HTF flow for the coalitions, that can group at most three loops. 
The coalitional controller is validated on the parabolic-trough solar collector field case study ACUREX. The results are compared with the nonlinear coalitional controller developed in \cite{masero_MarketbasedClusteringModel_2022b}. The NN-based coalitional controller \cite{masero_FastImplementationCoalitional_2023} shows a performance that is comparable with the one obtained in the nonlinear implementation \cite{masero_MarketbasedClusteringModel_2022b}, but providing a considerable reduction in the computation time needed to compute the control action and the partitioning of the network with a reduction up to the 99$\%$ w.r.t.\ the time required in the NLin-MPC implementation. The drawbacks of this strategy arise from the defining technical characteristics of NNs, which include the necessity of rich enough data to perform the training, the inability to provide suitable outputs when the operating conditions of the plant are distant from the training set, and the lack of guarantees for constraint satisfaction. 

\subsection{Further extensions}

Predicted topology transition is proposed in \cite{masero_RobustCoalitionalModel_2021} as an evolution of the work in \cite{fele_CoalitionalControlCooperative_2017b}. The method extends Coal-MPC by incorporating a transition horizon variable, which optimizes the timing of topology changes over the prediction horizon. Unlike previous coalitional control methods that switch coalition structures instantaneously, this approach gradually transitions between topologies, allowing agents to anticipate and optimize their control actions accordingly. The strategy also belongs to hierarchical coalitional control, where the upper layer, working at a lower rate, is designed to obtain the desired coalition and the transition horizon. This problem is formally a mixed-integer optimization program for which the solution space is reduced by defining a heuristic on the possible topology. 
The strategy in \cite{masero_RobustCoalitionalModel_2021} is applied for the control of an eight-tanks water system 
showing how the proposed approach can reduce communication and coordination costs of coalitional schemes while maintaining performance close to CMPC. 

{\CBb 
Pairwise clustering is proposed in \cite{masero_LightClusteringModel_2021} where the HTF flow coalitional predictive control for the parabolic-trough solar collector field ACUREX is considered. The control approach is hierarchical, and in the upper layer the partitioning of the network occurs at each time step. Specifically, at each time step $k$, the measurement of the flow rate in all loops at the previous time step is collected into a vector $\boldsymbol{q}^{\textrm{measured}}_{k-1}$. This vector is then sorted in ascending order, giving $\boldsymbol{q}^{\textrm{sorted}}_{k-1}$. Then, the partitioning of the plant is obtained by coupling together the first and last elements of  $\boldsymbol{q}^{\textrm{sorted}}_{k-1}$ and removing them from the vector until no further assignments are possible. This direct approach to partitioning is motivated by the fact that lops with a deficit of flow rate can benefit from those with excess flow. The approach has been proven to outperform Dec-MPC, approaching CMPC performance while significantly reducing computation time.  
}

The problem of resource sharing under partitioning is addressed in \cite{sanchez-amores_CoalitionalModelPredictive_2023}, which considers as case study the parabolic-trough solar collector field ACUREX. For this plant, a partition partitioning $\mathcal{P} = \{\mathcal{C}_1,\ldots,\mathcal{C}_{N_{\mathcal{P}}}\}$ is assumed to be given, e.g.\ using one of the techniques in \cite{masero_MarketbasedClusteringModel_2022b, masero_FastImplementationCoalitional_2023, masero_LightClusteringModel_2021}. Then, it is necessary to define the allocation of shared resource, which, for the specific case study, is total HTF flow. The problem of distributing the shared resource is solved using a population-dynamics-assisted resource allocation strategy \cite{barreiro-gomez_ConstrainedEvolutionaryGames_2018, martinez-piazuelo_NashEquilibriumSeeking_2022}, specifically a Smith population dynamics with carrying capacities \cite{barreiro-gomez_DistributedRobustPopulation_2018}. Following the hierarchical coalitional control methodology introduced in \cite{masero_MarketbasedClusteringModel_2022b}, the resource allocation (for a fixed partitioning) is performed at a slower time scale. The approach is validated on a 100-loop implementation of the parabolic-trough solar collector field ACUREX and compared with CMPC. The results show a negligible loss in performance while significantly reducing the computation time required to retrieve the control action.

\subsection{Partitioning for input-coupled systems} \label{subsec:coal-input-coupling}

The use of coalitional predictive control for systems with coupled input dynamics is found in \cite{masero_CoalitionalModelPredictive_2020a},  where the problem of controlling next-generation cellular networks \cite{auer_HowMuchEnergy_2012} is considered.
The dynamics of these system can be formulated as an input-coupled agent representation:
\begin{align}\label{eq:subsystem-input-coupled}
	& x_i(k+1) = A_{i}x_i(k) + \sum_{j\in\mathcal{C}_i}B_{ij}u_{ij}(k) + \omega_i(k)
\end{align}
where $u_{ij} = - u_{ji}$, and $\omega_i$ is a disturbance. The underlying coalitional formation approach is a hierarchical methodology of the form \cite{fele_CoalitionalControlCooperative_2017b}, where in the upper layer a new coalitional structure is assigned according to a fixed time step longer than the control sampling time. In this case, the computational complexity of evaluating the best topology is reduced by considering as candidate successors only the allocations $\mathcal{P}^{\textrm{next}}$ that have a Hamming distance of one from the current configuration $\mathcal{P}^{\textrm{current}}$, i.e.\ they differ from only one link allocation. The proposed strategy is applied to the case of a network of 37 base stations to optimize the number of served users and energy consumption. The approach is validated against the more traditional best-signal-level approach \cite{fletscher_EnergyawareResourceManagement_2019}, and decentralized and CMPC. Results show significant improvement of all the predictive control strategies w.r.t.\ the traditional approach, where coalitional control is the closest to CMPC in terms of performance while reducing the communication burden.

A further extension of coalitional predictive control for coupled input dynamics has been proposed in \cite{sanchez-amores_CoalitionalModelPredictive_2022}. In this work, couplings in the inputs among agents are decomposed into private and public variables, a feature detailed in \cite{labella_TwolayerModelPredictive_2019}.
This approach is used because it allows more flexibility in the computation of the control action w.r.t.\ robust approaches as tube-based MPC that is more conservative. The resulting architecture is validated using an eight-tank system coupled in the input, showing how varying implementation parameters allows to balance the communication burden with the performance loss.

An extension of \cite{sanchez-amores_CoalitionalModelPredictive_2022} is found in \cite{sanchez-amores_RobustCoalitionalModel_2023}, where a robust tube-based formulation of the controller is proposed \cite{mayne_RobustModelPredictive_2005a}.
Additionally, in \cite{sanchez-amores_RobustCoalitionalModel_2023} the presence of communication links is event-driven, i.e.\ communication links are activated only if scaling factors exceed predefined thresholds that allow to establish a trade-off between performances and communication burden. The approach is validated using an eight-tank water system against centralized and Dec-MPC. The simulation results show that coalitional control can outperform Dec-MPC while approaching CMPC performances with a reduction of 83$\%$ in terms of communication cost.

A further advancement in coalitional control for input-coupled dynamics is achieved in \cite{masero_RobustCoalitionalModel_2023}, where a robust strategy allowing plug-and-play capabilities is devised. The approach is based on an evolution of public and private factors introduced in \cite{trodden_DistributedPredictiveControl_2017} and already employed in \cite{sanchez-amores_CoalitionalModelPredictive_2022}.
Validation of the approach is performed through the control of a four-truck system in a coupled chain configuration as also tested in \cite{trodden_DistributedPredictiveControl_2017}. A fifth truck is added during the simulation to show the plug-and-play capabilities.

\subsection{Other applications}

In this section, we report applications of the coalitional predictive control schemes discussed above to case studies that have not been presented already, specifically: the control of irrigation canal, freeway transportation, vehicle platooning, and cyber-physical systems.

The first known contribution in coalitional predictive control is \cite{fele_CoalitionalModelPredictive_2014}, where the problem of controlling an irrigation canal is addressed. The aim of the strategy is to optimize water distribution by dynamically adjusting coalitions of control agents to balance control performance and communication cost. The framework is hierarchical: in the top layer, the partition of the system into coalition is achieved through topology optimization, where the optimal topology is selected from a predefined set of possible topologies. Decentralized feedback gains are associated with each topology, and the solution of an LMI problem guides the partition selection. Then, at a lower level, Dec-MPC is applied. Coalition formation and local optimization work at different time scales. The control methodology is validated through the SOBEK hydrodynamic simulator \cite{_SOBEKUserManual_2000} on a model of the Dez irrigation canal \cite{isapoor_DesigningEvaluatingControl_2011}, and compared against CMPC showing suboptimal but adequate performance, without the need of a complete communication topology.

A hierarchical formulation of coalitional predictive control has also been applied to nonlinear systems in \cite{chanfreut_CoalitionalModelPredictive_2021a}. In particular, this study focuses on freeway traffic control through ramp metering and variable speed limits \cite{papageorgiou_FreewayRampMetering_2002,papageorgiou_EffectsVariableSpeed_2008}. 
The solution proposed in \cite{chanfreut_CoalitionalModelPredictive_2021a} consists of a two-level structure: a top layer forms the coalitions, and at the bottom level, a DMPC strategy is deployed for the resulting coalitions, specifically feasible cooperation-based MPC \cite{venkat_DistributedMPCStrategies_2008b} with Genetic Algorithm solver (GA) \cite{goldberg_GeneticAlgorithmsSearch_1989}. Moreover, the two layers operate at different time scales, with the top one being slower, allowing more time to solve the coalition formation problem. 
The study proposes as a potential solution to the coalition formation the bargaining procedure based on the Shapley value \cite{fele_CoalitionalControlCooperative_2017b, muros_GameTheoreticalRandomized_2018b}, or the PageRank method \cite{maestre_PageRankBasedCoalitional_2017a}.
To simplify the problem, only a limited set of possible coalitions is considered. The approach is extensively validated against Dec-MPC, and feasible cooperation-based MPC on a 15 km freeway segment, with multiple ramps, and speed-limiting devices. The results show a reduction in communication and coordination costs.

An application of coalitional predictive control to cyber-physical systems \cite{lee_PresentFutureCyberphysical_2015,ding_SecureStateEstimation_2021} with chain architecture is proposed in \cite{maxim_CoalitionalDistributedModel_2021}. The key feature of this architecture is that the system first operates according to the non-cooperative DMPC strategy \cite{scattolini_ArchitecturesDistributedHierarchical_2009}, and when the feasibility of the solution fails, the system will switch to the coalitional predictive control formulation \cite{maxim_RobustCoalitionalDistributed_2018}. The switch occurs in cascade, triggered by one agent and propagating to its neighbors. Here, coalition formation is purely aggregative.
The procedure is applied to a four-agents system, showing that when the local feasibility of non-cooperative DMPC is lost, then the application of coalitional predictive control can still provide satisfactory performance.

Vehicle platooning is the application considered in \cite{maxim_CoalitionalDistributedModel_2022} for the robust coalitional control strategy of \cite{maxim_CoalitionalDistributedModel_2021}.
The approach is tested on a four-car platoon detailed in \cite{zhu_LMIbasedSynthesisStringstable_2020}, and string stability analysis \cite{dunbar_DistributedRecedingHorizon_2012} is performed. The simulation shows how dynamic coalition formation stabilizes the platoon's operation with reduced communication.
The work \cite{maxim_CoalitionalDistributedModel_2024} is proposed as an alternative approach to \cite{maxim_CoalitionalDistributedModel_2022, maxim_CoalitionalDistributedModel_2021} for coalitional control of vehicle platoons, distinguishing itself by the ability of individual agents to aggregate into coalitions autonomously. This objective is achieved by periodical evaluation of the string stability index \cite{dunbar_DistributedRecedingHorizon_2012}.
The approach is validated on a four identical vehicles platoon under three different testing conditions. The results show that an inversely proportional relationship exists between performance and string stability.

An eight-tank process is used as a case study to perform a comparative performance analysis between DMPC and coalitional control in \cite{maxim_DistributedModelPredictive_2023}.
In the paper, two non-cooperative DMPC formulations, one using a state-space model and the other an input-output model, are used to validate the performance of the coalitional control strategy based on a matrix gain feedback controller obtained through a gradient-based optimization previously introduced in \cite{maxim_AssessmentComputationMethods_2022}. The Coal-MPC methodology allows the switch between decentralized and distributed communication topologies according to performance satisfaction. This switching Coal-MPC method shows results that are comparable with the non-cooperative DMPC strategy while allowing for a reduction in the communication burden.

}
{\CBa
\section{Heuristic Partitioning} \label{sec:heuristic}
Partitioning for Dec-MPC of wide-area power systems is investigated in \cite{jain_DistributedWideareaControl_2018}. The technique is heuristic and based on the use of the modal participation matrix that highlights the effects of each generation on each dominant mode in low-frequency oscillations. 
The partitioning technique allows overlapping partitioning, giving both the nature of the dynamical couplings and the use of a DMPC strategy \cite{alessio_DecentralizedModelPredictive_2007a}. The approach is applied to the Northeast Power Coordinating Council nonlinear power system model \cite{rogers_PowerSystemOscillations_2000}, comprehending 48 electrical machines and 140 buses, showing the performance and the resilience of the network for two different partitionings compared to centralized control.

Partitioning for wind farms is proposed in \cite{ye_HierarchicalModelPredictive_2019}, where a HMPC strategy is proposed. The partitioning strategy is performed on the highest level of the hierarchy every $15$ minutes. Based on a forecast of the wind characteristics for the next $20$ minutes, an optimization strategy is deployed to cluster the wind turbines in one of 12 categories based on the possible load operating conditions the turbines can experience. The proposed HMPC strategy was validated over a modified version of the IEEE One Area RTS-96 network \cite{grigg_IEEEReliabilityTest_1999}, and compared with conventional dispatch and schedule allocation algorithms, achieving significantly better performance.

A strategy for partitioning vehicles platoons is implemented in \cite{liu_DistributedModelPredictive_2019a}, with the objective of deploying a noniterative two-level DMPC architecture ensuring closed-loop stability for an optimization problem with coupled cost functions and constraints. The partitioning strategy is based on the assumption that the cooperation set of vehicles $\mathcal{V}$ can be divided into groups that belong to two main conceptual categories, i.e.\ dominant and connecting clusters.
The algorithmic partitioning allows vehicles to perform the operations of joining and leaving a platoon on the basis of this group classification. The DMPC strategy is then designed around this partitioning approach, ensuring stability and feasibility. Validation of the approach is performed for a platoon of four vehicles, and compared against CMPC, showing minimal loss in performance.   

A strategy for event-triggered partitioning of microgrids is developed in \cite{ananduta_EventtriggeredPartitioningNoncentralized_2021}, where the economic dispatch problem for energy production is addressed. The power network is considered to be constituted of microgrids that are considered self-sufficient systems, i.e.\ they do not exchange energy with their neighbors in nominal operating conditions. However, if this generative autonomy is lost, re-partitioning of the network is triggered, leading to a new definition of the microgrids. This re-partitioning is performed through a communication protocol, which evaluates the best node exchange among the microgrids that minimizes the individual outcomes in economic terms, while ensuring self-sufficiency.
The approach is validated on the PG\&E 69-bus distribution network.
The simulation results show that during peak hours all microgrids should join into a single agent to satisfy the demand, whereas during off-peak hours they can split into multiple coalitions.  

The paper \cite{huanca_DesignDistributedSwitching_2023a} proposes a distributed Switching Model Predictive Control (SMPC) strategy for quadrotor UAV swarm aggregation incorporating collision avoidance. 
Teams of UAVs are selected using a clustering strategy, and local controllers solve the SMPC problem sequentially \cite{christofides_DistributedModelPredictive_2013b}. The clustering approach is based on the sphere packing problem \cite{conway_SpherePackingsLattices_1988}. A cluster of UAVs is selected according to the positions of UAVs in space \cite{gauci_SelforganizedAggregationComputation_2014}, assuming these are always available. In the sphere packing problem, the objective is to find the arrangement of non-overlapping spheres so that they occupy the largest possible fraction of space. Solutions are available in the literature for this problem \cite{conway_SpherePackingsLattices_1988}.
The approach is validated with a group of 150 UAVs, using both centralized and distributed control strategies for the aggregation. The proposed distributed SMPC can achieve comparable aggregation performance w.r.t.\ its centralized counterpart while drastically reducing computation time. 

}
\section{Applications and Case Studies}
\label{sec:applications}

{
In this section, we report a classification that relates the partitioning methodologies found in the literature to the systems used for their validation. In the resulting Tab.\ \ref{tab:applications}, for each application system, partitioning methodologies are classified according to Fig.\ \ref{fig:classification-partitioning}. When possible, we also provide references to more standardized test cases for their direct use in the development of further strategies.

From Tab.\ \ref{tab:applications}, we note that many works have been developed for power systems. However, if we consider standard generation and transmission systems, no specific case study has been consistently used to derive partitioning techniques. Therefore, it is difficult to quantitatively compare different works. An exception in this sector is the parabolic-trough plant ACUREX \cite{galvez-carrillo_NonlinearPredictiveControl_2009, gallego_AdaptativeStatespaceModel_2012}, for which many Coal-MPC strategies have been developed. 

Water network control is another field that has seen extensive application of partitioning strategies. In this case, we distinguish between water-tank systems, which are usually small-scale test cases used to validate the viability of the approaches, and large-scale water distribution networks, among which the Barcelona drinking water transport network \cite{ocampo-martinez_ImprovingWaterManagement_2009} is surely the most commonly used test case among different NC-MPC approaches. 

Chemical systems have been the subject of deep studies regarding partitioning, given the complexity of the associated dynamics. We report the presence of many system configurations involving CSTRs and separators, e.g.\ \cite{venkat_DistributedModelPredictive_2006, bakule_DecentralizedControlOverview_2008a,liu_DistributedModelPredictive_2009, stewart_CooperativeDistributedModel_2010a} among others. Also, in this case, no single benchmark system has been used consistently in the literature; rather, there are many different similar configurations that complicate the process of direct comparison of partitioning strategies.

For wind farms, we also report the presence of different studies in partitioning, with various topologies, turbine models, and operating conditions. 

Several other applications are reported in Tab.\ \ref{tab:applications}, all used in the development of a specific partitioning methodology for the application of non-centralized control. Especially for transportation networks, we observe a notable lack of studies in partitioning for NCen-MPC of urban traffic, freeway transportation, and railway networks \cite{luan_DecompositionDistributedOptimization_2020a}. The other case studies are, in general, smaller systems that can be used for the development of strategies, but do not stress the scalability of the approaches.

Several other large-scale application fields can be considered for studies in partitioning, such as swarms of mobile robots or autonomous maritime vehicles \cite{zhou_UAVSwarmIntelligence_2020}, automated agricultural systems, district heating \cite{blizard_OptimalityLossMinimization_2025}, satellite constellations \cite{curzi_LargeConstellationsSmall_2020}, and advanced industrial processes \cite{galloway_IntroductionIndustrialControl_2013}. Some of the applications listed can be found, for example, in the recent work \cite{pedroso_DistributedDesignUltra_2025a} about the design of large-scale systems, or in the set of benchmarks proposed in \cite{maestre_ControlSystemsBenchmarks_2025}. 

}

{
\onecolumn
\begin{longtable}{ll|l}
	\caption{Application fields of the partitioning techniques for NCen-MPC, classified by sector. When available, benchmark systems have been reported.
    } 
	\label{tab:applications}\\
	\toprule[1.5pt]
	\multirow{1}{*}{\textbf{Sector}} & \textbf{Specific application }& \textbf{Partitioning techniques }\\
	\midrule[1.5pt]
	\multirow{9}{*}{Power systems} 
	& Six-area power system & \cite{chen_CooperativeDistributedModel_2020b} \\  \cmidrule{2-3}
	& Smartgrids, 8 (check) prosumers: \cite{larsen_PowerSupplyDemand_2014,paauw_EnergyPatternGenerator_2009} & \cite{fele_CoalitionalControlCooperative_2017b} \\  \cmidrule{2-3} 
	& Wide area power network: \cite{chakrabortty_IntroductionWideareaControl_2013} & \cite{fele_CoalitionalControlSelforganizing_2018} \\  \cmidrule{2-3} 
	& The EEA-ENB:  \cite{riccardi_BenchmarkApplicationDistributed_2025a,riccardi_CodeUnderlyingPublication_2024d} & \cite{riccardi_GeneralizedPartitioningStrategy_2024a,riccardi_CodeUnderlyingPublication_2024c} \\  \cmidrule{2-3}
	& PG\&E 69-bus distribution network & \cite{ananduta_EventtriggeredPartitioningNoncentralized_2021} \\  \cmidrule{2-3}
	& IEEE 118-bus & \cite{labella_SupervisedModelPredictive_2022} \\ \cmidrule{2-3}
	& IEEE 123 node test feeder & \cite{wang_MPCbasedDecentralizedVoltage_2022} \\ \cmidrule{2-3}
	& Nonlinear power system: \cite{rogers_PowerSystemOscillations_2000}, 48 machines, 140 buses & \cite{jain_DistributedWideareaControl_2018} \\ \cmidrule{2-3}
	& Parabolic-trough plant: ACUREX model, 100 loops \cite{galvez-carrillo_NonlinearPredictiveControl_2009, gallego_AdaptativeStatespaceModel_2012} & \cite{masero_MarketbasedClusteringModel_2022b,masero_FastImplementationCoalitional_2023,masero_LightClusteringModel_2021,sanchez-amores_CoalitionalModelPredictive_2023,chanfreut_ClusteringbasedModelPredictive_2023} \\  \midrule 
	\multirow{7}{*}{Water systems} & 4-tanks system: \cite{alvarado_ComparativeAnalysisDistributed_2011} & \cite{wei_EventtriggeredDistributedModel_2020,segovia_DistributedModelPredictive_2021}  \\ \cmidrule{2-3}
	& 8-tanks system & \cite{masero_RobustCoalitionalModel_2021,sanchez-amores_CoalitionalModelPredictive_2022,sanchez-amores_RobustCoalitionalModel_2023,maxim_DistributedModelPredictive_2023}  \\ \cmidrule{2-3}
	& 16-tanks system: \cite{maestre_AssessmentCoalitionalControl_2015} & \cite{nunez_TimevaryingSchemeNoncentralized_2015,maestre_PageRankBasedCoalitional_2017a}   \\  \cmidrule{2-3}
	& Barcelona drinking water transport network: \cite{ocampo-martinez_ImprovingWaterManagement_2009} & \cite{barreiro-gomez_TimevaryingPartitioningPredictive_2019,ocampo-martinez_PartitioningApproachOriented_2011,ocampo-martinez_HierarchicalDecentralisedModel_2012,segovia_DistributedModelPredictive_2021,muros_GameTheoreticalRandomized_2018b} \\  \cmidrule{2-3}
	& Shanghai water distribution network & \cite{zhang_EnhancingCooperativeDistributed_2019} \\  \cmidrule{2-3}
	& Richmond water distribution network: \cite{vanzyl_OperationalOptimizationWater_2004} & \cite{arastou_OptimizationbasedNetworkPartitioning_2025} \\  \cmidrule{2-3}
	& Dez irrigation canal: \cite{isapoor_DesigningEvaluatingControl_2011,_SOBEKUserManual_2000} & \cite{fele_CoalitionalModelPredictive_2014} \\
	\midrule
	\multirow{6}{*}{Chemical systems} & 2 CSTR series: \cite{venkat_DistributedModelPredictive_2006, bakule_DecentralizedControlOverview_2008a} & \cite{kamelian_NovelGraphbasedPartitioning_2015,tang_RelativeTimeaveragedGain_2018,he_EnhancingTopologicalInformation_2023}  \\  \cmidrule{2-3}
	& 2 CSTR series and flash tank separator: \cite{liu_DistributedModelPredictive_2009, stewart_CooperativeDistributedModel_2010a,liu_TwotierControlArchitecture_2010,christofides_DistributedModelPredictive_2013b} & \cite{pourkargar_ImpactDecompositionDistributed_2017d,rocha_PartitioningDistributedModel_2018,tang_OptimalDecompositionDistributed_2018,wang_DistributedModelPredictive_2023} \\  \cmidrule{2-3}
	& Tennessee Eastman problem: \cite{downs_PlantwideIndustrialProcess_1993, lyman_PlantwideControlTennessee_1995}, five operation units & \cite{xie_GABasedDecomposition_2016b} \\  \cmidrule{2-3}
	& Benzene alkylation process: 4 CSTR and flash tank separator & \cite{pourkargar_DistributedEstimationNonlinear_2019,arastou_OptimizationbasedNetworkPartitioning_2025} \\  \cmidrule{2-3} 
	& Amine gas sweetening plant & \cite{moharir_DistributedModelPredictive_2018}  \\  \cmidrule{2-3} 
	& Air separation process & \cite{wang_DistributedModelPredictive_2023} \\
	\midrule
	\multirow{5}{*}{Wind farms} 
	& 12-turbine wind farm & \cite{zhao_OptimalSchedulingStrategy_2023} \\  \cmidrule{2-3} 
	& 20-turbine wind farm, NREL 5-MW & \cite{lin_HierarchicalClusteringbasedOptimization_2020} \\  \cmidrule{2-3} 
	& 25-turbine farm, 1.5 MW & \cite{changqing_FrequencyRegulationControl_2022} \\  \cmidrule{2-3} 
	& 42-turbine farm, NREL-5MW: \cite{jonkman_Definition5MWReference_2009a}, SimWindFarm \cite{grunnet_AeolusToolboxDynamics_2010} & \cite{siniscalchi-minna_NoncentralizedPredictiveControl_2020} \\  \cmidrule{2-3} 
	& IEEE One Area RTS-96 network: \cite{grigg_IEEEReliabilityTest_1999} & \cite{ye_HierarchicalModelPredictive_2019} \\  \midrule 
	\multirow{4}{*}{Transportation systems} 
	& 4-vehicles platoon: \cite{zhu_LMIbasedSynthesisStringstable_2020}& \cite{liu_DistributedModelPredictive_2019a,maxim_CoalitionalDistributedModel_2022} \\ \cmidrule{2-3}
	& Urban transportation network: \cite{deoliveira_MultiagentModelPredictive_2010}, 8 intersections  & \cite{chanfreut_FastClusteringMultiagent_2022} \\ \cmidrule{2-3}
	& Jinan road network & \cite{guo_DynamicIdentificationUrban_2019} \\ \cmidrule{2-3}
	& 15 km freeway stretch: \cite{messmer_METANETMacroscopicSimulation_1990}, METANET model   & \cite{chanfreut_CoalitionalModelPredictive_2021a}\\ \midrule
	\multirow{4}{*}{Mechanical systems} 
	& Mass-spring-damper chain, 4 elements  & \cite{baldiviesomonasterios_CoalitionalPredictiveControl_2021} \\ \cmidrule{2-3}
	& (4+1)-tracks, connected with springs and dumpers: \cite{trodden_DistributedPredictiveControl_2017} & \cite{masero_RobustCoalitionalModel_2023} \\ \cmidrule{2-3}
	& 12-tracks, connected with springs and dumpers: \cite{riverso_TubebasedDistributedControl_2012,trodden_DistributedPredictiveControl_2017} & \cite{chanfreut_DistributedModelPredictive_2022} \\ \midrule
	\multirow{2}{*}{Smart buildings} 
	& 8 rooms temperature regulation & \cite{zheng_CouplingDegreeClusteringbased_2018b} \\ \cmidrule{2-3}
	& 20 thermal zones control: \cite{chandan_OptimalPartitioningDecentralized_2013} & \cite{atam_OptimalPartitioningMultithermal_2021} \\ \midrule
	\multirow{2}{*}{Abstract networks}
	& 43 agents flow system: \cite{koeln_StabilityDecentralizedModel_2017} & \cite{wang_HierarchicalClusteringConstrained_2023} \\ \cmidrule{2-3} 
	& Random 50 systems, modular 64 systems, hybrid & \cite{riccardi_GeneralPartitioningStrategy,riccardi_CodePublicationGeneral_2025} \\ \midrule
	\multirow{1}{*}{Railway networks} & Dutch railway network: \cite{kersbergen_RailwayTrafficManagement_2016} & \cite{kersbergen_DistributedModelPredictive_2016} \\ \midrule 
	\multirow{1}{*}{Telecommunication systems} & Next generation cellular networks: \cite{auer_HowMuchEnergy_2012}& \cite{masero_CoalitionalModelPredictive_2020a} \\ \midrule 
	\multirow{1}{*}{Industrial plants} & Walking beam reheating furnace system & \cite{chen_CooperativeDistributedModel_2020b} \\ \midrule 
	\multirow{1}{*}{Process plant} &  Refinery: gas-to-liquid process, hydrocracking process& \cite{tang_AutomaticDecompositionLargescale_2023} \\ \midrule 
	\multirow{1}{*}{Aerial vehicles} & Group of 150 UAVs & \cite{huanca_DesignDistributedSwitching_2023a} \\ \midrule 
	\multirow{1}{*}{Cyber-physical systems} & 4-agents chain& \cite{maxim_CoalitionalDistributedModel_2021} \\ \bottomrule[1.5pt]
\end{longtable}
}
\twocolumn

\section{Conclusions and Future Work}
\label{sec:conclusions}

{
This survey presents the first systematic classification and in-depth analysis of partitioning techniques for non-centralized predictive control. The scope of this work is both to unify the approaches currently present in the literature under a single framework, and to lay solid methodological foundations for future developments.

{\CBf These objectives are achieved through the novel contributions of this work, which we summarize in the following. First, we introduce a formal reformulation of the partitioning problem in terms of mixed-integer programming, showing how, in the context of predictive control, the problem requires the solution of a bi-level optimization program, where network control performance is the cost functional of the partitioning problem. This aspect is at the basis of the complexity of network partitioning for control. Developing this framework, we introduce the concept of predictive partitioning, which uses predicted topology behavior to obtain the optimal network partitioning over the prediction horizon. Given the inherent NP-hard nature of these problems, their optimization-based solution would be prohibitive in real time; therefore, developing such a framework using greedy or heuristic algorithms or data-driven approaches would be advisable. Moreover, we introduce the concept of multi-topological network representations, which can serve as a basis for applying partitioning methodologies on networks whose topology and dynamical coupling are driven by different factors, such as events, time, network dynamics, or stochastic phenomena. Additionally, we provide a systematization of the key performance indicators to assess the quality of a partitioning for network control. On this basis, we establish an evaluation methodology that allows the direct comparison of different partitioning strategies. Such an approach can be the basis of further systematic development in this field, providing solid quantitative metrics for performance assessment. 

In addition, this survey proposes several other ways to analyze and organize the literature in partitioning for predictive control. We start by presenting a systematization of network equivalents based on graphs. Then we introduce a classification of the partitioning techniques based on five main classes: optimization-based, algorithmic, community-detection-based, game-theoretic-oriented, and heuristic partitioning. For each class we discuss its level of optimality, scalability, complexity of computation and implementation, technical requirements, and other specific features it might exert. Further we introduce a functional sub-classification of the partitioning techniques, introducing cross-methodological partitioning objectives. We conclude the survey by discussing the known applications of the partitioning techniques  proposing, when possible, reference systems for further developments and comparison.

}

Future work in the field of partitioning for non-centralized predictive control should focus on the following areas. First, there is space for further practical and theoretical developments in the time-varying (and hierarchical) partitioning approaches, especially considering predictive partitioning. Methodological approaches in this direction should also explore the use of data-driven, evolutionary, or reinforcement learning techniques to obtain the partition, which are strategies rarely deployed so far. Considering instead the theoretical developments, only the framework of coalitional control currently offers solid guarantees of satisfying the properties of feasibility, stability, and robustness when partitioning is involved, with few studies addressing these issues in general. Therefore, such properties might be established for time-varying partitioning approaches under different non-centralized control frameworks. The use of unified evaluation metrics should be extended to allow for cross-disciplinary and cross-methodological evaluation, with centralized model predictive control as a reference strategy to always include in the study. Finally, we stress that there is a limited selection of large-scale application benchmarks, which include at least 10 000 agents with different connection topologies for the validation and scalability assessment of current strategies. Future work should focus on addressing the indicated aspects to reach a level of sophistication for the partitioning strategies such that they can adapt online to topological changes while ensuring the stability of the network, the feasibility of the control actions, robustness with respect to unexpected events, and minimal losses in terms of global optimality.

}

\bibliographystyle{elsarticle-num} 
{\footnotesize \bibliography{bibliography.bib}}
\appendix

\section{Analytical Classification Table}

In the following table Tab.\ \ref{tab:analytical classification table}, we report the references presenting the partitioning strategies that have been investigated throughout the survey. They are listed in chronological order, which allows us to further understand the order of development of the techniques. Additionally, we report the control methodology that has been deployed in the study, essential details about the partitioning method developed, and the application considered for the validation of the overall architecture.

\onecolumn
\begin{longtable}{p{.05\linewidth}|p{.05\linewidth}p{.15\linewidth}p{.30\linewidth}p{.30\linewidth}}
	\caption{Analytical Classification Table} 
	\label{tab:analytical classification table}\\
	\toprule[1.5pt]
	\textbf{Work} & \textbf{Year} & \textbf{Control Method} & \textbf{Partitioning method} & \textbf{Application} \\
	\midrule[1.5pt]
\cite{ocampo-martinez_PartitioningApproachOriented_2011} & 2011 & H-Dec-MPC & Graph-partitioning-based ordering algorithm (GPB) & Barcelona DWN \\ \hline
\cite{ocampo-martinez_HierarchicalDecentralisedModel_2012} & 2012 & H-Dec-MPC & Nested epsilon decomposition & Barcelona DWN \\ \hline
\cite{fele_CoalitionalModelPredictive_2014} & 2014 & H-Coal-MPC & Coalition formation based on topology optimization from a predefined set & Irrigation canal networks \\ \hline
\cite{nunez_TimevaryingSchemeNoncentralized_2015} & 2015 & Dec-, D-, and H- MPC & MI optimization partitioning & 16 tanks water system \\ \hline
\cite{kamelian_NovelGraphbasedPartitioning_2015} & 2015 & Dec-NLin-MPC & Algorithmic partitioning & Two-reactor (CSTR) chain followed by a flash separator with recycle \\ \hline
\cite{xie_GABasedDecomposition_2016b} & 2016 & DMPC & Genetic algorithm minimization of input-output coupling between subsystems & Chemical plant: Tennessee Eastman problem. Five operation units:  a reactor, a condenser, a compressor, a separator, and a stripper. \\ \hline
\cite{kersbergen_DistributedModelPredictive_2016} & 2016 & DMPC & MIQP optimization for constraints decomposition & Dutch railway network \\ \hline
\cite{pourkargar_ImpactDecompositionDistributed_2017d} & 2017 & CMPC, iterative and sequential DMPC & Community detection through modularity maximization & Reactor-separator process \\ \hline
\cite{fele_CoalitionalControlCooperative_2017b} & 2017 & Coal-MPC & Game theoretic coalition formation based on Shapley value & Smart grids \\ \hline
\cite{maestre_PageRankBasedCoalitional_2017a} & 2017 & Coal-MPC & Coalition formation based on an algorithm to handle aid requests sorted using distributed PageRank & 16 tanks water system \\ \hline
\cite{fele_CoalitionalControlSelforganizing_2018} & 2018 & Coal-MPC & Coalition formation based on bargaining procedure and TU-games & Wide-area control of power grids \\ \hline
\cite{zheng_CouplingDegreeClusteringbased_2018b} & 2018 & Dual mode DMPC & Algorithmic partitioning based on coupling degree & Building thermal management: eight rooms \\ \hline
\cite{tang_RelativeTimeaveragedGain_2018} & 2018 & DMPC (noncooperative and iterative) & Relative Time-Averaged Gain Array (RTAGA)-based algorithmic modularity maximization over weighted IO bipartite graph using fast unfold & Reactor-separator process: 2CSTRs \\ \hline
\cite{tang_OptimalDecompositionDistributed_2018} & 2018 & DMPC-ADMM for nonlinear systems & Community-based decomposition of the optimization problem based on bipartite and unipartite representations, and fast unfold algorithm & Reactor-separator process: 2CSTRs \\ \hline
\cite{rocha_PartitioningDistributedModel_2018} & 2018 & Linearized cooperative and non-cooperative DMPC for nonlinear systems & Algorithmic partitioning based on variables matching and controllability check & Reactor-separator process: 2CSTRs \\ \hline
\cite{jain_DistributedWideareaControl_2018} & 2018 & Dec-MPC & Heuristic partitioning based on ad-hoc performance index (modal participation matrix) & Northeast Power Coordinating Council nonlinear power system model \\ \hline
\cite{moharir_DistributedModelPredictive_2018} & 2018 & DMPC (iterative) & Modularity-based partitioning (iterative division) & Amine gas sweetening plant \\ \hline
\cite{muros_GameTheoreticalRandomized_2018b} & 2018 & Coal-MPC & Coalition formation based on estimation of Shapley value and randomized methods & Barcelona DWN \\ \hline
\cite{zhang_EnhancingCooperativeDistributed_2019} & 2019 & Enhancing DMPC & Data-driven partitioning using $k$-Shape & Shanghai WDN \\ \hline
\cite{ye_HierarchicalModelPredictive_2019} & 2019 & HMPC & Heuristic partitioning (optimization-based) & Modified IEEE One Area RTS-96 network with wind turbines \\ \hline
\cite{liu_DistributedModelPredictive_2019a} & 2019 & HMPC & Heuristic partitioning (algorithmic based on dominant and connecting clusters) & Four vehicles platoon \\ \hline
\cite{pourkargar_DistributedEstimationNonlinear_2019} & 2019 & DMPC & Modularity-based partitioning (iterative division) & Benzene alkylation process: four continuous stirred-tank reactors, and a flash tank separator \\ \hline
\cite{barreiro-gomez_TimevaryingPartitioningPredictive_2019} & 2019 & DMPC based on density-dependent population games & Multiobjective optimization, computed through distributed algorithm for graph partitioning & Barcelona DWN \\ \hline
\cite{guo_DynamicIdentificationUrban_2019} & 2019 & DMPC for perimeter control & Modularity-based paritioning based on dynamic traffic estimation & Road network in downtown Jinan, China \\ \hline
\cite{chen_CooperativeDistributedModel_2020b} & 2020 & Cooperative DMPC, over a sequential hierarchical down-stream of solutions & Hierarchical interpretive structural modeling (ISM) & Walking beam reheating furnace system, six-area power system \\ \hline
\cite{wei_EventtriggeredDistributedModel_2020} & 2020 & DMPC (Cooperative) & Algorithmic partitioning based on treshold given by coupling sensitivity analysis & Four-tanks water systems \\ \hline
\cite{siniscalchi-minna_NoncentralizedPredictiveControl_2020} & 2020 & H-NCen-MPC & MIP optimization using ad hoc indicator (wake effect) & 42 turbines farm (NREL-5MW) \\ \hline
\cite{lin_HierarchicalClusteringbasedOptimization_2020} & 2020 & HMPC & Frequency-based fuzzy $c$-means algorithmic partitioning & 20 turbines farm (NREL-5MW) \\ \hline
\cite{masero_CoalitionalModelPredictive_2020a} & 2020 & Coal-MPC & Hierarchical time-varying & Next-generation cellular networks with 37 base stations \\ \hline
\cite{baldiviesomonasterios_CoalitionalPredictiveControl_2021} & 2021 & Coal-MPC & Coalition formation based on consensus optimization and potential games & Mass-spring-damper planar chain \\ \hline
\cite{chanfreut_CoalitionalModelPredictive_2021a} & 2021 & H-Coal-MPC & Coalition formation based on bargaining procedure and TU-games, or PageRank method & Freeway transportation network, METANET model \\ \hline
\cite{masero_RobustCoalitionalModel_2021} & 2021 & H-Coal-MPC & Coalition formation based TU-games, and mixed-integer selection of the coalitions with predicted topologies & Eight tanks water system \\ \hline
\cite{maxim_CoalitionalDistributedModel_2021} & 2021 & Coal-MPC and DMPC & Coalition formation based on cooperative game & Theoretical four agents chain system \\ \hline
\cite{masero_LightClusteringModel_2021} & 2021 & H-Coal-MPC & Loop-pair clustering & Parabolic-trough solar collector fields with 100 loops \\ \hline
\cite{segovia_DistributedModelPredictive_2021} & 2021 & DMPC based on optimality condition decomposition (OCD) & Modularity-based partitioning of the optimization problem & Quadruple-tank benchmark; Barcelona DWN \\ \hline
\cite{atam_OptimalPartitioningMultithermal_2021} & 2021 & Dec-MPC & MI optimization, robust and stochastic & 5 and 20 zones thermal buildings \\ \hline
\cite{ananduta_EventtriggeredPartitioningNoncentralized_2021} & 2021 & Dec-MPC for economic dispatch & Heuristic partitioning based on communication protocol (algorithmic) & PG\&E 69-bus distribution network \\ \hline
\cite{chanfreut_DistributedModelPredictive_2022} & 2022 & Coal-MPC and Dec-MPC & Coalition formation based on cooperative game and invariant sets & 12 tracks system \\ \hline
\cite{maxim_CoalitionalDistributedModel_2022} & 2022 & Coal-MPC and DMPC & Coalition formation based on cooperative game & Autonomous vehicle platooning \\ \hline
\cite{sanchez-amores_CoalitionalModelPredictive_2022} & 2022 & Coal-MPC & Coalition formation based on private and public factors & 8 tanks input-coupled water system \\ \hline
\cite{masero_MarketbasedClusteringModel_2022b} & 2022 & H-NLin-Coal-MPC & Market-based coalition formation strategy & Parabolic-trough solar collector fields with 100 loops \\ \hline
\cite{wang_MPCbasedDecentralizedVoltage_2022} & 2022 & Dec-MPC & Modularity-based partitioning using ad-hoc performance indicators & IEEE 123 node test feeder \\ \hline
\cite{labella_SupervisedModelPredictive_2022} & 2022 & HMPC & $k$-way partitioning using METIS on a flow graph & IEEE 118-bus \\ \hline
\cite{changqing_FrequencyRegulationControl_2022} & 2022 & HMPC & $k$-means clustering for wake-effect interaction minimization & 25 turbines farm (1.5 MW) \\ \hline
\cite{chanfreut_FastClusteringMultiagent_2022} & 2022 & Dec-MPC & Binary quadratic programming (BQP) & Urban traffic network with 8 intersections \\ \hline
\cite{chanfreut_ClusteringbasedModelPredictive_2023} & 2023 & DMPC, ADMM- or ALADIN-based & $k$-means clustering & Solar parabolic trough plants \\ \hline
\cite{he_EnhancingTopologicalInformation_2023} & 2023 & Lyapunov-based DMPC & Hierarchical multiway spectral community detection & Reactor-separator process \\ \hline
\cite{huanca_DesignDistributedSwitching_2023a} & 2023 & Distributed Switching MPC & Sphere packing clustering combined with MPC & Quadrotor UAV swarm control \\ \hline
\cite{masero_RobustCoalitionalModel_2023} & 2023 & H-Coal-MPC with PnP capabilities & Coalition formation based on invariant sets and dynamic scaling factors & 4 + 1 trucks system \\ \hline
\cite{masero_FastImplementationCoalitional_2023} & 2023 & H-NLin-Coal-MPC based on neural networks & Neural-networks-based market-based coalition formation strategy & Parabolic-trough solar collector fields with 100 loops \\ \hline
\cite{maxim_DistributedModelPredictive_2023} & 2023 & Coal-MPC with switching topologies & Coalition formation based on cooperative game & 8 tanks water system \\ \hline
\cite{sanchez-amores_RobustCoalitionalModel_2023} & 2023 & Coal-MPC & Coalition formation based on private and public factors & 8 tanks input-coupled water system \\ \hline
\cite{sanchez-amores_CoalitionalModelPredictive_2023} & 2023 & H-Coal-MPC & Arbitrary partitioning & Parabolic-trough solar collector fields with 100 loops \\ \hline
\cite{wang_DistributedModelPredictive_2023} & 2023 & DMPC & Modularity-based partitioning using frequency metric, and gap metric & Reactor separator process (2CSTR and a flash separator); and air separation process \\ \hline
\cite{changqing_FrequencyRegulationControl_2022} & 2023 & HMPC & $k$-means clustering (crowd search) using a set of key performance indicators & 12 turbines farm \\ \hline
\cite{wang_HierarchicalClusteringConstrained_2023} & 2023 & Dec-MPC & Agglomerative hierarchical clustering based on minimal robust positively invariant sets & 43 agents flow-based network \\ \hline
\cite{tang_AutomaticDecompositionLargescale_2023} & 2023 & DMPC & Modularity-based partitioning (iterative division) & Crude distillation process for a refinery, gas-to-liquid process, and a hydrocracking process \\ \hline
\cite{maxim_CoalitionalDistributedModel_2024} & 2024 & Coal-MPC with switching topologies & Coalition formation based on string stability condition & Autonomous vehicle platooning \\ \hline
\cite{arastou_OptimizationbasedNetworkPartitioning_2025} & 2025 & DMPC & Algorithmic (Kernighan-Lin) partitioning using computational complexity metric & Richmond water distribution network; Barcelona DWN \\ \hline
\cite{jogwar_DistributedControlArchitecture_2019} & 2025 & DMPC & Spectral community detection for modularity based on time-varying graph representation & Benzene alkylation process: 4CSTR, and a flash tank separator \\ \hline
\cite{riccardi_GeneralPartitioningStrategy} & 2025 & DMPC-ADMM for hybrid systems & Bi-level partitioning; algorithmic selection of system units, and algorithmic or optimization-based (BQP) partitioning; balancing intra- and inter-agent interactions, with granularity parameter & Modular network with 64 agents, random network of hybrid systems with 50 agents \\
	 \bottomrule[1.5pt]
\end{longtable}

\
\end{document}